\documentclass[11pt,reqno]{amsart}

\usepackage{amsmath,amssymb,amsthm,amsfonts,bm,amscd}
\usepackage{mathrsfs}
\usepackage{setspace}
\usepackage{verbatim}
\usepackage{enumerate}
\usepackage{cite}
\usepackage{epsfig}
\usepackage{subcaption}
\usepackage{color}
\usepackage{tikz}
\usetikzlibrary{shapes.misc}
\usetikzlibrary{arrows}

\usepackage{psfrag}

\usepackage[top=1in, bottom=0.7in, left=0.7in, right=0.7in]{geometry}

\newtheorem{thm}{Theorem}[section]

\newtheorem{lem}[thm]{Lemma}

\theoremstyle{definition}

\theoremstyle{remark}
\newtheorem{remark}{Remark}[section]
\newtheorem{prob}{Problem}[section]

\newcommand{\xh}{\hat{x}}

\newcommand{\Kxt}{\widetilde{K}_x}
\newcommand{\Psih}{\hat{\Psi}}
\newcommand{\Psit}{\widetilde{\Psi}}
\newcommand{\wt}{\widetilde{w}}
\newcommand{\xt}{\widetilde{x}}
\newcommand{\xit}{\widetilde{\xi}}
\newcommand{\Tr}{\mathrm{T}}

\newcommand{\RR}{{\mathbb R}}

\newcommand{\cP}{{\mathcal P}}

\newcommand{\ol}{\overline}

\newcommand\tqed{\leavevmode\unskip\penalty9999 \hbox{}\nobreak\hfill\quad\hbox{$\triangleleft$}}

\title{\Large \bf Adaptive Internal Models: Explaining the Oculomotor System and the Cerebellum}
\author{Mireille E. Broucke
\thanks{The author is with the Dept. of Electrical and Computer Engineering, University of Toronto, 
Canada (e-mail: broucke@control.utoronto.ca). Supported by the Natural Sciences and Engineering 
Research Council of Canada (NSERC). }%
}

\begin{document}

\begin{center}
{\Large \bf Adaptive Internal Models: Explaining the Oculomotor System and the Cerebellum} \\ ~ \\
{\large Mireille E. Broucke} \\
{Department of Electrical and Computer Engineering} \\
{University of Toronto} \\
\today 
\end{center}

\begin{abstract}
We propose a new model of the oculomotor system, particularly the vestibulo-ocular reflex, 
gaze fixation, and smooth pursuit. Our key insight is to exploit recent developments on adaptive 
internal models. The outcome is a simple model that includes the interactions between the brainstem 
and the cerebellum. In addition, we put forward a thesis that the cerebellum embodies internal models 
of all persistent, exogenous reference and disturbance signals acting on the body and observable 
through the error signals it receives. Our proposed architecture is compared to feedback error 
learning, a variant of the computed torque method in robotics. 
\end{abstract}

\section{Introduction}
\label{sec:intro}

This report presents a control-theoretic model of the oculomotor system, particularly the 
vestibulo-ocular reflex, gaze fixation, and the smooth pursuit system, and including the 
interactions between the brainstem and the cerebellum. Parts of this report have been published
in an IFAC conference paper \cite{IFAC20} and as an IEEE TAC paper \cite{MEBTAC20}. 

We show that developments on adaptive internal models \cite{MARINO11,NIKIFOROV98,SERRANI00,SERRANI01} 
provide a compelling framework to explain this system. We obtain a model that is simple yet 
is able to explain more behaviors than previously proposed models. In addition, we make 
a proposal about the function of the cerebellum. A computational model of the cerebellum is one of the 
great open problems of neuroscience today. Our model suggests that the cerebellum embodies adaptive 
internal models of persistent, exogenous disturbance and reference signals observable through the error 
signals arriving at the cerebellum. 

Control theory has been well accepted as a mathematical basis to explain motor control systems for many 
decades. A number of unified control-theoretic models of the oculomotor system, in particular, have 
been proposed in \cite{GLASAUER03,POLA02,ROBINSON86,YOUNG63,YASUI75,ZHANG01}, among others. However, 
these models are limited in the behaviors they capture, and further, certain behaviors such as the 
so-called predictive capability of the smooth pursuit system have not yet been fully characterized. 
Meanwhile, since the 1990's neuroscientists have explored internal models as a means to explain the 
function of the cerebellum. And there is mounting interest in the control community, as witnessed 
by a session on internal models in neuroscience in the 2018 IEEE Conference on Decision and Control. 
Despite mounting interest in both communities, a computational model of the cerebellum that includes 
the internal model principle has never been formalized, to date. In sum, to the best of our 
knowledge, we present here the first control theoretic model of the oculomotor system and the 
cerebellum that incorporates the internal model principle of control theory \cite{IMP75,IMP76}. 

The oculomotor system comprises several eye movement systems: the {\em vestibulo-ocular reflex} (VOR), 
{\em optokinetic reflex} (OKR), the {\em saccadic system}, the {\em gaze fixation system}, 
the {\em smooth pursuit system}, and the {\em vergence system}. The VOR serves to keep the gaze 
(sum of eye and head angles) stationary when the head is moving. The OKR reduces image motion 
across the retina when a large object or the entire visual surround is moving. The saccadic 
system provides rapid, discrete changes of eye position in order to place an object of interest 
on the fovea. The gaze fixation system stabilizes the gaze on a stationary object. The smooth 
pursuit system keeps a moving object centered on the fovea. The vergence system coordinates the
movement of the two eyes. 

The oculomotor system anatomy includes the oculomotor plant consisting of the eyeball, muscles 
moving the eye, and oculomotor neurons that stimulate the muscles; the brainstem which provides 
the main feedback loop by receiving the retinal and vestibular (from the semicircular canals of 
the ear) signals and issuing the oculomotor command to the eye muscles; and the cerebellum which 
regulates eye movements as a top up to the main control loop through the brainstem. 

The cerebellum is a purely feedforward, uniform, laminated brain structure that is divided into 
functional zones; e.g. locomotion, posture control, eye movement, arm movement, speech regulation, etc. 
Our concern here is with the vestibulocerebellum or floccular complex, which is responsible for 
regulating eye movements. Each cerebellar zone receives two types of inputs on {\em mossy fibers} 
and {\em climbing fibers}. The sole output of the cerebellum is through the {\em Purkinje cells} 
of the cerebellar cortex. The {\em cerebellar microcircuit}, consisting of Purkinje cells, basket cells, 
Gogli cells, granule cells, stellate cells, mossy fibers, climbing fibers, and parallel fibers, 
has been fully characterized \cite{ECCLES69}. 

This report focuses on the VOR, gaze holding, and smooth pursuit. For simplicity we consider 
only horizontal movement of a single eye; other aspects not covered by our model are discussed 
in Section~\ref{sec:discuss}. Next we highlight features of our model in addition to our use 
of the internal model principle. 

{\bf Error Signals.} \
Each of the eye movement systems has driving signals, signals required for computation of ongoing 
eye movement. Head velocity is a driving signal for the VOR. {\em Retinal error}, the difference 
between the target and fovea positions on the retina, drives the saccadic system \cite{POLA02}. 
{\em Retinal slip velocity}, the time derivative of retinal error, is often assumed to be the 
driving signal for the smooth pursuit system (despite the mathematical dilemma of how positional 
errors can be driven to zero using only velocity errors). It is known that in primates, the 
VOR, gaze holding, and smooth pursuit systems share the same neural pathways in the brainstem and 
cerebellum\footnote{When we use the term {\em cerebellum}, we refer more specifically to the 
floccular complex, comprising the flocculus and the ventral paraflocculus \cite{LISBERGER09}.}, 
so it is plausible these systems share certain driving signals \cite{BUTTNER84,LISBERGER15}. 
We assume that a common visual driving signal shared by the VOR, gaze holding, and smooth pursuit
systems is the retinal error. This signal is believed to arise in the superior colliculus of the 
brainstem \cite{BASSO00,GLASAUER03,KRAUZLIS97,KRAUZLIS04}. 

Evidence for the relevance of retinal error as a driving signal of the VOR, gaze holding, and smooth 
pursuit is reported in \cite{BERTHOZ88,EGGERS03,SHELHAMER95, SHELHAMER94,ZHOU01}. A series of studies 
by Pola and Wyatt \cite{POLA80,WYATTPOLA81, WYATT83} showed that retinal slip velocity is 
inadequate to explain all the behaviors of the smooth pursuit system. Other studies used 
strobe-reared cats, who never experience retinal slip velocity \cite{MANDL81,MANDL79}. 
Finally, direct experimental evidence that retinal errors drive the smooth pursuit system 
was given in \cite{BLOHM05}; they used a flashing visual target for which no velocity 
information could be perceived directly. 

{\bf Brainstem v.s. Cerebellum.} \
There has been considerable research both to understand how the VOR, OKR, gaze holding, and smooth 
pursuit systems interact, as well as to differentiate which computations arise in the cerebellum versus 
the brainstem. For the VOR, several authors have proposed that there is a switching or gating mechanism 
that chooses between vestibular (head movement) and retinal error signals 
\cite{BUIZZA82,LISBERGERFUCHS78A}. 

In our model, the control input generated in the brainstem-only pathway is a linear combination of eye 
movement information and vestibular inputs. Specifically, we assume the brainstem cancels a part of the 
vestibular signal (to generate the VOR) and a part of the disturbance introduced by the oculomotor 
plant itself. Instead, the cerebellum receives only visual information. Its role is to provide a top 
up to the disturbance supression activities of the brainstem. This view is consistent with the 
{\em flocculus central vestibular neuron complementary hypothesis} of \cite{BUTTNER84}. It postulates 
that the cerebellum will be modulated if the signal provided by central vestibular neurons 
(the brainstem) is not sufficient to achieve the objectives of the VOR, OKR, or smooth pursuit. 

Finally, we assume that when the visual driving signal is removed, as in darkness, the cerebellum 
falls inactive.  Numerous studies support the idea that the cerebellum (the flocculus) is relatively 
inactive without visual input \cite{LISBERGER15}. This interpretation is corroborated by experiments 
in which a sudden change in oculomotor behavior known to be mediated by the cerebellum occurs when 
the lights are turned on. 

{\bf Corollary Discharge.} \
A long-standing debate in the neuroscience community regards how eye position information becomes available
to the brain. One theory dating to the 1800's proposed that the brain receives an {\em efference
copy} of an internal signal carrying eye position information. An opposing theory argues that 
{\em proprioception} of eye muscle activity provides eye movement information, obviating the need for 
efference copies. In the 1950's, the term {\em corollary discharge} was coined to characterize a copy 
of the motor command that informs the brain of ongoing eye movement. 

It has been proven experimentally that proprioception from the eye muscles plays a negligible role in 
eye movement \cite{CARPENTER72,GUTHRIE83,KELLER71}. Consonant with these findings, our model assumes 
no proprioception. The {\em brainstem neural integrator} is now regarded to be the mechanism that 
provides the eye position to the brainstem \cite{POLA02}. In this work, we write the neural integrator 
in the form of an observer of the oculomotor plant. Our observer equation is identical to a leaky 
integrator in the Laplace domain, therefore matching experimental findings \cite{SKAVENSKI73}.

Since in our model the cerebellum only receives visual information, ongoing eye movement information is 
not directly supplied to the cerebellum. Our proposal is that residing in the cerebellum is an internal 
model of all exogenous disturbances acting on the oculomotor system and observable through the retinal 
error signal. The states of the internal model provide the signals for ongoing activity of the eye in 
the cerebellum, even with zero retinal slip. In our model, such extraretinal signals arise in the 
cerebellum by using a corollary discharge of the motor command. 

{\bf Internal Models.} \ 
Theories on the function of the cerebellum have been dominated by internal models for at least 
25 years \cite{GOMI92,KAWATO92,KAWATO99,MIALL96,WOLPERT95,WOLPERT98A,WOLPERT98B}. One view is that the 
cerebellum provides a {\em forward model} of the system to be controlled \cite{KAWATO99,MIALL96}. Another 
theory called {\em feedback error learning} (FEL) argues the cerebellum provides {\em inverse models} 
\cite{GOMI92,KAWATO92}. Another is that multiple forward and inverse models reside in the cerebellum 
\cite{WOLPERT98A}. These theories are related to notions in robotics on forward and inverse kinematics; 
indeed FEL is a variant of the computed torque method in robotics. We notice the term 
``internal model'' in the neuroscience literature is distinct from the internal model principle of
control theory \cite{IMP75,IMP76}. 

It seems reasonable that the brain would require kinematic models of the body, both as forward and 
inverse models. But we do not relegate this role to the cerebellum. Rather, our work here mathematically 
formalizes the idea that the role of the cerebellum is to realize the internal model principle: 
{\em to provide internal models of persistent, exogenous signals acting on a biological system}. 

The idea that the cerebellum or other regions of the brain may be involved in generating internal models 
of exogenous signals has already been suggested \cite{CERMINARA09,CHURCHLAND03,LISBERGER09}. Particularly,
in the review article \cite{LISBERGER09}, Lisberger presents three theories about the type of internal 
model that may reside in the cerebellum to support the oculomotor system. His first theory is that the 
cerebellum provides a model of the inertia of realworld objects - we can interpret his statement as an 
instance of the internal model principle.

Experimental evidence from the oculomotor system for the existence of internal models of exogenous signals 
comes in four forms. First, there is the so-called predictive capability of the smooth pursuit system -
to track moving targets with zero steady-state error \cite{BAHILL83A,DENO95,WYATT88}. Second, it has 
been shown experimentally that exogenous signals that can be modeled by low-order linear exosystems are 
easily tracked, while unpredictable signals are not \cite{BAHILL83B,COLLEWIJN84,DENO95,MICHAEL66}. Third, 
in an experiment called {\em target blanking}, a moving target is temporarily occluded, yet the eye 
continues to move \cite{CERMINARA09,CHURCHLAND03}; researchers postulate the brain has an internal 
model of the motion of the target.  The fourth evidence comes from an experiment called the {\em error 
clamp}, in which the retinal error is artificially clamped at zero using an experimental apparatus 
that places the target image on the fovea \cite{BARNES95,MORRIS87,STONE90}. Despite zero retinal 
error, the eye continues to track the target, suggesting that extraretinal signals drive the pursuit 
system. 

{\bf Organization}. \ 
This report is organized as follows. In the next section we derive the open-loop model of the 
oculomotor system. In Section~\ref{sec:problem} we derive the error model, formulate the 
disturbance rejection problem, and solve the problem using the theory of adaptive internal models. 
In Section~\ref{sec:simulations} we present simulation results. 
In Section~\ref{sec:discuss} we compare our model architecture to current architectures 
involving the cerebellum. Concluding remarks are presented in Section~\ref{sec:conclude}. 

\section{Open-loop Model}
\label{sec:ol}

The horizontal motion of the eye is modeled by considering the eyeball as a sphere that is suspended in 
fluid and subjected to viscous drag, elastic restoring forces, and the pulling of two muscles 
\cite{ROBINSON81,SYLVESTRECULLEN99}. A reasonable appproximation is obtained by assuming that the 
inertia of the eyeball is insignificant. Letting $x$ be the horizontal eye angle and $u$ be the net 
torque imparted by the two muscles, we obtain a first order model 
\begin{equation}
\label{eq:plant}
\dot{x} = - K_x x + u \,. 
\end{equation}
The parameter $K_x > 0$ is constant (or very slowly varying) such that the time constant of the eye is 
$\tau_x := 1/K_x \simeq 0.2$s \cite{ROBINSON81}. 
This first order model may be compared with the model of an ocular motoneuron. Let $f$ be the firing 
rate, and let $f_0$ be the baseline firing rate when the eye is stationary at $x = 0$. 
A commonly used model of neuronal firing rate is $f = f_0 + c_1 x + c_2 \dot{x}$, 
where $c_1$ and $c_2 \neq 0$ are constants \cite{ROBINSON70,ROBINSON81,SYLVESTRECULLEN99}. 
Comparing this model with \eqref{eq:plant}, we observe that $K_x = c_1/c_2$ and 
$u = \frac{1}{c_2} ( f - f_0 )$. That is, the torque is proportional to the firing rate, modulo 
a constant offset of $f_0$.

Next consider a reference signal $r$ representing the angle of a target moving in the horizontal plane. 
Let $x_h$ and $\dot{x}_h$ be the horizontal head angular position and angular velocity, respectively. 
The {\em retinal error} is defined to be 
\begin{equation}
\label{eq:e}
e := \alpha_e ( r - x_h - x ) \,. 
\end{equation}
Notice that $r - x_h - x$ is the target angle $r$ relative to the {\em gaze angle} $x_h + x$. 
For sufficiently distant targets, this relative angle is proportional (through the scale factor 
$\alpha_e \in \RR$) to a linear displacement on the retina from the fovea to the target. 
Since the goal of the VOR, OKR, gaze holding, and smooth pursuit is to drive $e$ to zero, 
for the purposes of the present paper we set $\alpha_e = 1$, since for $\alpha_e \neq 1$ we 
can always redefine the error to be $e' = e / \alpha_e$. 
 
We assume that the control input $u$ takes the form 
\[
u = u_b + u_c \,,
\]
where the brainstem component $u_b$ is generated through a brainstem-only pathway, while the 
cerebellar component $u_c$ is generated by a side pathway through the cerebellum. The reference 
signal $r$ is treated as a persistent unmeasurable disturbance acting on the oculomotor system. 
The eye position $x$ is assumed to be unavailable for direct measurement 
\cite{CARPENTER72, GUTHRIE83,KELLER71}. The vestibular system provides a measurement of the head 
angular velocity $\dot{x}_h$ to the brainstem but not directly to the cerebellum \cite{GERRITS89,ROBINSON81}, 
and it does not provide the head position $x_h$ \cite{ROBINSON81}. 
Finally, we assume that both the brainstem and the cerebellum receive a measurement of the retinal 
error $e$ (or a scaled version of it) based on retinal information supplied by a brain region such
as the superior colliculus \cite{BASSO00,GLASAUER03,KRAUZLIS97}. 

Because we assume the retinal error is available for measurement, it is unnecessary to have measurements 
of $x$ or $x_h$ since, in theory, these signals can be reconstructed based on observability through $e$. 
In practice, the phylogenetically older brainstem likely evolved without the benefits of observability; 
therefore, it receives certain measurements (such as $\dot{x}_h$) directly. Moreover, there is reason to 
believe a brainstem-only pathway serves the VOR to cancel voluntary head movements, while the cerebellum 
serves to cancel exogenous, involuntary head movements and target motion. 

To model the brainstem, we start from Robinson's parallel pathway model \cite{SKAVENSKI73}
consisting of two parallel pathways that combine to form the motor command; 
that is, $u = u_v + u_n$, where $u_v$ is carried on the direct pathway, 
and $u_n$ corresponds to the indirect pathway. The signal $u_n$ is the 
output of the {\em brainstem neural integrator} \cite{SKAVENSKI73}. Invoking equation (3) in 
\cite{ROBINSON74B}, the neural integrator is modeled as a leaky integrator:
\begin{equation}
\label{eq:xhat0} 
\dot{\xh} = - \Kxt \xh + u_v \,, \quad u_n = \alpha_x \xh \,,
\end{equation}  
where $\alpha_x$ and $\Kxt$ are constants (or very slowly varying).
Using the fact that $u_v = u - \alpha_x \xh$, this model can be re-expressed as 
\begin{equation}
\label{eq:xhat}
\dot{\xh} = - \widehat{K}_x \xh + u \,,
\end{equation}
where $\widehat{K}_x := \Kxt + \alpha_x$. Finally, we incorporate the idea from \cite{GALIANA84} 
that $\widehat{K}_x \simeq K_x$ (henceforth we drop the hat); see also \cite{DALE15,GREEN04}. 
In sum, we deduce that the brainstem neural integrator forms an {\em observer} of the oculomotor plant. 
If we define the estimation error $\xt := x - \hat{x}$, then $\xt$ evolves according to 
$\dot{\xt} = - K_x \xt$, implying that $\hat{x}(t)$ converges exponentially to $x(t)$. 
Aside from a momentary perturbation (a push on the eyeball), $\hat{x}(t)$ well approximates $x(t)$. 

\begin{remark}
The neural integrator contributes to a distributed gaze holding function, resulting
in three time constants for gaze holding: $\tau_x = 0.2$s is the time constant of the
oculomotor plant; $\tilde{\tau}_x = 1 / \Kxt = 2$s is the time constant of the combined 
neural integrator and plant; and the time constant induced by the top up from the cerebellum 
is $\tau = 25$s \cite{GLASAUER03}.
\tqed
\end{remark}

To complete the modeling of the brainstem, we consider the components of the signal $u_v$. 
In our model $u_v = u_c - \alpha_h \dot{x}_h$, where $u_c$ contains visual information and 
the output of the cerebellum, and $\alpha_h \dot{x}_h$ is the vestibular measurement of head 
angular velocity representing the direct feedthrough from the semicircular canals to the 
oculomotor plant. The overall motor command is 
\[
u = u_v + u_n = \alpha_x \xh - \alpha_h \dot{x}_h + u_c \,, 
\]
so the brainstem-only pathway of the control input is 
\begin{equation}
\label{eq:ub1}
u_b = \alpha_x \hat{x} - \alpha_h \dot{x}_h \,,
\end{equation}
where $\alpha_x \in \RR$ and $\alpha_h \in \RR$ are constant (or slowly varying) parameters; and once
again, $\dot{x}_h$ is the head angular velocity, and $\hat{x}$ is an estimate of the eye 
position. We can see that the role of $u_b$ is to supress a portion of the head velocity disturbance 
and to partially cancel the drift term in the oculomotor plant dynamics.

\section{Disturbance Rejection Problem}
\label{sec:problem}

We approach the derivation of a model of the cerebellum as a problem of control synthesis:
to design a controller $u_c$ to drive the error $e(t)$ to zero. Assuming that $\hat{x}(t) \simeq x(t)$ 
for $t \ge 0$, we obtain the {\em error model}
\begin{equation}
\label{eq:edot1}
\dot{e} = - \Kxt e - u_c + \dot{r} + \Kxt r - (1 - \alpha_h) \dot{x}_h - \Kxt x_h \,,
\end{equation}
where $\Kxt = K_x - \alpha_x$.
We assume that the reference signal $r$ as well as the head position $x_h$ are modeled as the outputs 
of a linear exosystem. Let $\eta \in \RR^q$ be the exosystem state and define the exosystem 
\begin{subequations}
\label{eq:exosystem}
\begin{eqnarray}
\label{eq:exoa}
\dot{\eta} & = & S \eta  \\
\label{eq:exob}
r          & = & D_1 \eta \,, \qquad x_h = D_2 \eta \,,
\end{eqnarray}
\end{subequations}
where $S \in \RR^{q \times q}$, $D_1 \in \RR^{1 \times q}$, and $D_2 \in \RR^{1 \times q}$. 
Then \eqref{eq:edot1} takes the form
\begin{equation}
\label{eq:edot2}
\dot{e} = - \Kxt e - u_c + E \eta 
\end{equation}
where $E := D_1 S + \Kxt D_1 - (1 - \alpha_h ) D_2 S - \Kxt D_2 \in \RR^{1 \times q}$. 

It is useful to transform the exosystem using the technique in \cite{NIKIFOROV98}. 
Let $(F,G)$ be a controllable pair with $F$ Hurwitz. Specifically, we take
\small
\begin{equation}
\label{eq:FG}
F = 
\left[ 
\begin{array}{ccccc}
0         & 1         & \cdots & 0      & 0 \\
\vdots    &           &        & \ddots &   \\
0         & 0         & \cdots & 0      & 1 \\
-\lambda_1 & -\lambda_2 & \cdots &        & -\lambda_{q}
\end{array} 
\right], ~~ 
G = 
\left[ 
\begin{array}{c}
0 \\ \vdots \\ 0 \\ 1
\end{array} 
\right],  
\end{equation}
\normalsize
where the polynomial $s^q + \lambda_{q} s^{q-1} + \cdots + \lambda_1$ is Hurwitz. 
Define the coordinate transformation $w = M \eta$, with $M \in \RR^{q \times q}$ nonsingular and
satisfying the Sylvester equation $MS = FM + GE$ (without loss of generality we can
assume $(E,S)$ is observable and the spectra of $S$ and $F$ are disjoint) \cite{NIKIFOROV98}. Also 
define $\Psi := E M^{-1} \in \RR^{1 \times q}$. In new coordinates, the exosystem 
model is
\begin{equation}
\label{eq:wdot}
\dot{w} = (F + G \Psi ) w \,.
\end{equation}
Because $E \eta = \Psi w$, we can write the error dynamics \eqref{eq:edot2} in terms of the 
new exosystem state:
\begin{equation}
\label{eq:edot}
\dot{e} = - \Kxt e - u_c + \Psi w \,.
\end{equation}
The parameters $( \Kxt, \Psi^{\Tr} ) \in \RR^{q+1}$ capture all unknown model and disturbance 
parameters. 

\begin{prob}
\label{prob1}
Consider the error dynamics \eqref{eq:edot}. Suppose the unknown parameters $( \Kxt, \Psi^{\Tr} )$ 
belong to a known compact set $\cP \subset \RR^{q+1}$. We want to find an error feedback controller
\begin{eqnarray*}
\dot{\xi} & = & F_c ( \xi, e) \\
u_c       & = & H_c \xi + K_c e  
\end{eqnarray*}
such that for all initial conditions $(e(0), w(0), \xi(0))$ and for all $( \Kxt, \Psi^{\Tr} ) \in \cP$, 
the solution $( e(t), w(t), \xi(t) )$ of the closed loop system  
\begin{eqnarray*}
\dot{e}   & = & - (\Kxt + K_c) e - H_c \xi + \Psi w \\
\dot{\xi} & = & F_c ( \xi, e ) 
\end{eqnarray*}
satisfies $\lim_{t \rightarrow \infty} e(t) = 0$. \tqed
\end{prob}
We invoke the design approach of \cite{SERRANI00,SERRANI01}. 
The controller takes the form of an {\em adaptive internal model} consisting of an internal model of the 
disturbances acting on the oculomotor system combined with a parameter estimation process to recover the 
unknown parameters. Let $\hat{w}$ and $\Psih$ be estimates of $w$ and $\Psi$, respectively. The controller is 
\begin{eqnarray}
\dot{\hat{w}} & = & F \hat{w} + G u_c \\ 
 u_c          & = & u_{imp} + u_{s} \,.
\end{eqnarray}
The controller $u_{imp}$ is selected to satisfy the internal model principle: 
$u_{imp} = \Psih \hat{w}$. 
The controller $u_{s}$ is selected to make the closed-loop system asymptotically stable. We choose 
$u_{s} = K_e e$, with $K_e > 0$ sufficiently large. Based on a Lyapunov argument, 
the adaptation law for the parameter estimates is $\dot{\Psih} = e \hat{w}^{\Tr}$. 

In summary, the overall model is 
\begin{subequations}
\label{eq:control}
\begin{eqnarray}
\label{eq:xhatdot}
\dot{\hat{x}}  & = & - K_x \hat{x} + u \\
\label{eq:whatdot}
\dot{\hat{w}}  & = & F \hat{w} + G u_c \\
\label{eq:Psihatdot}
\dot{\Psih}    & = & e \hat{w}^{\Tr} \\
\label{eq:ub}
u_b            & = & \alpha_x \hat{x} - \alpha_h \dot{x}_h \\
\label{eq:uc}
u_c            & = & \Psih \hat{w} + K_e e \\
\label{eq:u}
u              & = & u_b + u_c \,. 
\end{eqnarray}
\end{subequations}
A proof of correctness of this design is provided in the Appendix.

\section{Simulation Results}
\label{sec:simulations}

In this section we simulate our model under a number of experimental scenarios involving the VOR, OKR, 
gaze holding, and smooth pursuit. The parameter values for the simulations are: $q = 2$, $K_x = 5$, 
$\alpha_x = 0.95 K_x$, $\alpha_h = 0.65$, $K_e = 5$, $\lambda_1 = 1$, and $\lambda_2 = 1$. In a few cases 
noted below, different parameters are used to exaggerate certain transient phenomena.

\subsection{VOR} 

We consider the VOR in which the eye must track a fixed target while the head is moving.
First, we consider what happens when the head is rotated in darkness. It is known that 
the cerebellum is relatively inactive due to a lack of visual input \cite{LISBERGER15}. As such, we
assume in darkness $u_c = 0$, so the eye dynamics evolve according to a brainstem-only 
control input. Assuming that $x(t) \simeq \hat{x}(t)$, we have 
\begin{equation}
\label{eq:dark}
\dot{x} = - \Kxt x - \alpha_h \dot{x}_h \,.
\end{equation}
Suppose $x_h(t) = a_h \sin ( \beta_h t )$, and let $\Omega$ be the limit set of any solution 
of \eqref{eq:dark}. Assuming $\Kxt > 0$, a solution $x_{\infty}(t)$ in $\Omega$ has the form
\[
x_{\infty}(t) = 
-\alpha_h a_h \frac{\beta_h}{\Kxt^2 + \beta_h^2} 
\biggl( \beta_h \sin (\beta_h t) - \Kxt \cos (\beta_h t) \biggr) \,. 
\]
Generally $\Kxt \ll \beta_h$, so 
\[
x_{\infty}(t) \simeq - \alpha_h a_h \sin ( \beta_h t ) = - \alpha_h x_h(t) \,.
\]
That is, the eye moves relative to the head with a scale factor of -$\alpha_h$. The parameter $\alpha_h$ 
is called the {\em VOR gain} since it well approximates the ratio of head velocity to eye velocity 
measured in darkness. We note that our model predicts that the VOR in the dark is unaffected by 
disabling the cerebellum, as reported experimentally \cite{ROBINSON81,ZEE81}.

\begin{figure}[t!]
\centering
\begin{subfigure}[b]{.25\textwidth}
\centering
\includegraphics[width=.9\linewidth]{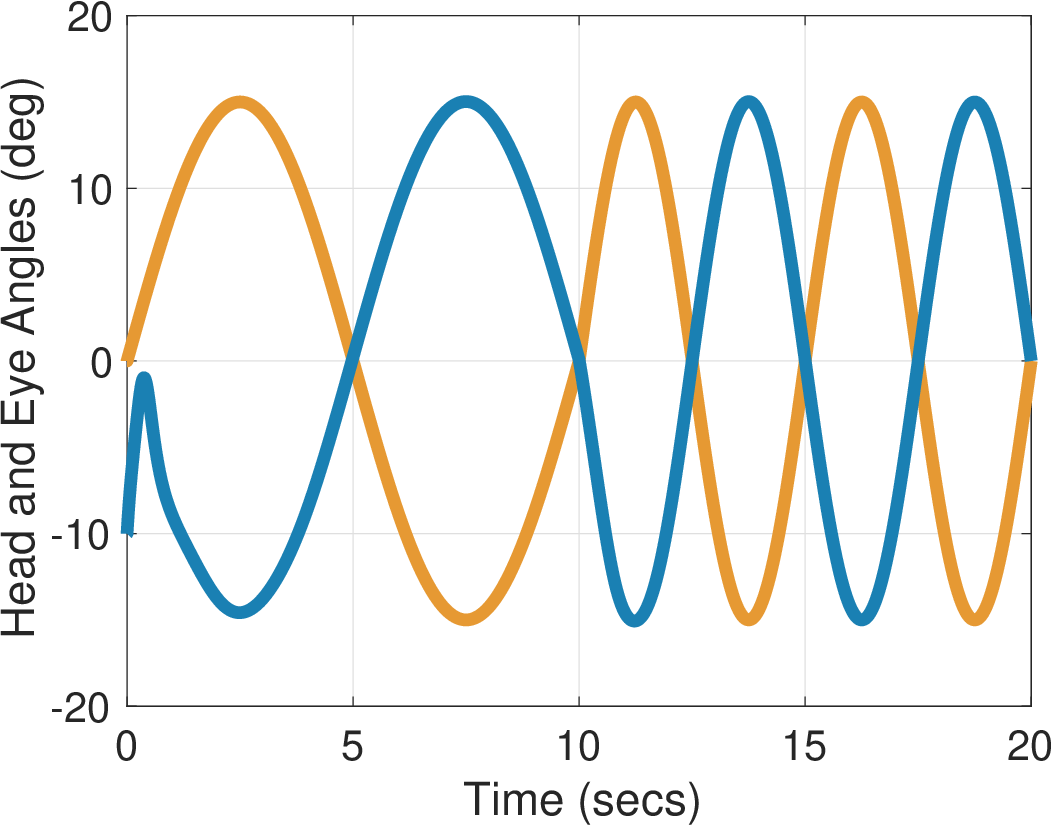}
\end{subfigure}
\begin{subfigure}[b]{.25\textwidth}
\centering
\includegraphics[width=.9\linewidth]{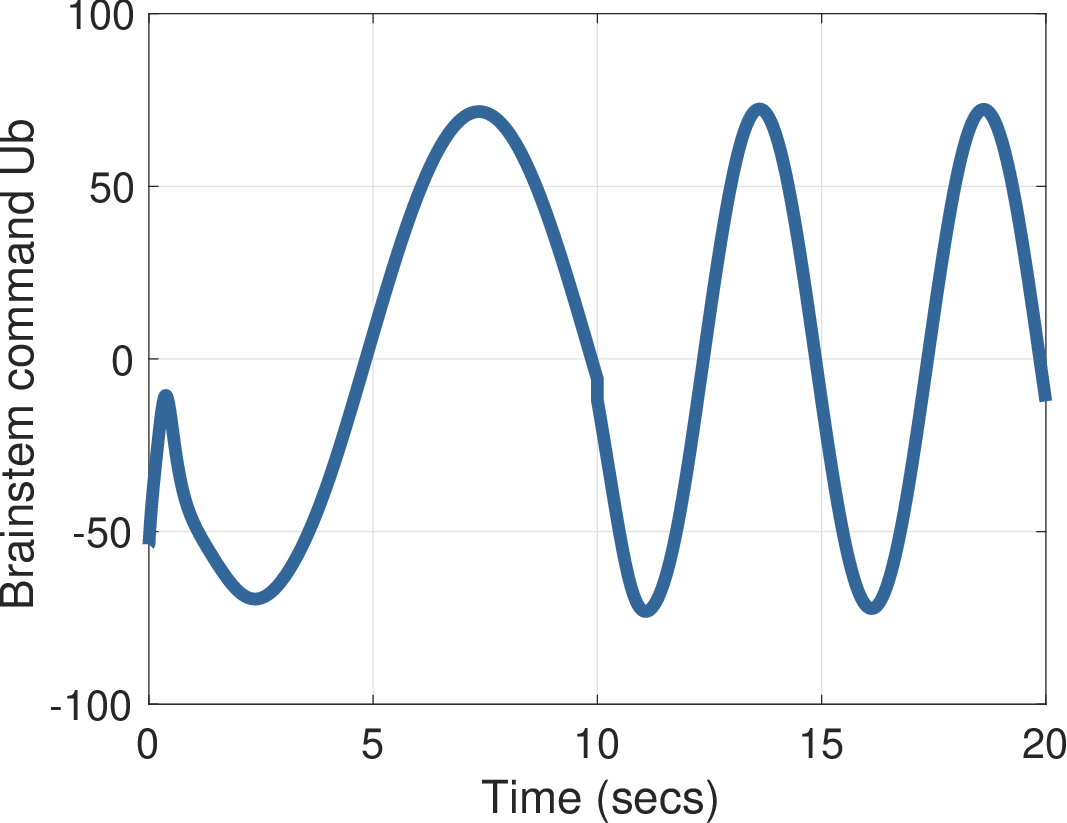}
\end{subfigure} 
\begin{subfigure}[b]{.25\textwidth}
\centering
\includegraphics[width=.9\linewidth]{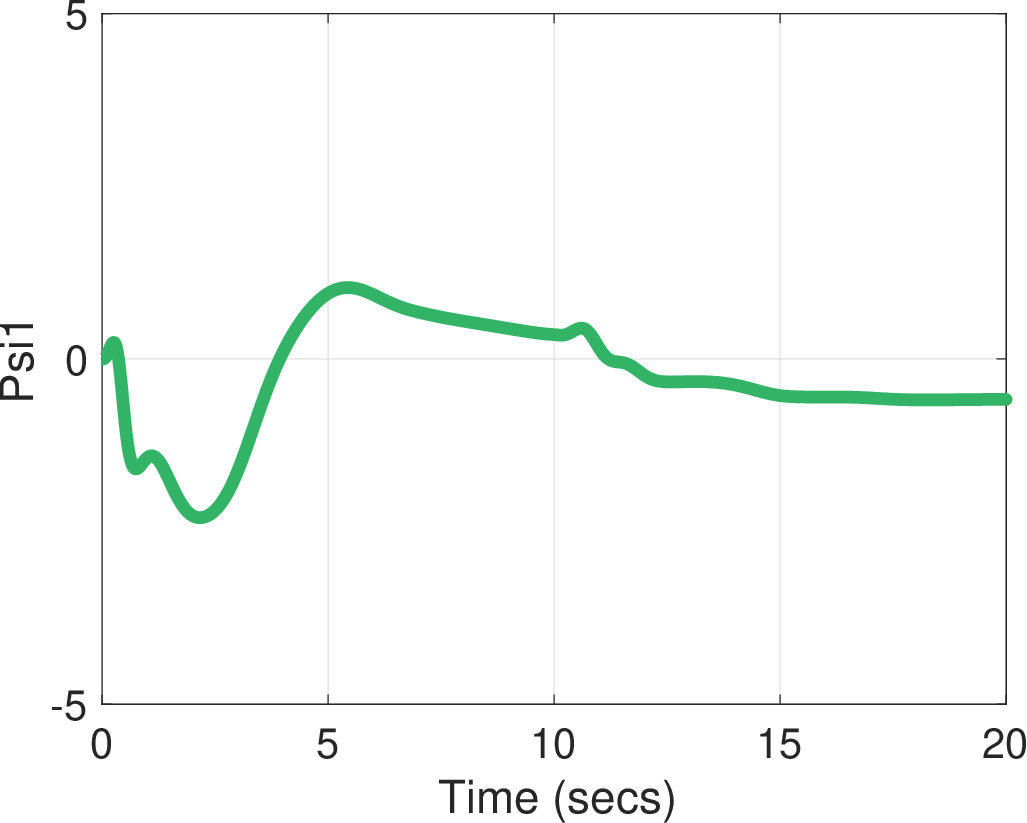}
\end{subfigure} \\ ~ \\
\begin{subfigure}[b]{.25\textwidth}
\centering
\includegraphics[width=.9\linewidth]{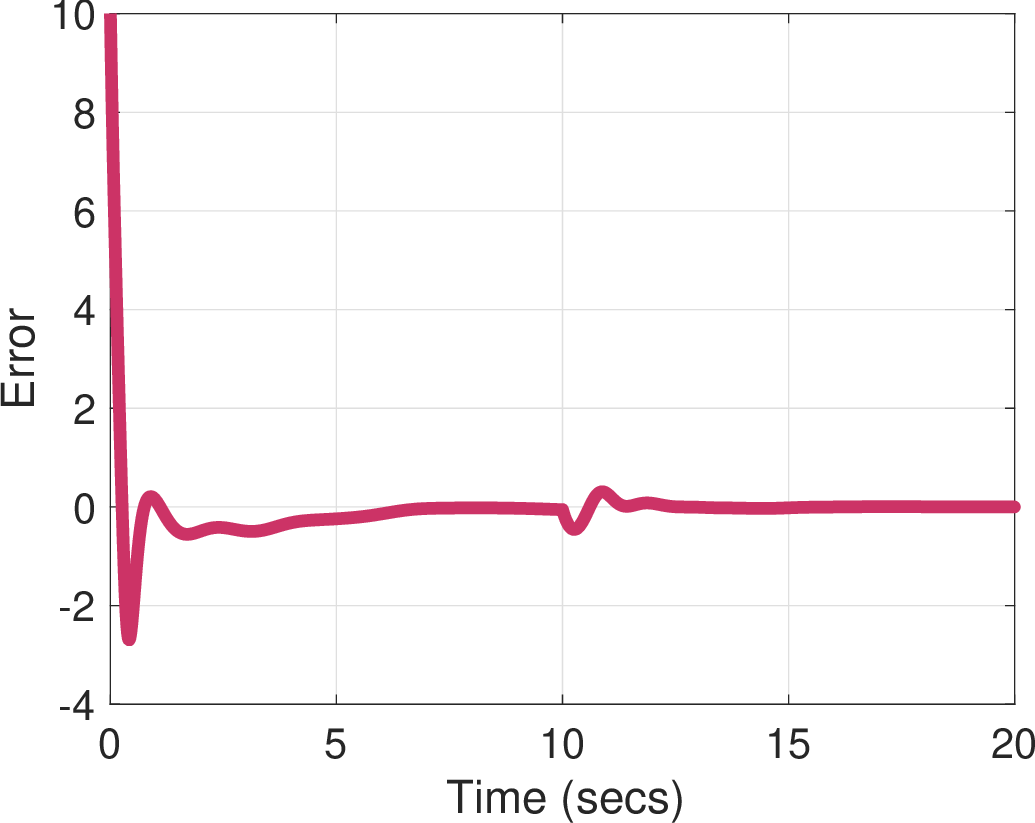}
\end{subfigure}
\begin{subfigure}[b]{.25\textwidth}
\centering
\includegraphics[width=.9\linewidth]{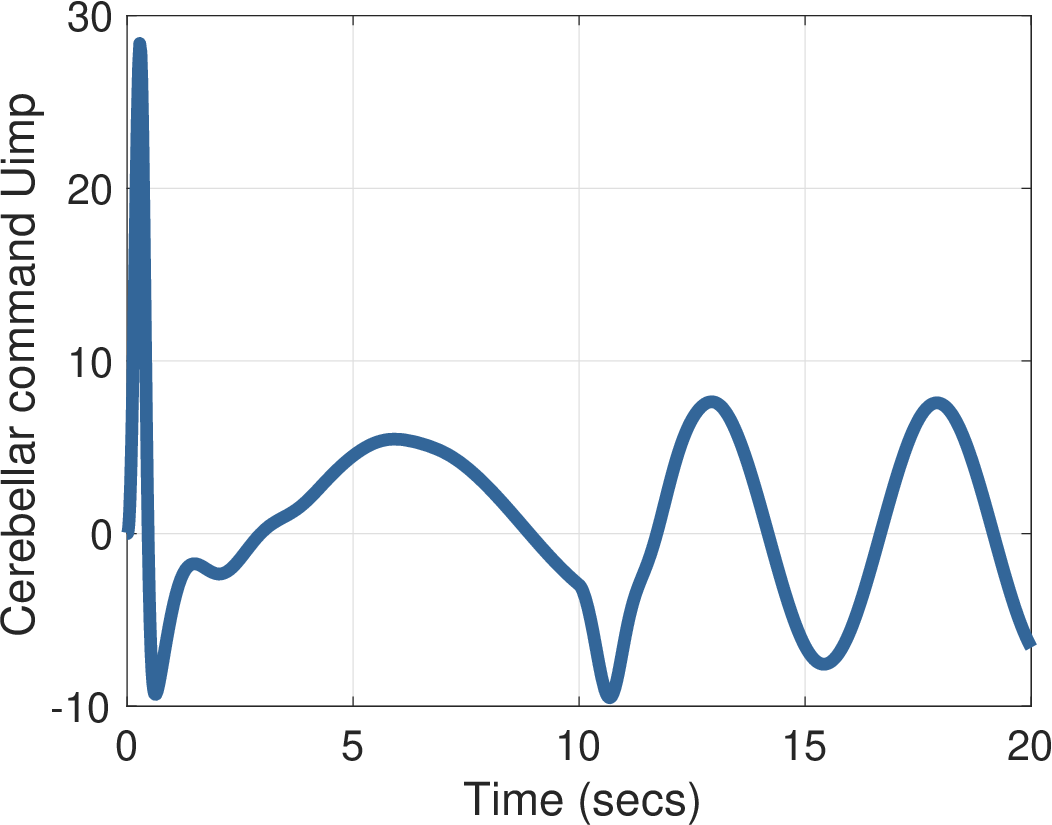}
\end{subfigure} 
\begin{subfigure}[b]{.25\textwidth}
\centering
\includegraphics[width=.9\linewidth]{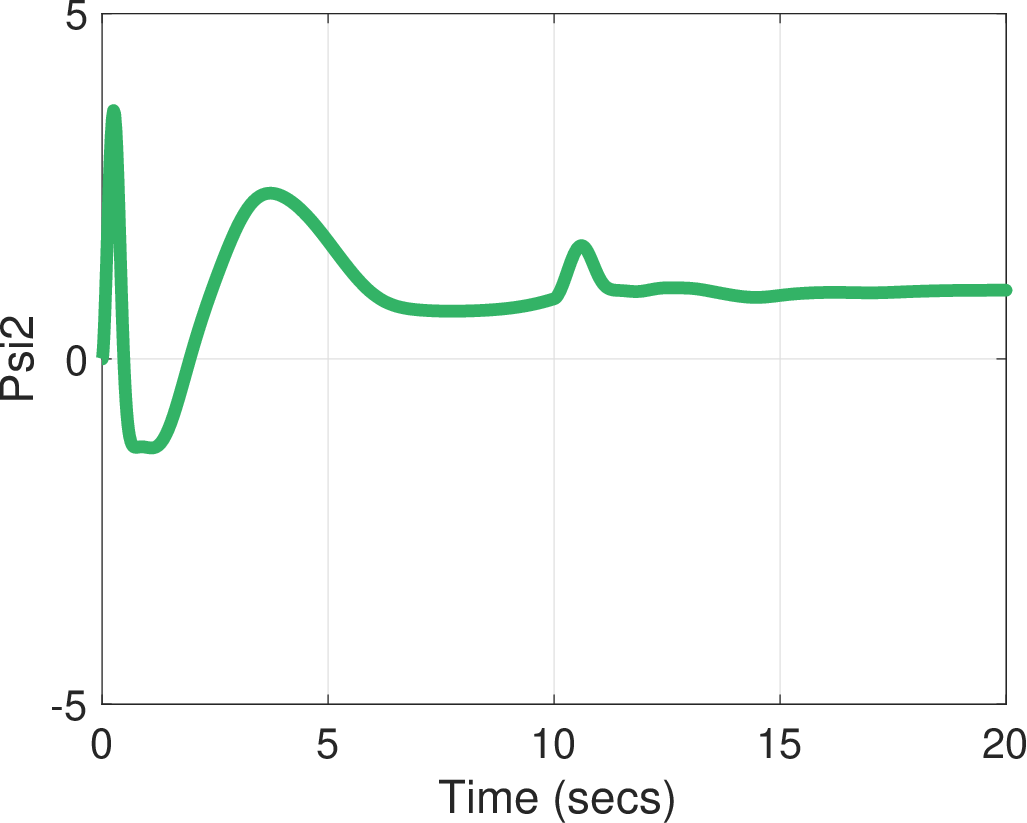}
\end{subfigure} 
\caption{VOR with a sinusoidal head rotation. The top left figure shows the head (yellow) and 
eye (blue) angles. The bottom left is the retinal error (red). The middle figures are $u_b$ and 
$u_{imp}$, and the right figures are the parameter estimates $\Psih_1$ and $\Psih_2$.}
\label{fig:VOR1}
\end{figure}

The standard VOR experiment is to apply an involuntary sinusoidal head rotation: 
$x_h(t) = a_h \sin ( \beta_h t )$, where $a_h, \beta_h > 0$. 
Figure~\ref{fig:VOR1} shows simulation results for the values $a_h = 15$, $\beta_h = 0.1$Hz for 
$t \in [0,10]$, and $\beta_h = 0.2$Hz for $t \in [10,20]$. The initial condition on all states 
is zero except the eye angle, which starts at $x(0) = -10^{\circ}$. We also plot the retinal error $e$, 
the cerebellar output $u_{imp}$, the brainstem component $u_b$, and the parameter estimates 
$\Psih_1$ and $\Psih_2$. As expected, the eye moves opposite to the head rotation, and it adapts 
to the frequency of the sinusoidal disturbance. 

\begin{figure}[t!]
\centering
\begin{subfigure}[b]{.25\textwidth}
\centering
\includegraphics[width=.9\linewidth]{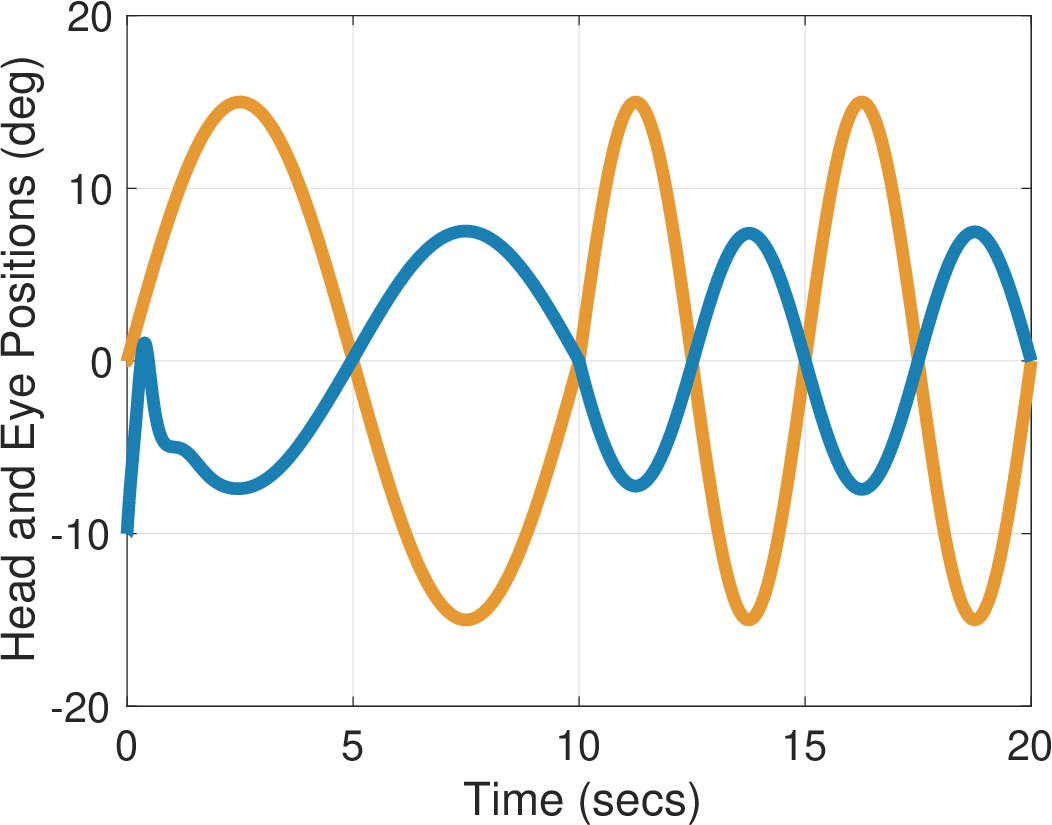}
\end{subfigure}
\begin{subfigure}[b]{.25\textwidth}
\centering
\includegraphics[width=.9\linewidth]{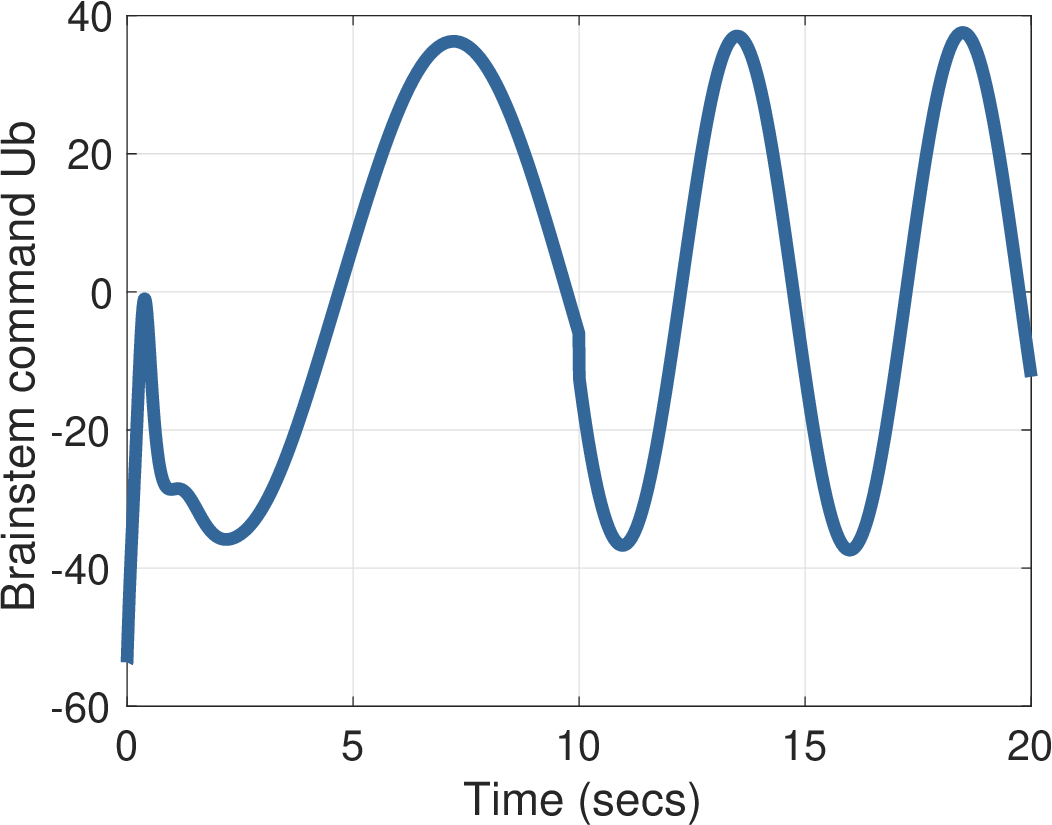}
\end{subfigure} 
\begin{subfigure}[b]{.25\textwidth}
\centering
\includegraphics[width=.9\linewidth]{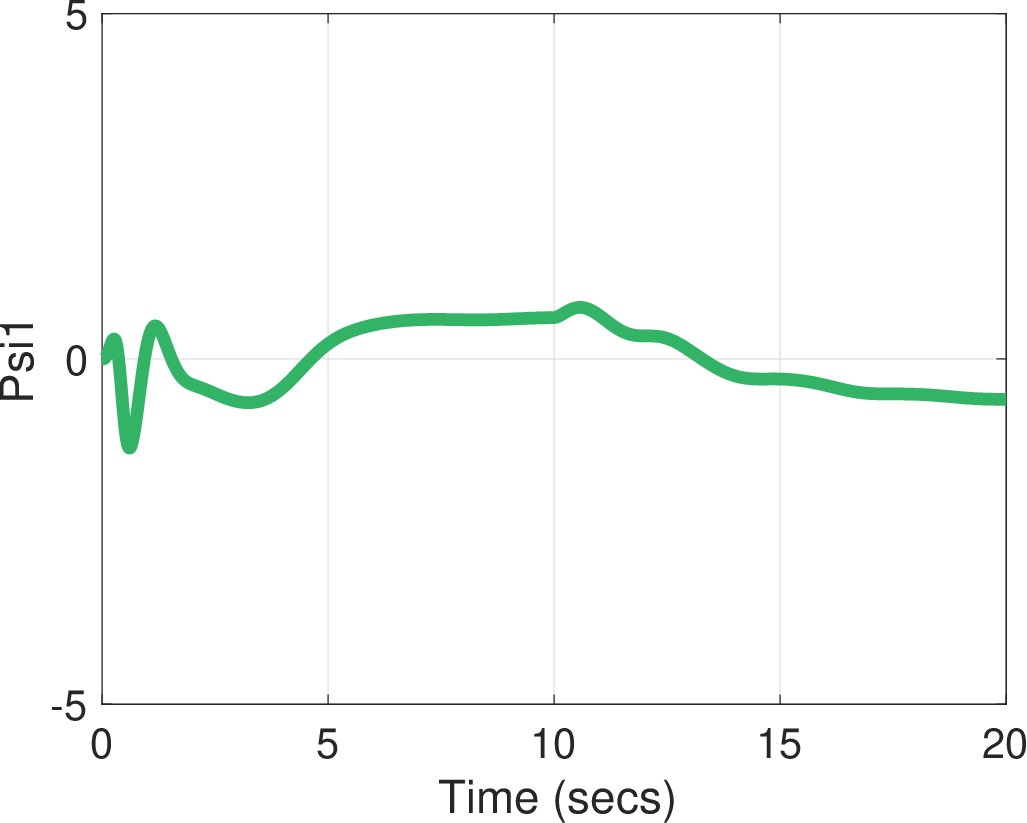}
\end{subfigure} \\ ~ \\
\begin{subfigure}[b]{.25\textwidth}
\centering
\includegraphics[width=.9\linewidth]{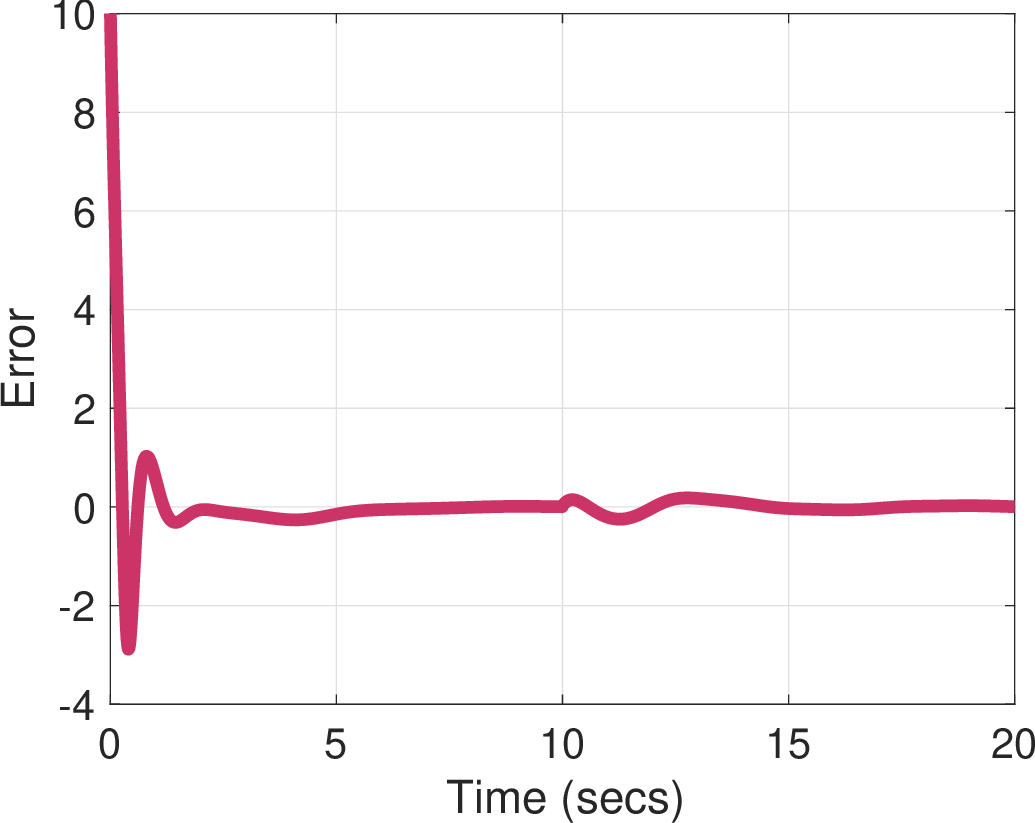}
\end{subfigure}
\begin{subfigure}[b]{.25\textwidth}
\centering
\includegraphics[width=.9\linewidth]{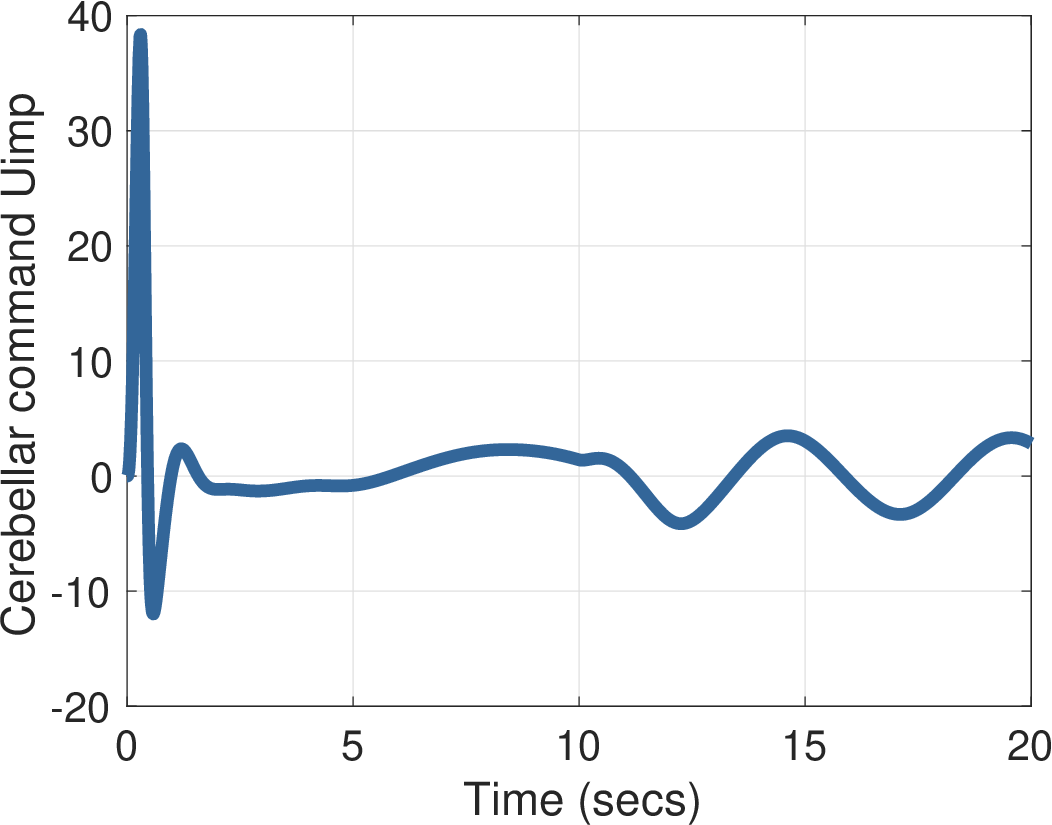}
\end{subfigure} 
\begin{subfigure}[b]{.25\textwidth}
\centering
\includegraphics[width=.9\linewidth]{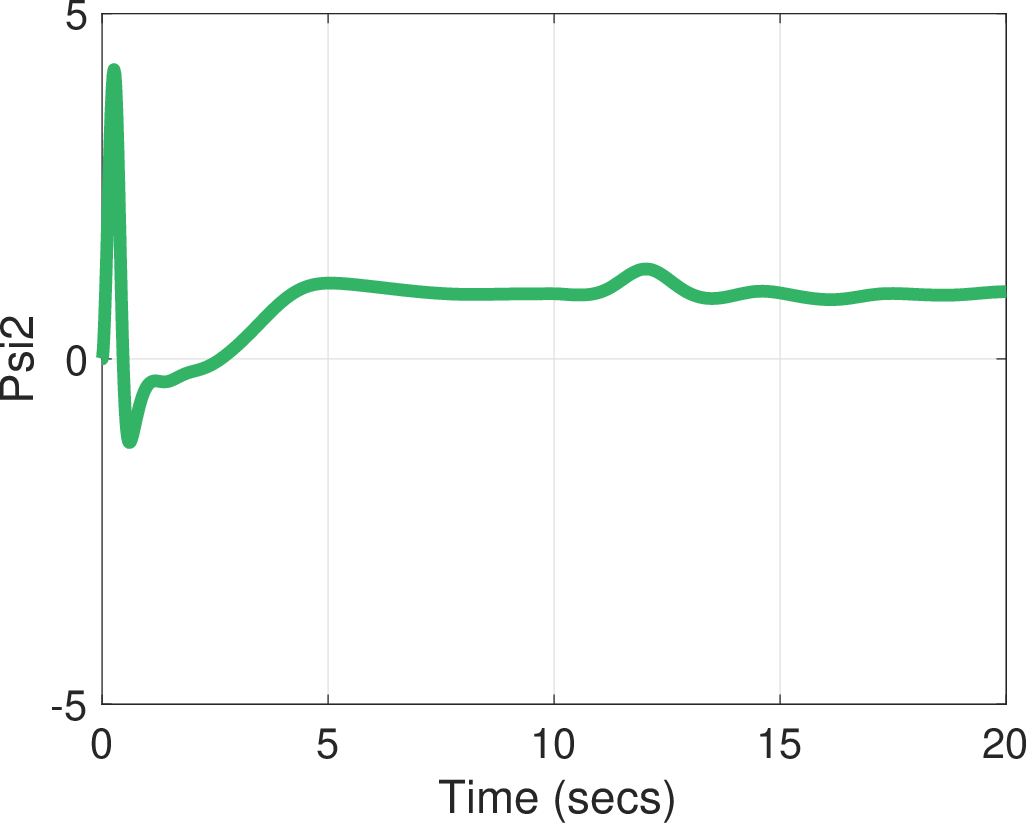}
\end{subfigure} 
\caption{VOR while tracking a target moving relative to the head rotation. 
The top left figure shows the head (yellow) and eye (blue) angles. The bottom 
left is the retinal error (red). The middle figures are $u_b$ and 
$u_{imp}$, and the right figures are the parameter estimates $\Psih_1$ and $\Psih_2$.}
\label{fig:VOR_ST}
\end{figure}

A second standard experiement is to evoke {\em short-term adaptation} of the VOR. For example,
suppose an involuntary sinusoidal head rotation is applied 
$x_h(t) = a_h \sin ( \beta_h t )$, where $a_h, \beta_h > 0$, while at the same time
the subject must track a target $r(t) = \alpha_r x_h(t)$, where $\alpha_r$ is a constant. 
Figure~\ref{fig:VOR_ST} shows simulation results for $\alpha_r = 0.5$, $a_h = 15$, 
$\beta_h = 0.1$Hz for $t \in [0,10]$, and $\beta_h = 0.2$Hz for $t \in [10,20]$. 
The initial condition on all states is zero except the eye angle, which starts at 
$x(0) = -10^{\circ}$. We also plot the retinal error $e$, 
the cerebellar output $u_{imp}$, the brainstem component $u_b$, and the parameter estimates 
$\Psih_1$ and $\Psih_2$. The eye moves opposite to the head rotation, but only
with half the amplitude. 

\begin{figure}[t!]
\centering
\begin{subfigure}[b]{.25\textwidth}
\centering
\includegraphics[width=.9\linewidth]{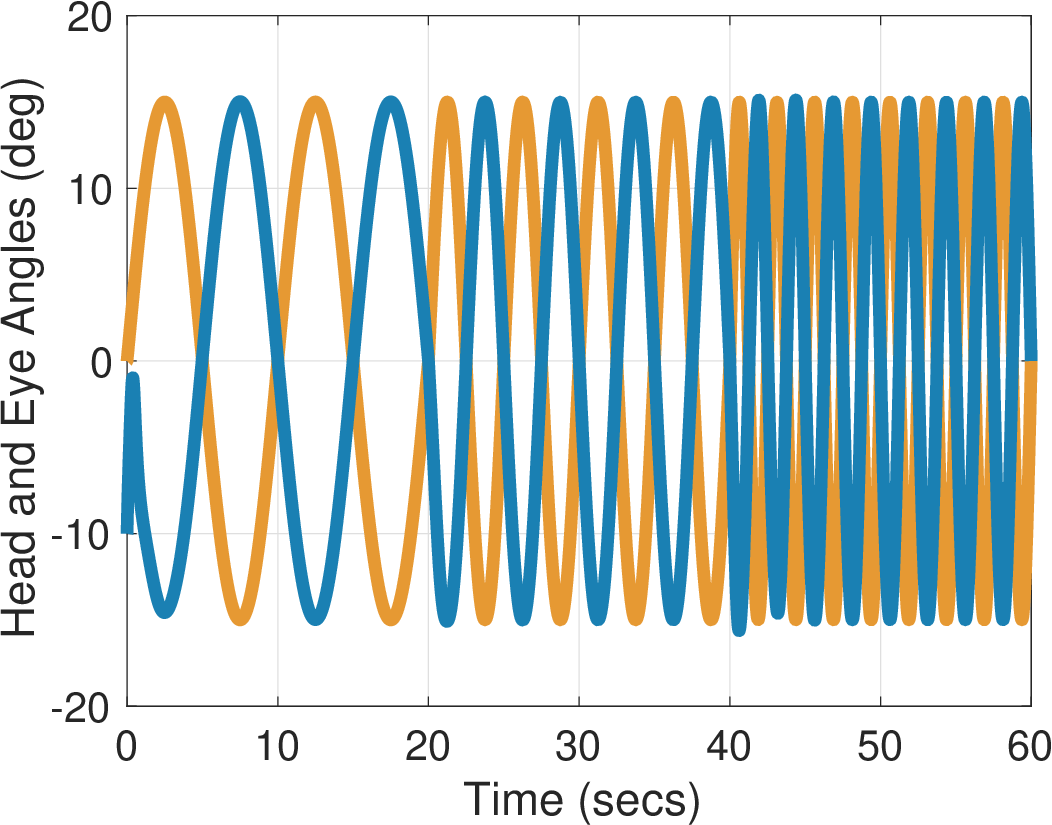}
\end{subfigure}
\begin{subfigure}[b]{.25\textwidth}
\centering
\includegraphics[width=.9\linewidth]{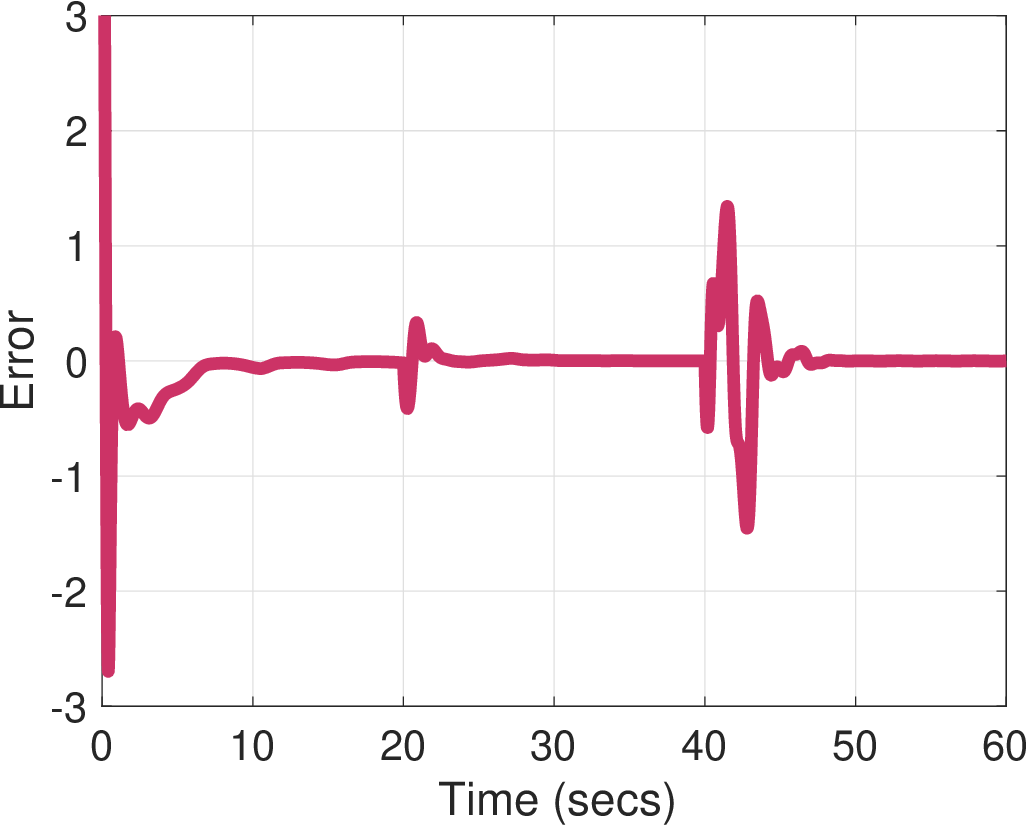}
\end{subfigure} 
\begin{subfigure}[b]{.25\textwidth}
\centering
\includegraphics[width=.9\linewidth]{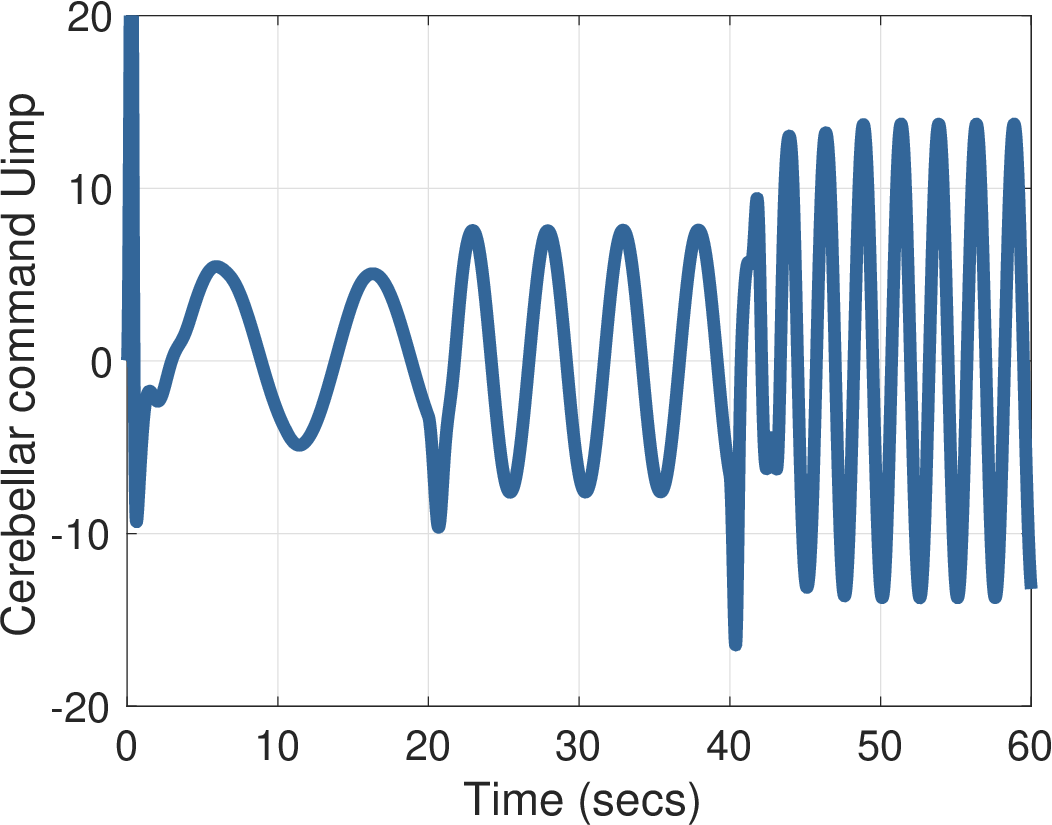}
\end{subfigure} 
\caption{Effect of the frequency of oscillations of the head on the depth of modulation of the 
cerebellar output $u_{imp}$. From left to right, the head (yellow) and eye (blue) angles, the 
retinal error $e$, and the cerebellar output $u_{imp}$.}
\label{fig:VOR2}
\end{figure}

An experiment reported in \cite{LISBERGER78} demonstrated that the depth of 
firing rate of the output of the cerebellum, $u_{imp}$ in our model, 
increases with the frequency of head rotation. This behavior is predicted by our model 
because $u_{imp} = \Psih \hat{w}$ must build an estimate of $- (1 - \alpha_h) \dot{x}_h - \Kxt x_h$.
In particular, the term $\dot{x}_h = a_h \beta_h \cos ( \beta_h t)$ is proportional to $\beta_h$. This 
behavior is depicted in Figure~\ref{fig:VOR2} by simulating our model with the values 
$a_h = 15$, $\beta_h = 0.1$Hz for $t \in [0,20]$, $\beta_h = 0.2$Hz for $t \in [20,40]$, 
and $\beta_h = 0.5$Hz for $t \in [40,60]$. We see in the right figure of Figure~\ref{fig:VOR2} 
that the amplitude of $u_{imp}$ increases as the frequency of the head rotation increases. 

\begin{figure}[t!]
\centering
\begin{subfigure}[b]{.25\textwidth}
\centering
\includegraphics[width=.9\linewidth]{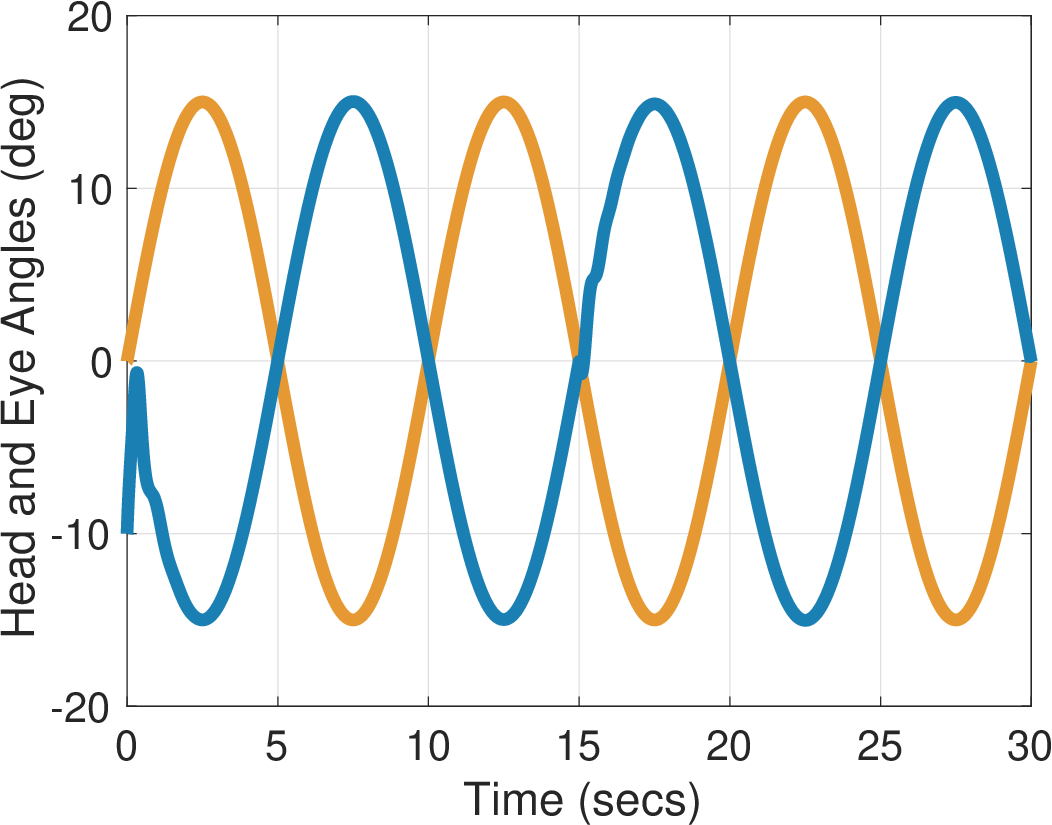}
\end{subfigure}
\begin{subfigure}[b]{.25\textwidth}
\centering
\includegraphics[width=.9\linewidth]{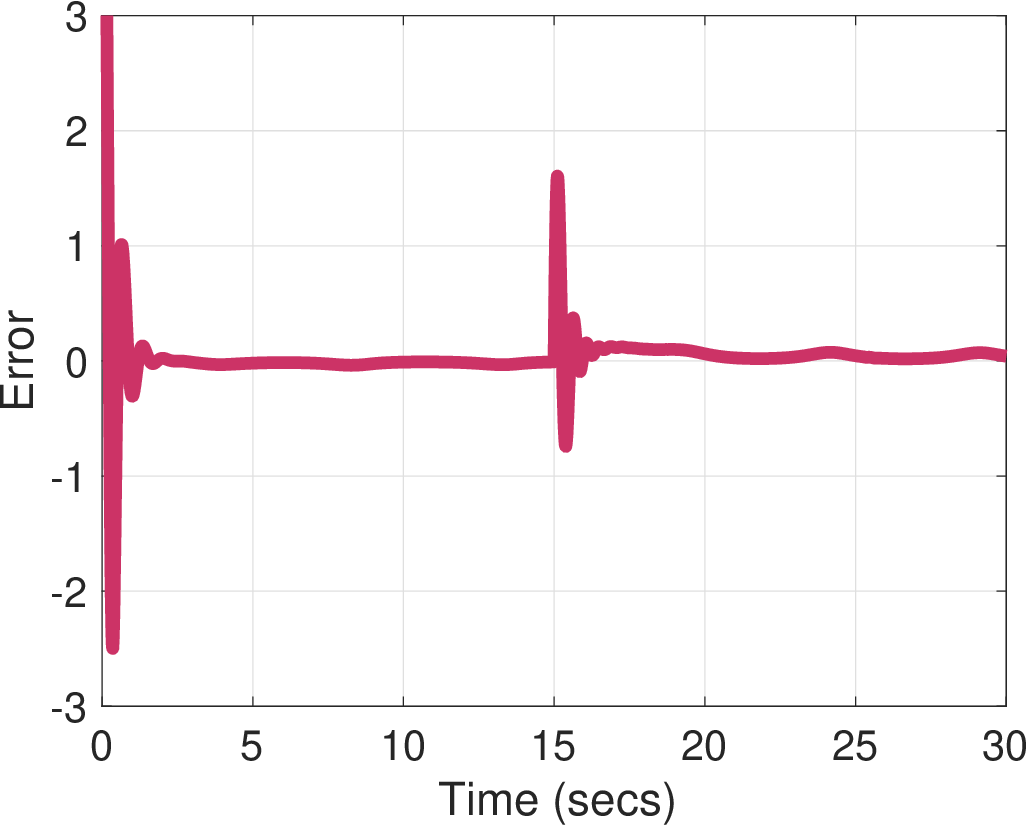}
\end{subfigure} 
\begin{subfigure}[b]{.25\textwidth}
\centering
\includegraphics[width=.9\linewidth]{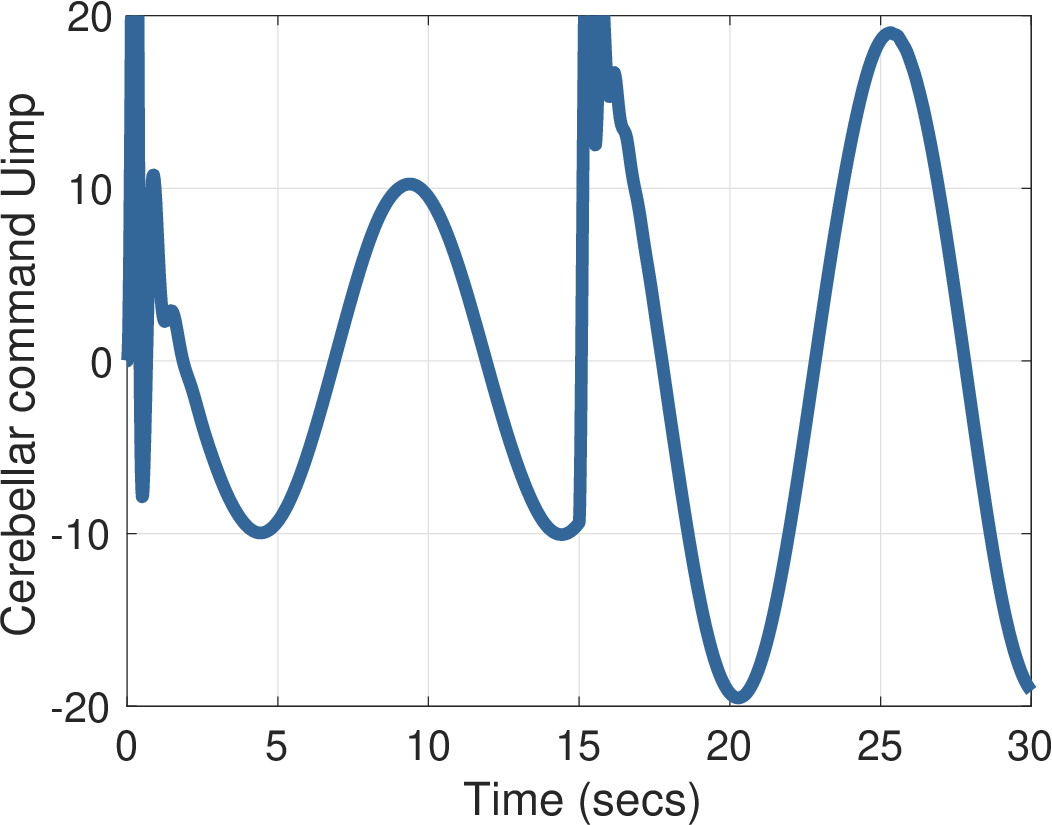}
\end{subfigure} 
\caption{Effect of $\alpha_h$ on the VOR. 
From left to right, the head (yellow) and eye (blue) angles, the retinal error $e$, and the 
cerebellar output $u_{imp}$.}
\label{fig:VOR3}
\end{figure}

It has been demonstrated that the VOR in the light is unaffected by changes in the VOR gain 
\cite{MILES80}. Figure~\ref{fig:VOR3} shows this experimental behavior with our model, where 
$\alpha_h = 2$ for $t \in [0,15]$ and $\alpha_h = -1$ for $t \in [15,30]$. It is clear from 
the left figure that our model predicts that in steady-state, the VOR in the light is unaffected 
by changes in the VOR gain. 

\begin{figure}[t!]
\centering
\includegraphics[width=.4\linewidth]{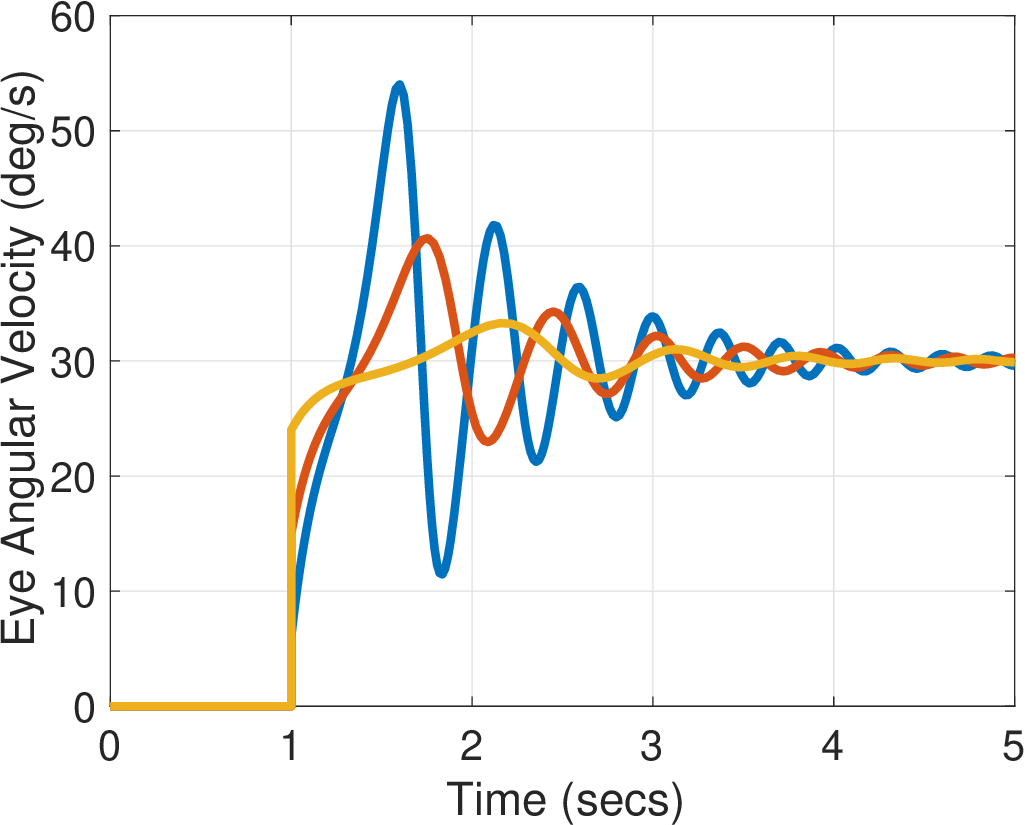}
\caption{VOR with a step input in head velocity for the values $\alpha_h = 0.3, 0.5, 0.8$ 
(blue, red, yellow). The size of the overshoot in the eye velocity is inversely proportional to the value of $\alpha_h$.}
\label{fig:VOR5}
\end{figure}

An experiment investigating the transients of the VOR in monkeys was reported in \cite{LISBERGERPAVELKO86}.
It was discovered that the overshoot in the eye velocity to a sudden rotation of the head was larger when 
the VOR gain is smaller. In the experiment, a light spot at $r = 0$ on which the monkey fixates 
(in another otherwise dark room) is strobed. Here we assume the subject attempts to continuously fixate 
the eyes on a target at $r = 0$, even when the light spot is extinguished. 
The head position is a ramp function: $x_h(t) = 0$ for $t \in [0,1]$ and $x_h(t) = -30 t$ for 
$t \in [1,5]$, resulting in a head angular velocity of -$30^{\circ}$/s. 
Figure~\ref{fig:VOR5} illustrates that our model recovers the behavior in \cite{LISBERGERPAVELKO86}. 
The blue curve is the eye angular velocity for $\alpha_h = 0.3$, red is with $\alpha_h = 0.5$, and 
yellow is with $\alpha_h = 0.8$. We see clearly that smaller VOR gains result in larger overshoots. 

\begin{figure}[t!]
\centering
\begin{subfigure}[b]{.25\textwidth}
\centering
\includegraphics[width=.9\linewidth]{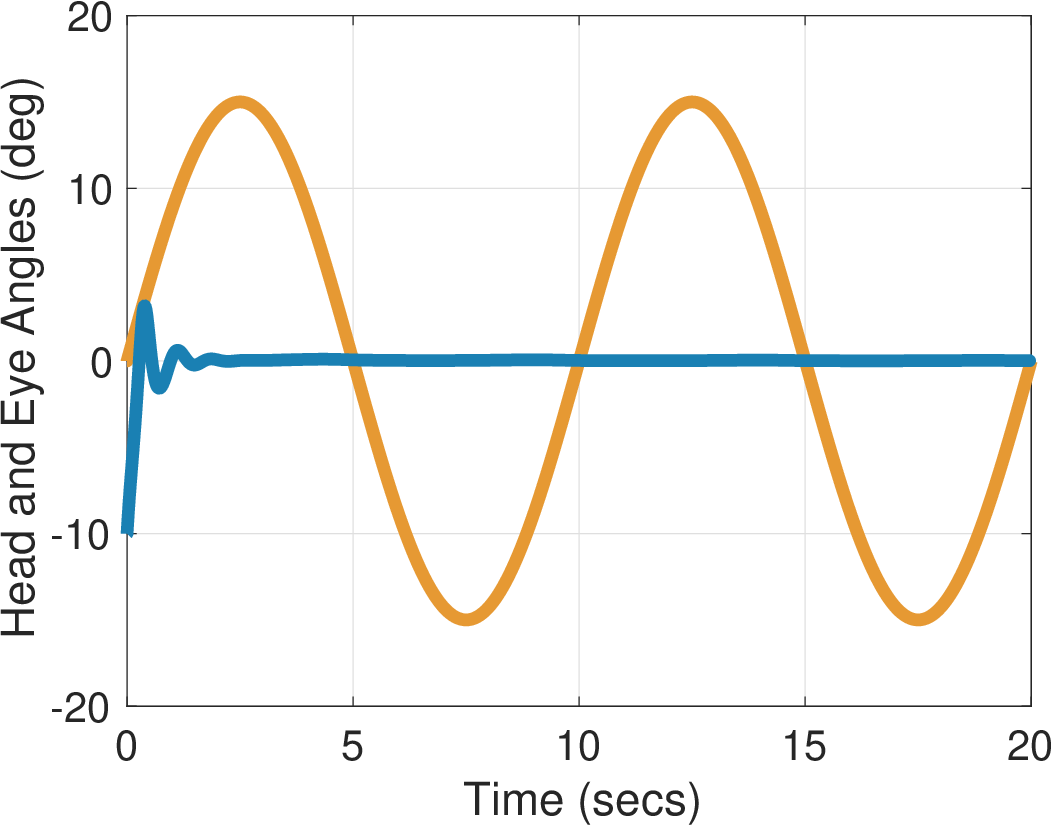}
\end{subfigure}
\begin{subfigure}[b]{.25\textwidth}
\centering
\includegraphics[width=.9\linewidth]{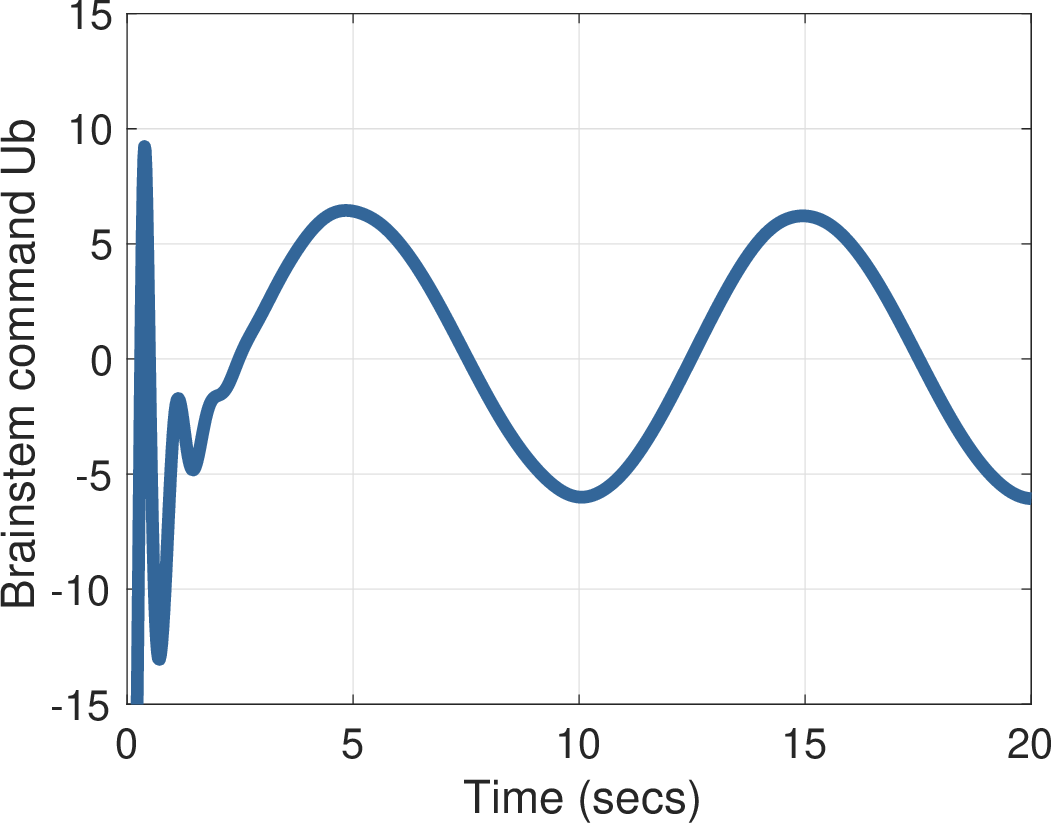}
\end{subfigure} 
\begin{subfigure}[b]{.25\textwidth}
\centering
\includegraphics[width=.9\linewidth]{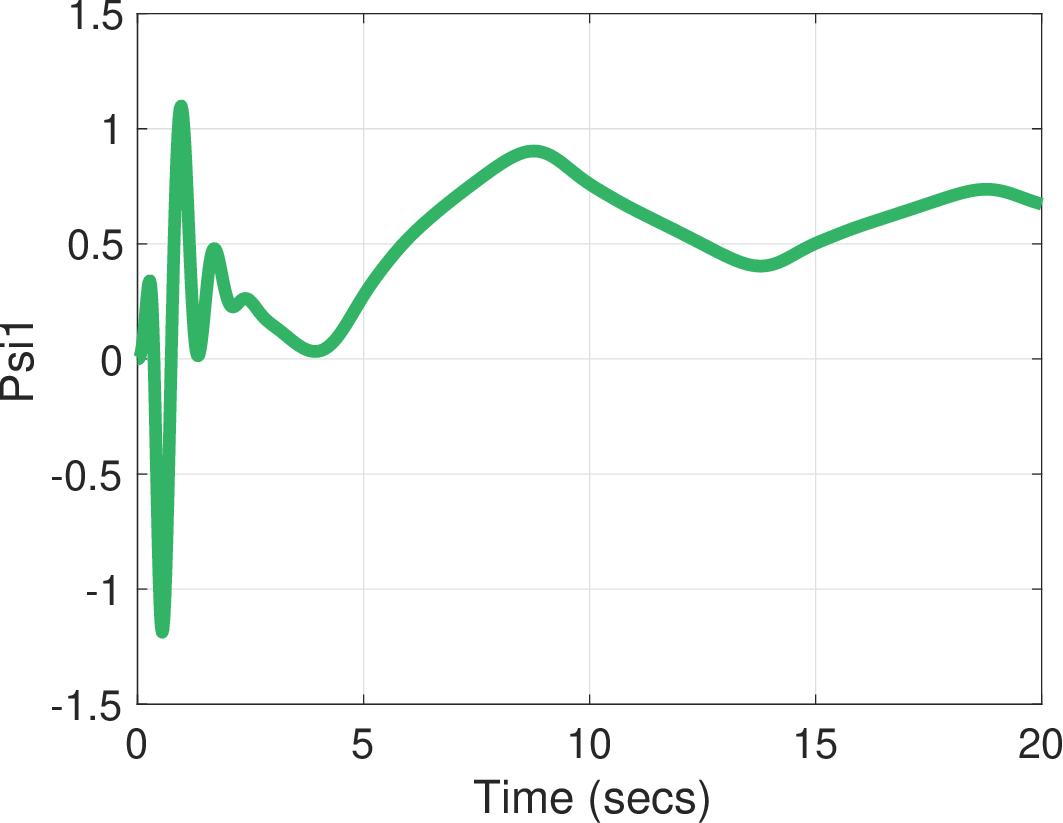}
\end{subfigure} \\ ~ \\
\begin{subfigure}[b]{.25\textwidth}
\centering
\includegraphics[width=.9\linewidth]{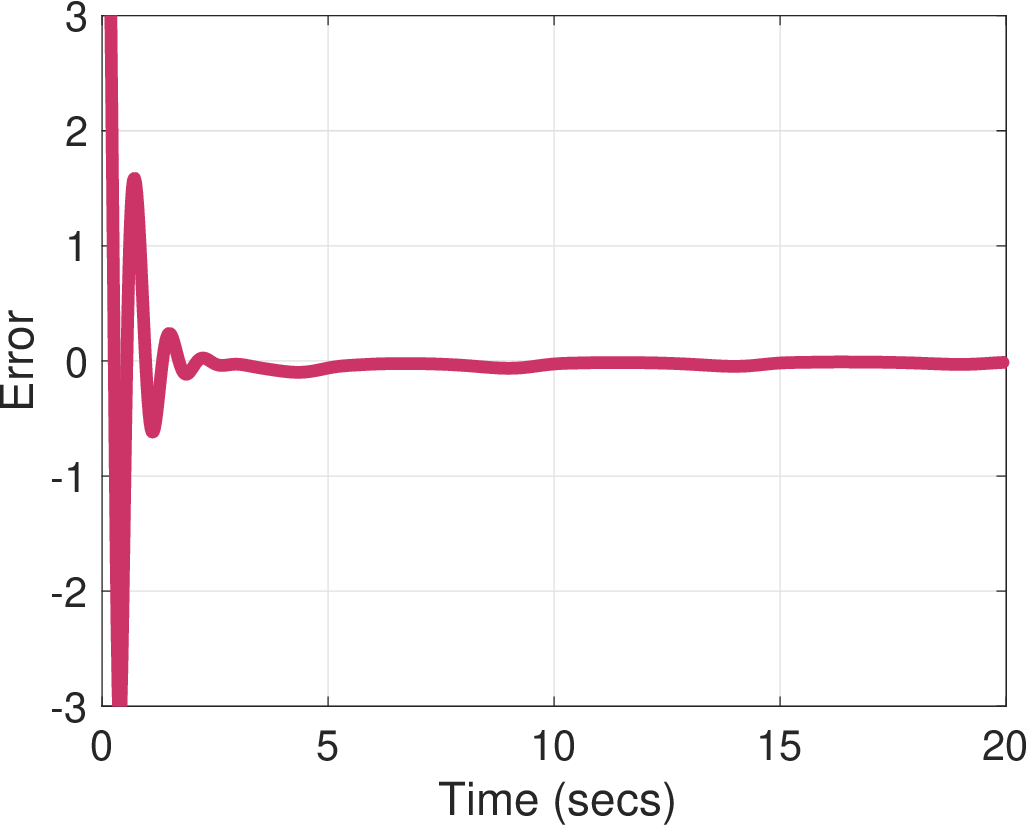}
\end{subfigure}
\begin{subfigure}[b]{.25\textwidth}
\centering
\includegraphics[width=.9\linewidth]{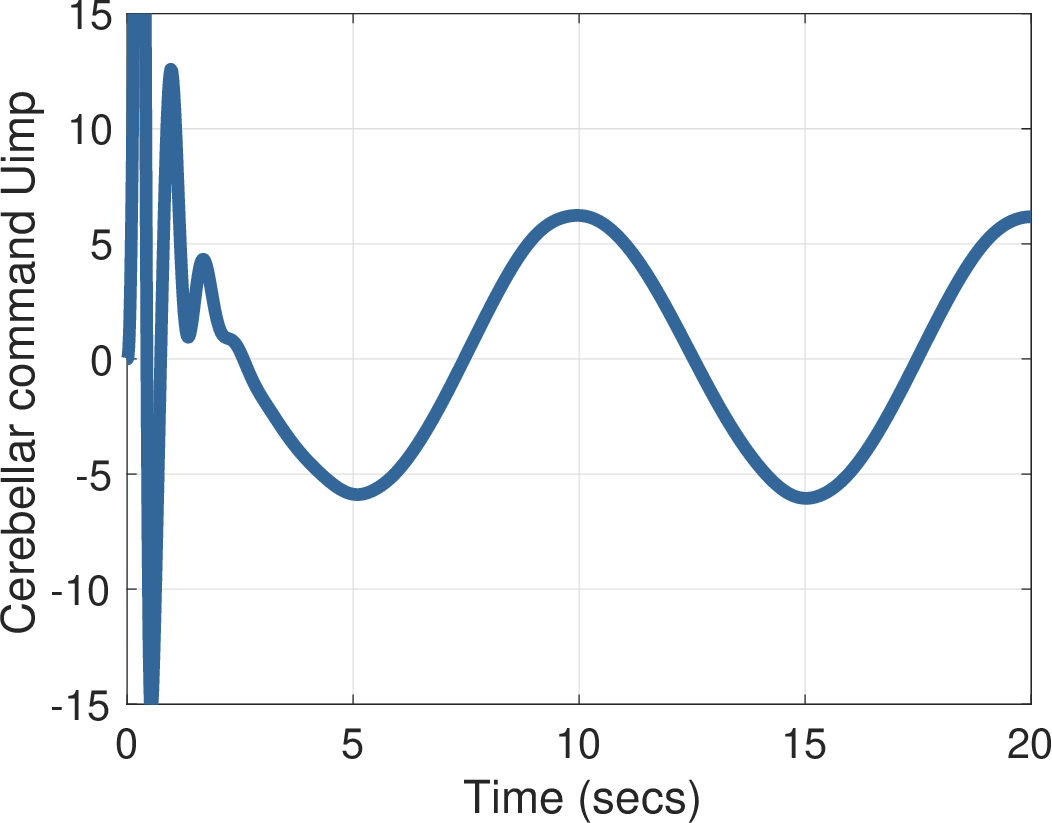}
\end{subfigure} 
\begin{subfigure}[b]{.25\textwidth}
\centering
\includegraphics[width=.9\linewidth]{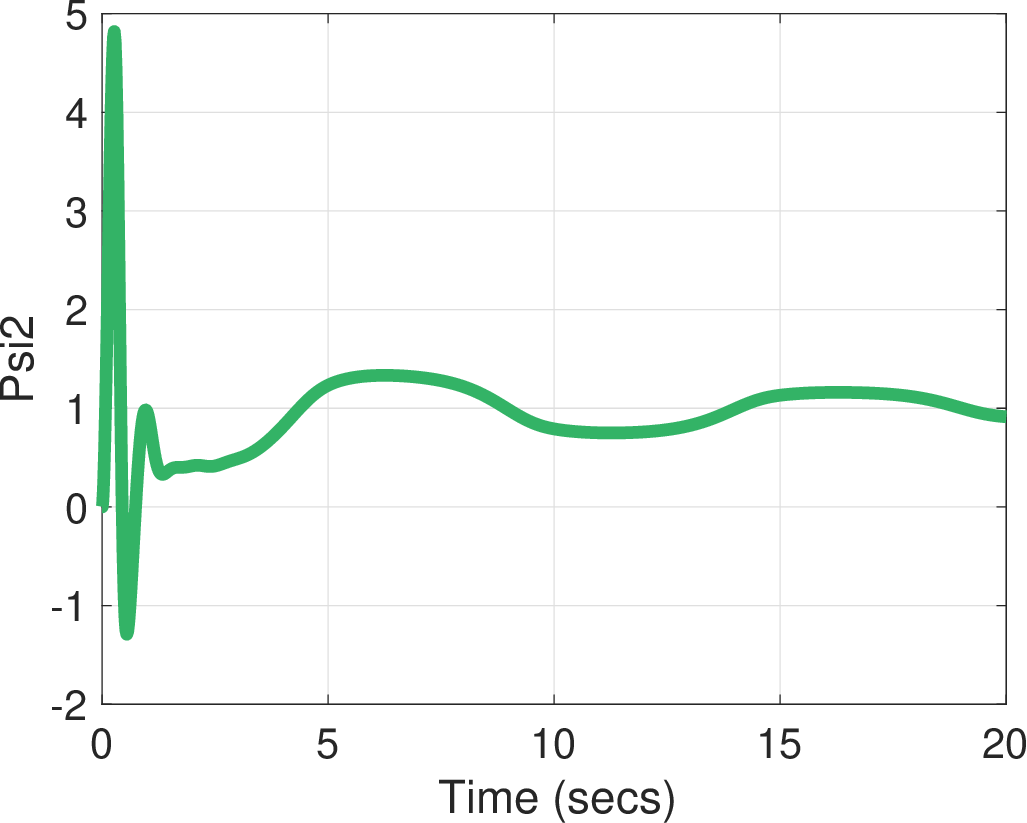}
\end{subfigure} 
\caption{VOR cancellation. The signals are the same as in Figure~\ref{fig:VOR1}.}
\label{fig:VOR6}
\end{figure}

In an experiment called {\em VOR cancellation}, the head is rotated involuntarily while the eyes must 
track a head-fixed target \cite{BUTTNER84}. Suppose the head angle is $x_h(t) = a_h \sin ( \beta_h t )$ 
with $a_h, \beta_h > 0$, and the target angle is $r(t) = x_h(t)$. Then the error is given by $e = - x$. 
The role of $u_{imp}$ in this case is to cancel the disturbance $\alpha_h \dot{x}_h$ introduced by the 
brainstem component $u_b$. Figure~\ref{fig:VOR6} illustrates the results for VOR cancellation using 
our model. Particularly, we note that the response amplitude of the brainstem component is not reduced 
during VOR cancellation, as experimentally confirmed in \cite{BUETTNER79,KELLER75}.

\begin{figure}[t!]
\centering
\begin{subfigure}[b]{.25\textwidth}
\centering
\includegraphics[width=.9\linewidth]{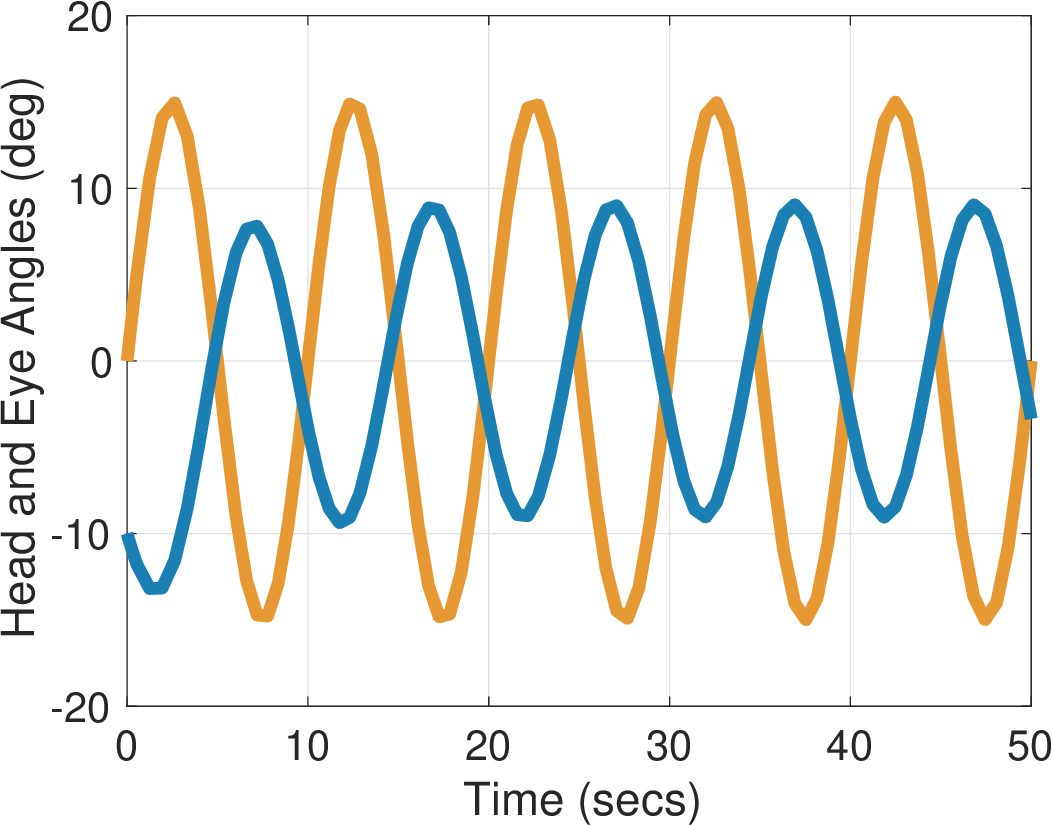}
\end{subfigure}
\begin{subfigure}[b]{.25\textwidth}
\centering
\includegraphics[width=.9\linewidth]{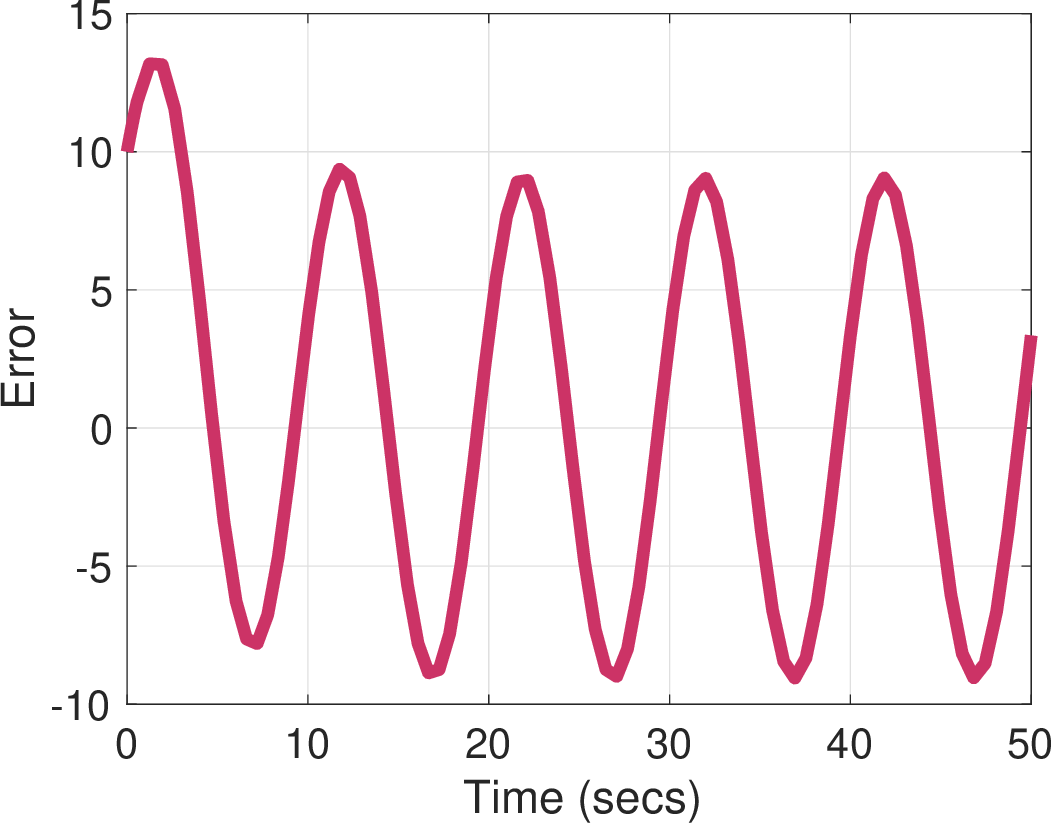}
\end{subfigure} 
\begin{subfigure}[b]{.25\textwidth}
\centering
\includegraphics[width=.9\linewidth]{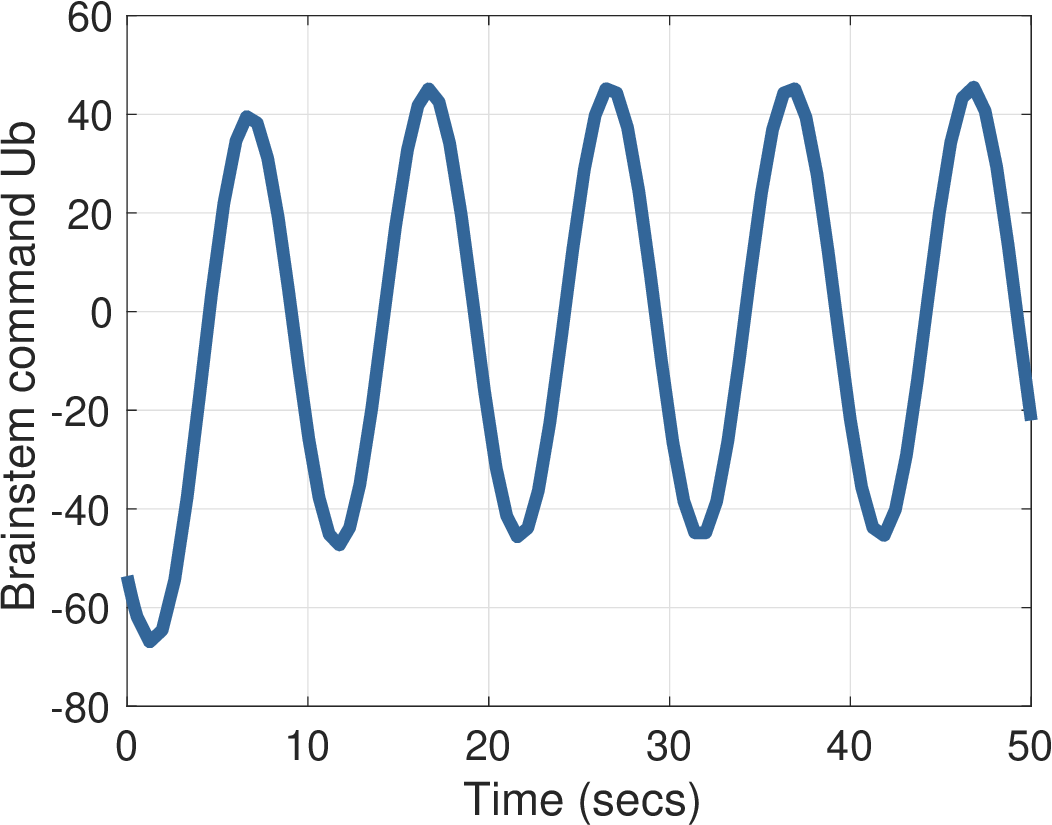}
\end{subfigure} 
\caption{VOR cancellation with the cerebellum disabled.
From left to right, the head (yellow) and eye (blue) angles, the retinal error $e$, and the 
brainstem component $u_b$.}
\label{fig:VOR7}
\end{figure}

A number of researchers have studied the VOR in the situation when the cerebellum is disabled 
either due to disease or cerebellectomy \cite{CARPENTER72,ZEE76,ZEE81}. We illustrate this effect 
for the previous scenario of VOR cancellation, but now with $u_c = 0$. Simulation results are 
shown in Figure~\ref{fig:VOR7}. What we observe in the left figure is that the subject is no 
longer able to suppress the VOR - the blue curve shows that the eye position is not stabilized, 
despite a head-fixed target. This result corroborates the experimental findings in \cite{ZEE81}. 

A careful study of the effects of disabling the neural integrator on the VOR, OKR, gazing holding, 
and smooth pursuit appeared in \cite{CANNON87}. Here we discuss the VOR in the dark. In our model, 
disabling the neural integrator corresponds to disabling the observer \eqref{eq:xhatdot}. This 
means the brainstem component of the control input no longer includes the estimate $-K_x \hat{x}$. 
Since the VOR is being tested in darkness, the cerebellum makes no compensation for this missing 
estimate of the oculomotor plant drift term. Therefore, without the neural integrator, the eye 
position evolves according to the dynamics
\begin{equation}
\label{eq:dark2}
\dot{x} = - K_x x - \alpha_h \dot{x}_h \,.
\end{equation}
Comparing with \eqref{eq:dark}, we see the difference is in the constant $K_x$, which is larger
than $\Kxt$. For instance, if the head angular velocity is a constant $\dot{x}_h = v$, then eye 
position converges exponentially to $\ol{x} = - \alpha_h v / K_x$, rather than approximately 
tracking a ramp (with a very slow exponential decay). This is precisely the behavior recovered 
in experiments \cite{CANNON87}: a step of constant head velocity in total darkness evoked a step 
change in eye position, not in eye velocity. The author's of \cite{CANNON87} interpreted this 
behavior by saying ``the step in head velocity was not integrated in the brainstem to produce 
a ramp of eye position''. 

\begin{figure}[t!]
\centering
\begin{subfigure}[b]{.25\textwidth}
\centering
\includegraphics[width=.9\linewidth]{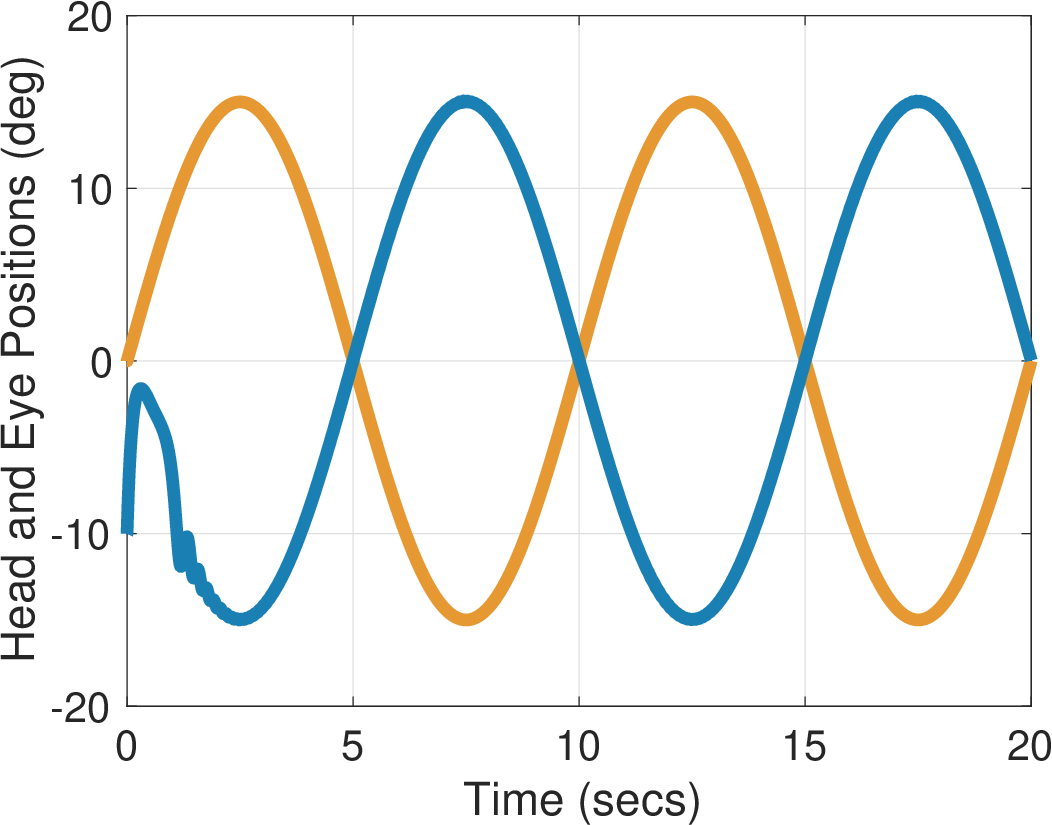}
\end{subfigure}
\begin{subfigure}[b]{.25\textwidth}
\centering
\includegraphics[width=.9\linewidth]{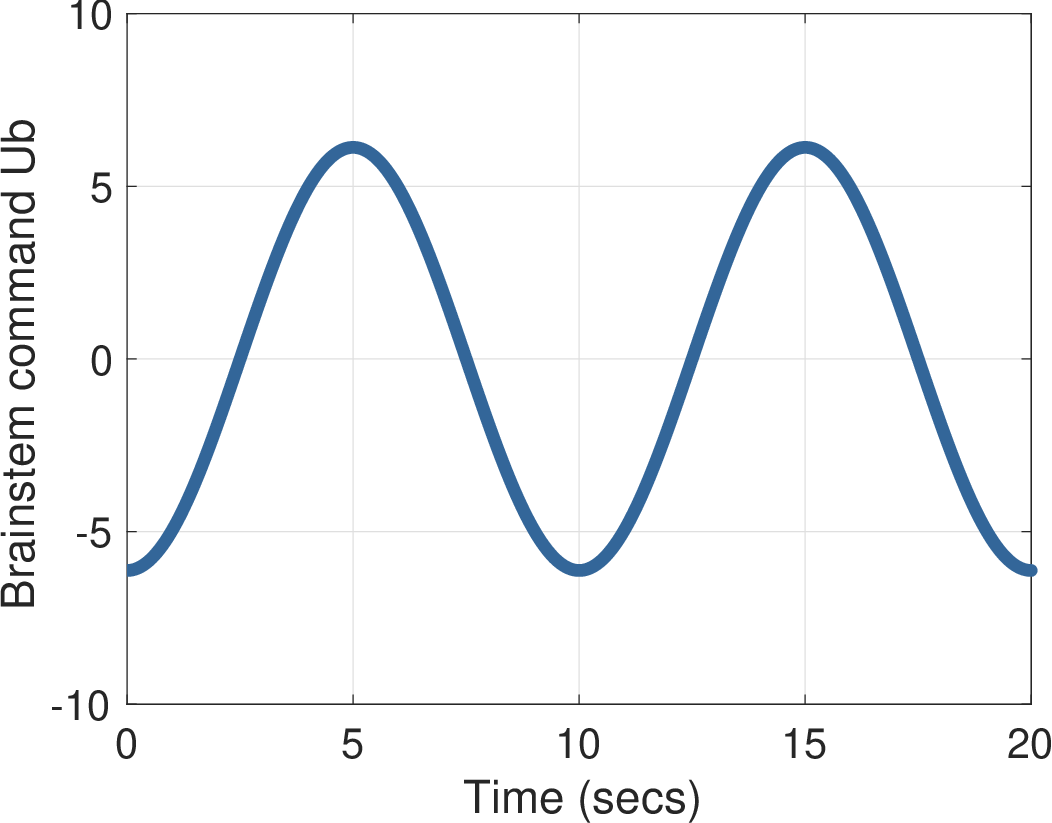}
\end{subfigure} 
\begin{subfigure}[b]{.25\textwidth}
\centering
\includegraphics[width=.9\linewidth]{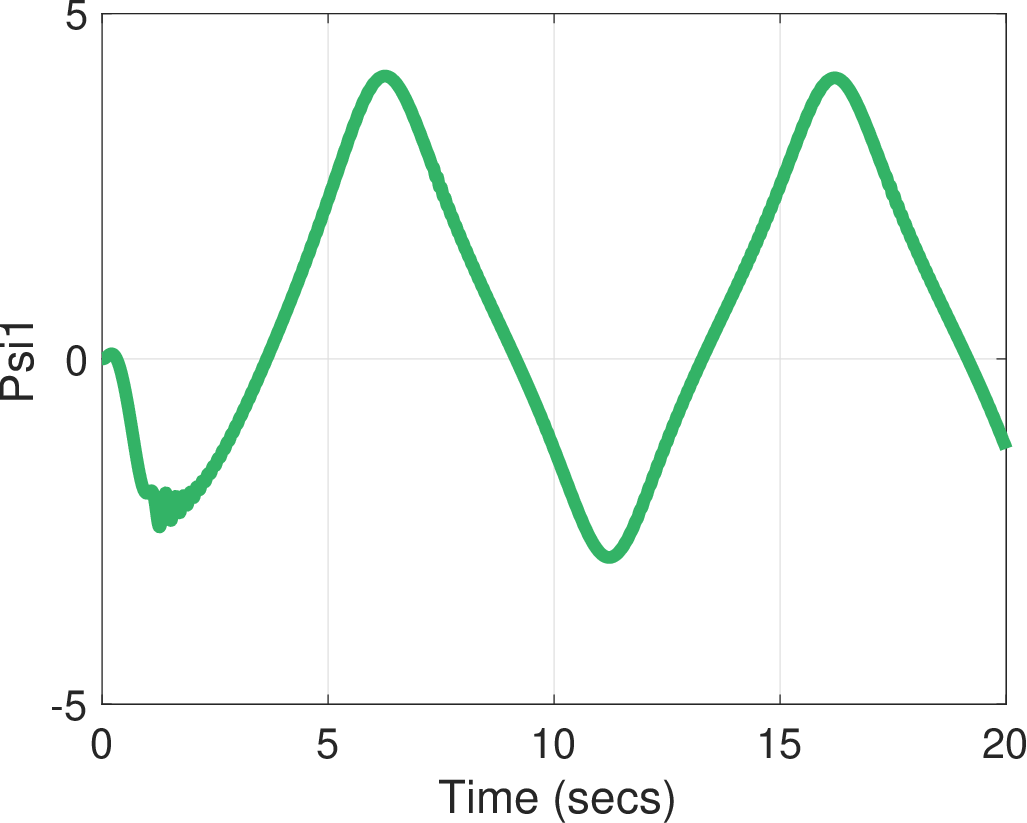}
\end{subfigure} \\ ~ \\
\begin{subfigure}[b]{.25\textwidth}
\centering
\includegraphics[width=.9\linewidth]{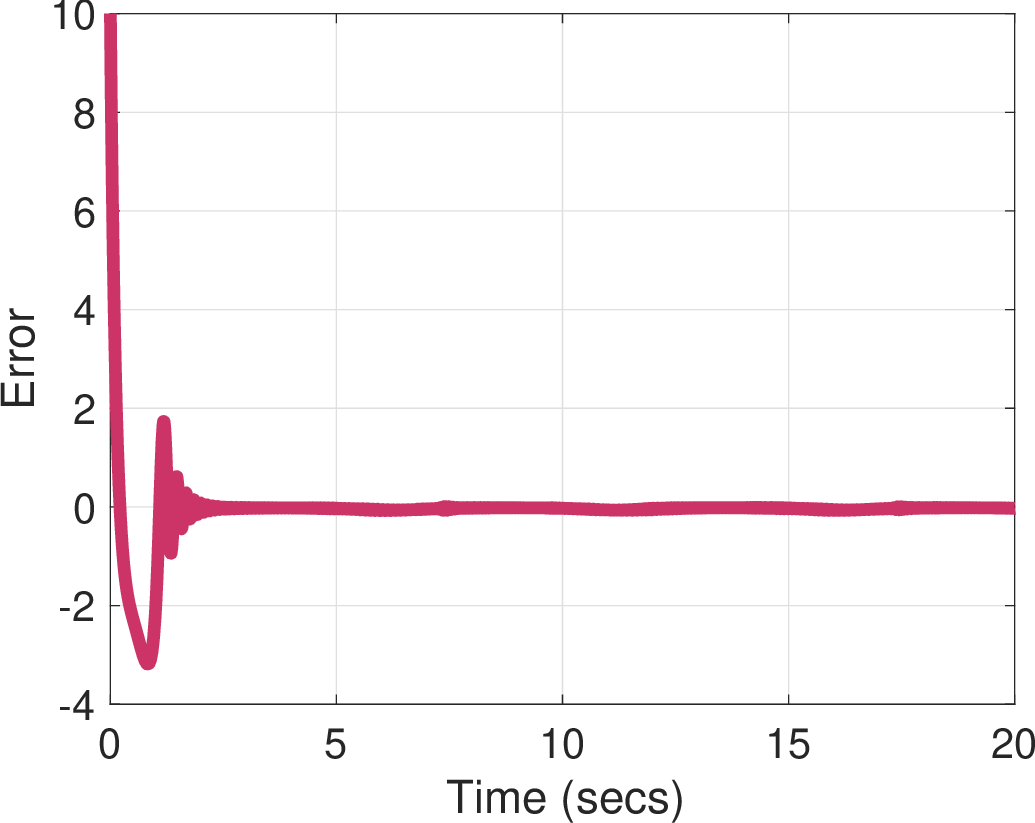}
\end{subfigure}
\begin{subfigure}[b]{.25\textwidth}
\centering
\includegraphics[width=.9\linewidth]{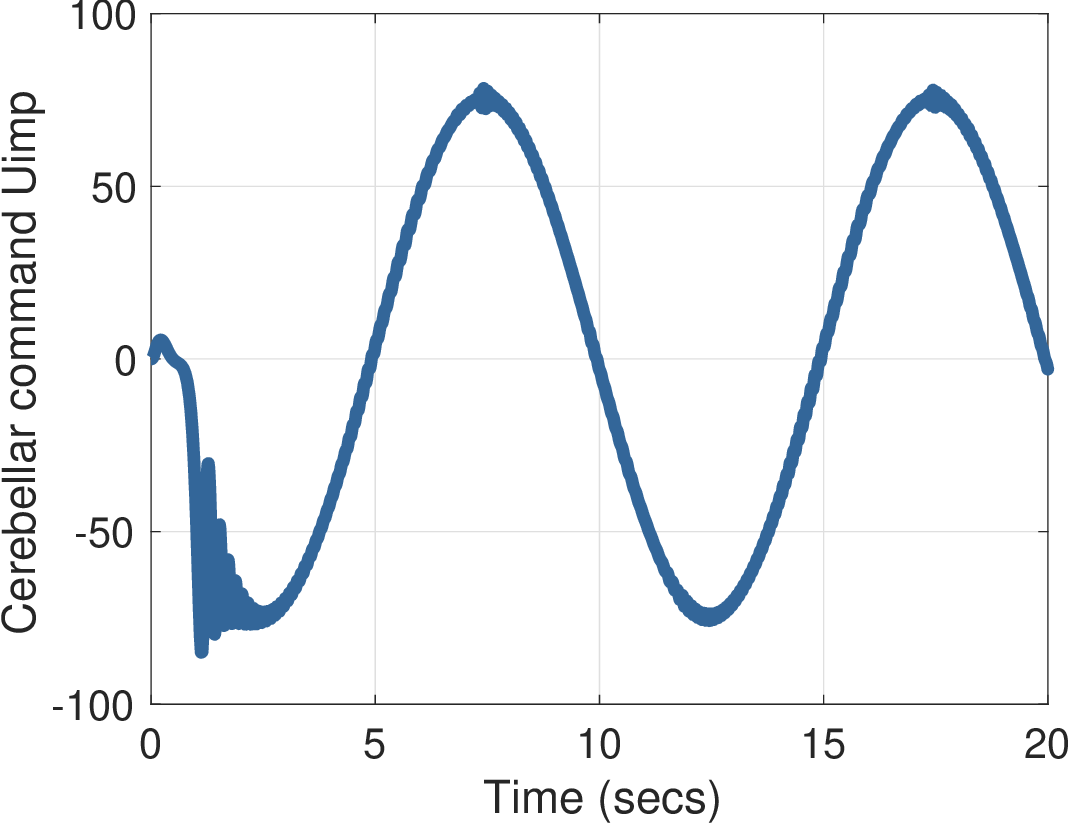}
\end{subfigure} 
\begin{subfigure}[b]{.25\textwidth}
\centering
\includegraphics[width=.9\linewidth]{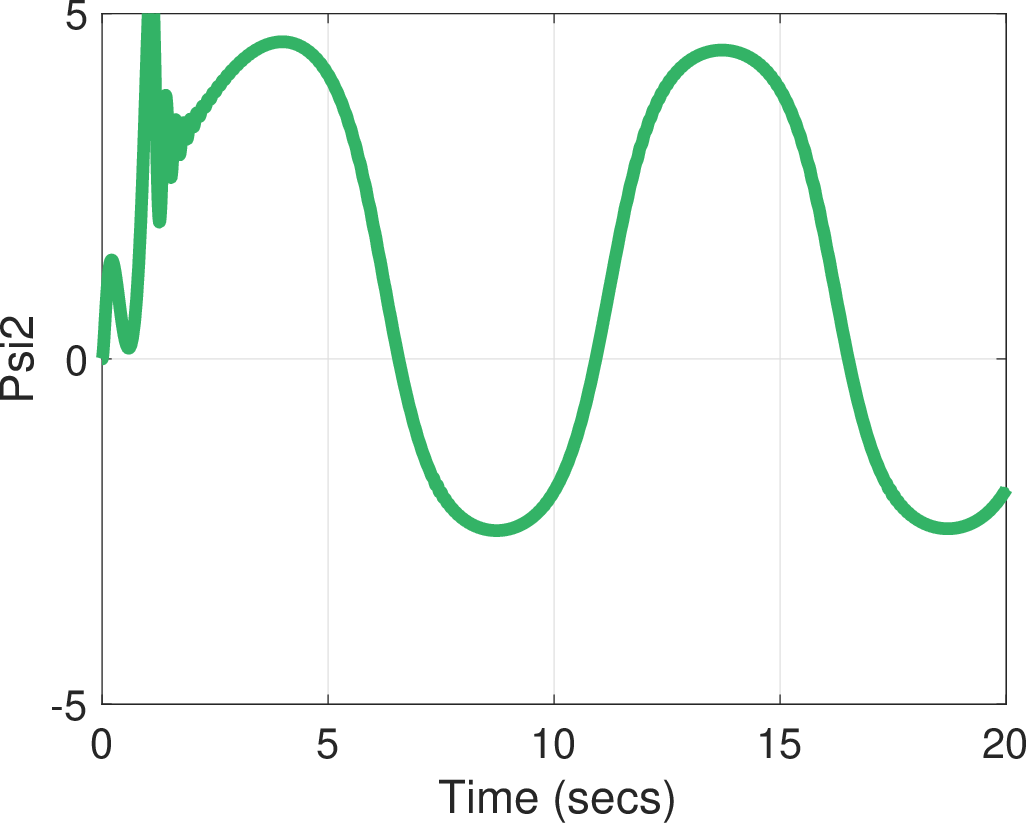}
\end{subfigure} 
\caption{VOR in the light with the neural integrator disabled. The signals are the same as in 
Figure~\ref{fig:VOR1}.}
\label{fig:VOR8}
\end{figure}

A further study of the effects of disabling the neural integrator on the VOR, OKR, and smooth pursuit 
in monkeys appeared in \cite{KANEKO99}. They found these systems are minimally affected after a 
recovery period.  Our model predicts that in the light, the cerebellum will compensate for the additional 
disturbances arising from the removal of the term $-\alpha_x \hat{x}$, such that the VOR is only mildly 
affected, as reported in \cite{KANEKO99}. Figure~\ref{fig:VOR8} shows the behavior of the VOR in the 
light with the neural integrator disabled, $x_h(t) = a_h \sin ( \beta_h t )$, $a_h = 15$, and 
$\beta_h = 0.1$Hz for $t \in [0,20]$. We observe the eye moves opposite to the head rotation, as expected. 

\subsection{OKR}

The optokinetic reflex is elicited by movement of large objects in the visual field or movement of the 
visual surround; it operates in tandem with the VOR. We consider the case of the visual surround rotating 
sinusoidally, $r_{vs}(t) = - a_{v} \sin ( \beta_{v} t)$, for example by using an optical 
drum \cite{COLLEWIJN74}. The head may be stationary, moving with the visual surround, or moving 
independently but involutarily. The eyes may be fixating on a stationary
target, a head-fixed target, a drum-fixed target, or a target moving within the moving visual field.

The motion of the visual surround may induce in the subject a perception of a stationary background, with 
the head and target moving with respect to (w.r.t.) a stationary background. If $r(t)$ and $x_h(t)$ are 
the target and head angles w.r.t. a fixed inertial frame, then the apparent head and target motion w.r.t. 
the visual surround are given by $r^{vs}(t) = r(t) - r_{vs}(t)$ and $x_h^{vs}(t) = x_h(t) - r_{vs}(t)$. 
The perceived error is given by $e = r^{vs} - x_h^{vs} - x = r - x_h - x$. We see that the retinal 
error is unaffected. Mathematically speaking, the situation is the same as the VOR with a fixed visual 
surround. 

In many experiments with the OKR, the eyes must track a drum-fixed light slit with the
head stationary and the optical drum rotating sinusoidally. In this case the error is $e = r - x$, 
where $r(t) = a_h \sin ( \beta_h t )$. We treat this situation as being the same as smooth pursuit, 
to be discussed below. In an experiment called {\em OKR cancellation}, a light spot at $r = 0$ 
is placed in front of a moving striped optical drum. In this case, the pursuit system appears to 
override the OKR, as the eyes fixate on the fixed light spot, and the error is $e = - x$. If 
there is no head rotation, then this situation is the same as gaze holding, discussed in the 
next subsection.    

In an experiment called {\em visual-vestibular conflict} the head and the optokinetic drum are 
mechanically coupled so that they rotate together, and the eyes must track a light strip on the 
drum \cite{COLLEWIJN74}. Therefore, we have $r(t) = x_h(t) = a_h \sin ( \beta_h t )$, so 
$e = r - x_h - x = -x$. From the point of view of our mathematical model, this situation is no 
different than VOR cancellation. It has been reported that under such stimulation, the modulation 
of the firing rate of the cerebellum is larger than when the drum is not rotated \cite{WAESPEHENN78}; 
that is, when $r(t) = 0$, $x_h(t) = a_h \sin ( \beta_h t )$, and $e = -x_h - x$. 

\begin{figure}[t!]
\centering
\begin{subfigure}[b]{.25\textwidth}
\centering
\includegraphics[width=.9\linewidth]{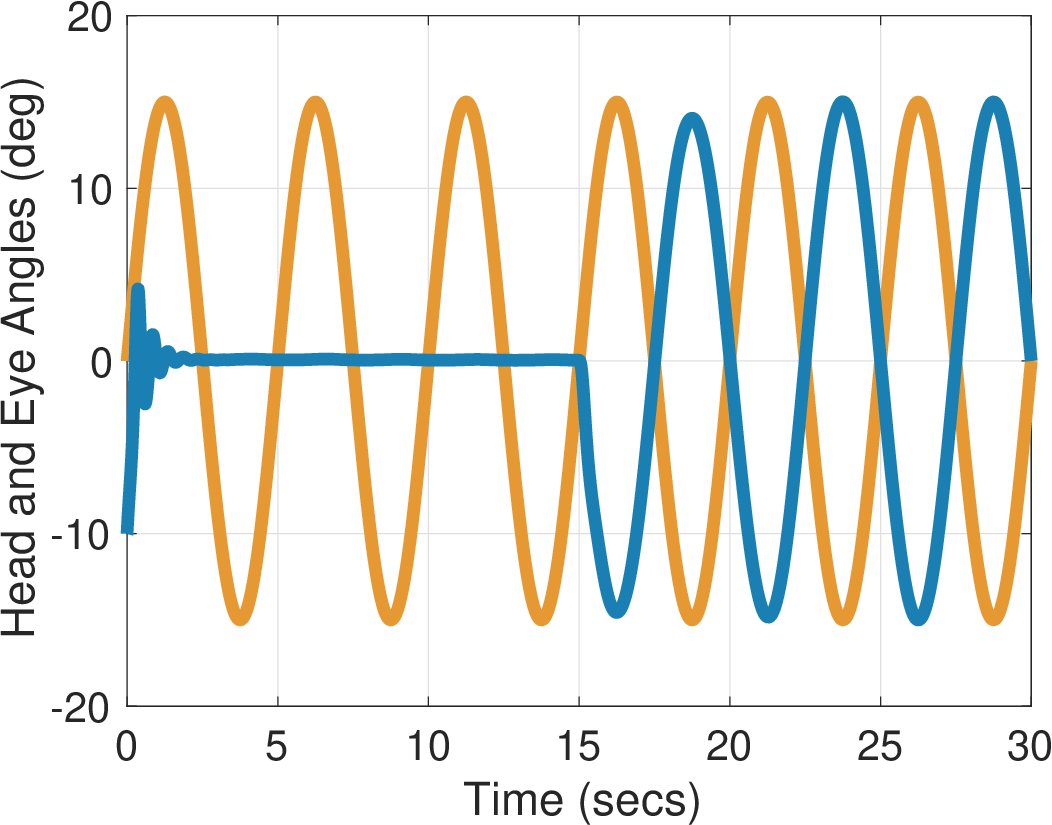}
\end{subfigure}
\begin{subfigure}[b]{.25\textwidth}
\centering
\includegraphics[width=.9\linewidth]{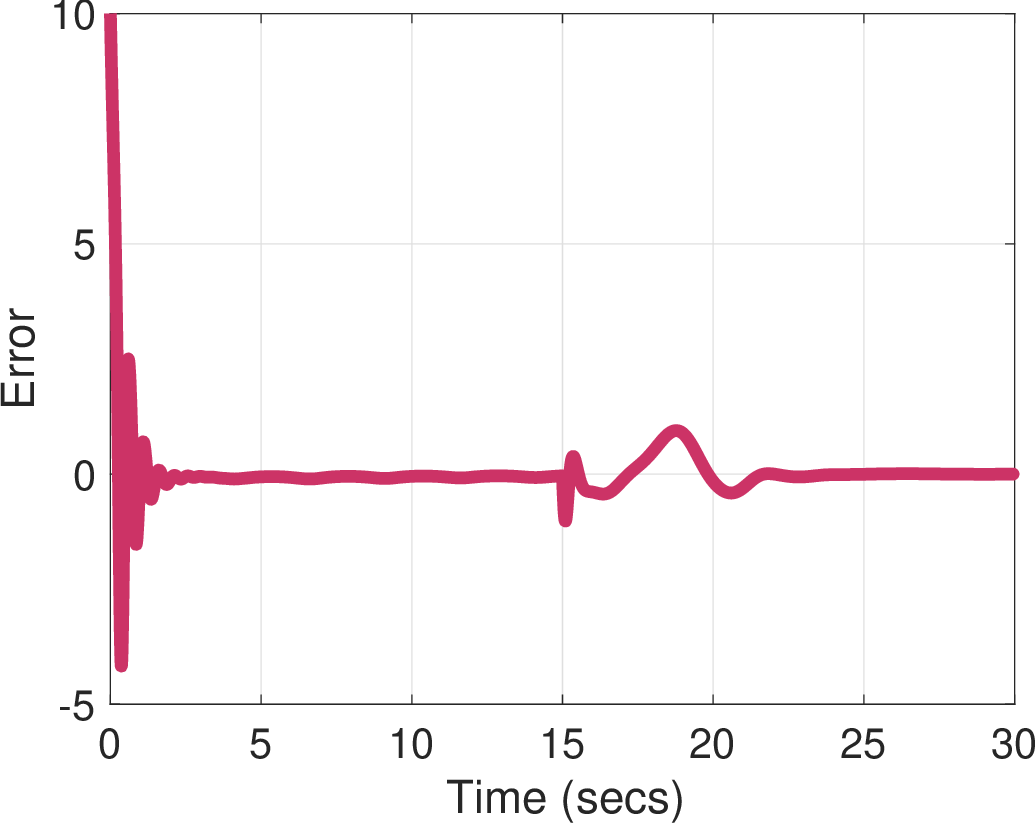}
\end{subfigure} 
\begin{subfigure}[b]{.25\textwidth}
\centering
\includegraphics[width=.9\linewidth]{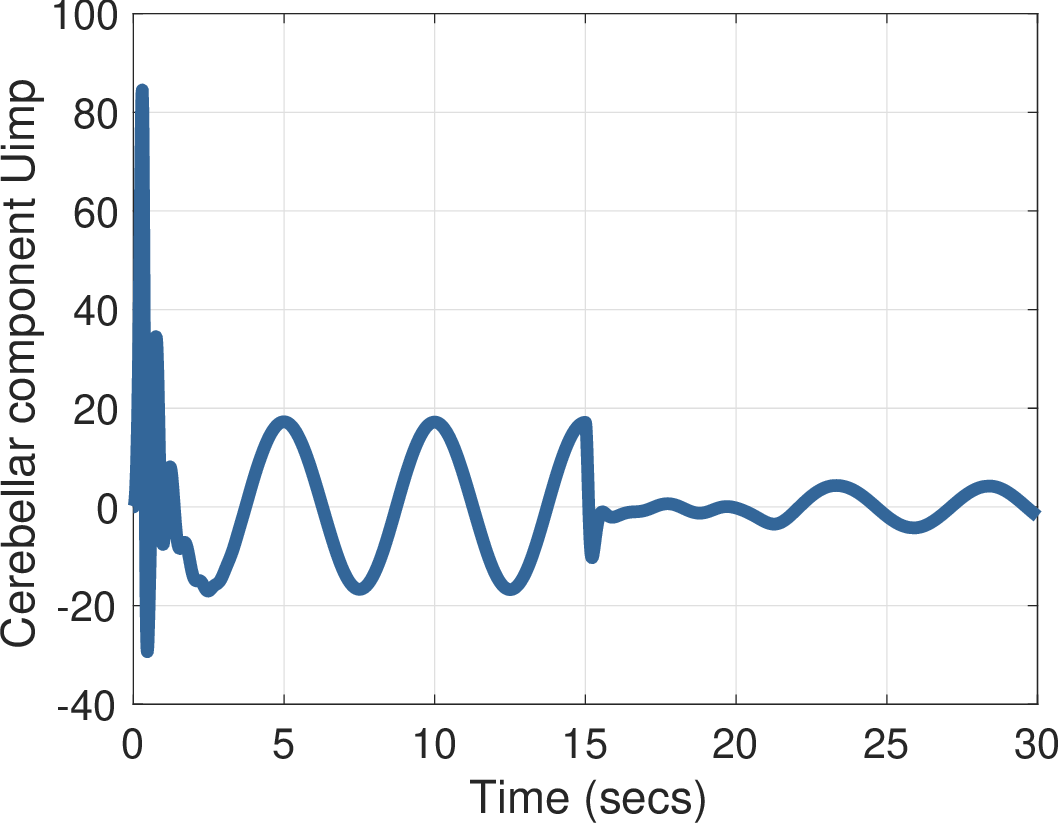}
\end{subfigure} 
\caption{Visuo-vestibular conflict in the OKR and its effect on the depth of modulation of the 
cerebellar output $u_{imp}$. From left to right, the head (yellow) and eye (blue) angles, 
the retinal error $e$, and the cerebellar output $u_{imp}$.}
\label{fig:OKR1}
\end{figure}

In the context of our model, this finding makes sense. In the first case, the role of $u_{imp}$ 
is to cancel the term $\alpha_h \dot{x}_h$. In the second case, the role of $u_{imp}$ is to 
cancel the term $- (1 - \alpha_h) \dot{x}_h - \Kxt x_h$. Assuming that $\alpha_h$ is not close 
to $0.5$ and that $\Kxt$ is close to zero, the amplitude of the latter term is larger than the 
amplitude of the former. Figure~\ref{fig:OKR1} illustrates this comparison for values $\alpha_h = 0.9$; 
$a_h = 15$; $\beta_h = 0.2$Hz; $r = x_h = a_h \sin ( \beta_h t)$ for $t \in [0,15]$; 
and $r = 0$, $x_h = a_h \sin ( \beta_h t)$ for $t \in [15,30]$. 

\subsection{Gaze Fixation}

\begin{figure}[t!]
\centering
\begin{subfigure}[b]{.25\textwidth}
\centering
\includegraphics[width=.9\linewidth]{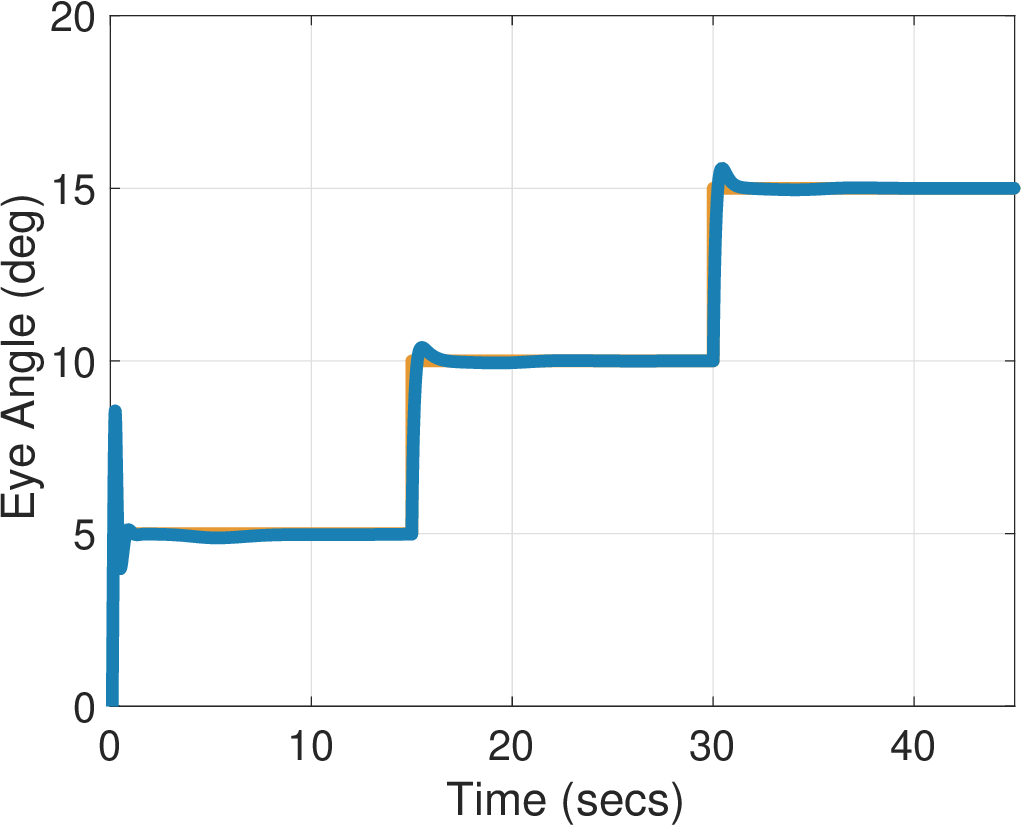}
\end{subfigure}
\begin{subfigure}[b]{.25\textwidth}
\centering
\includegraphics[width=.9\linewidth]{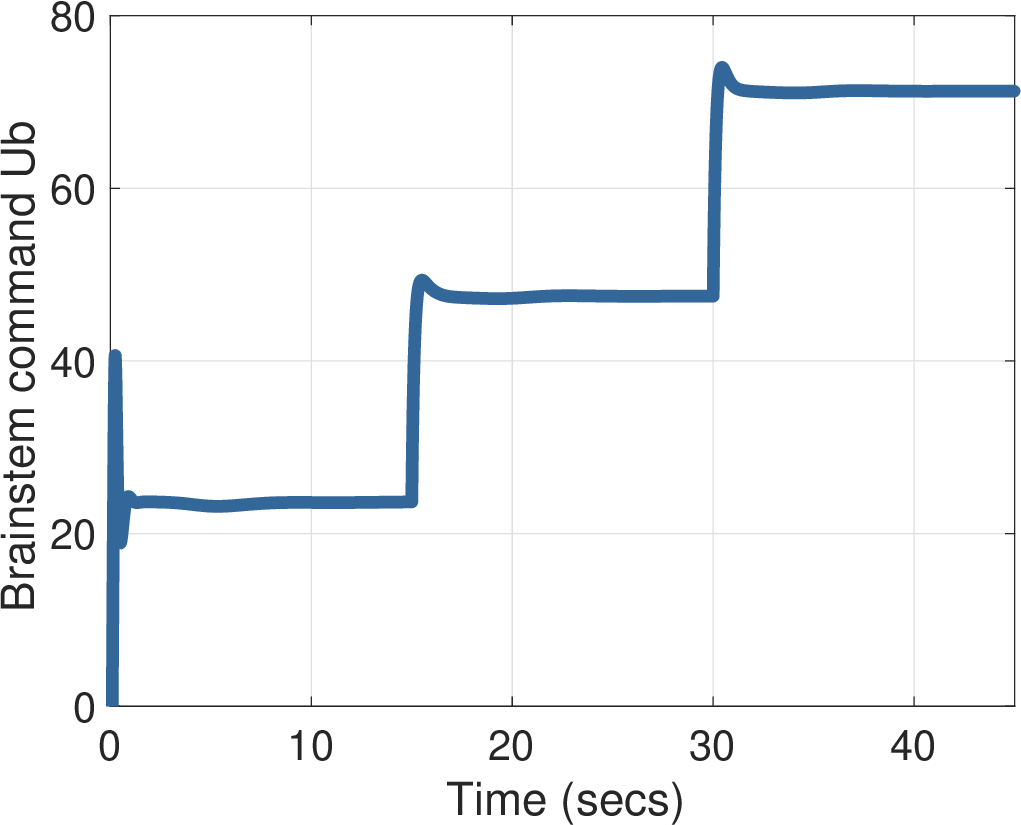}
\end{subfigure} 
\begin{subfigure}[b]{.25\textwidth}
\centering
\includegraphics[width=.9\linewidth]{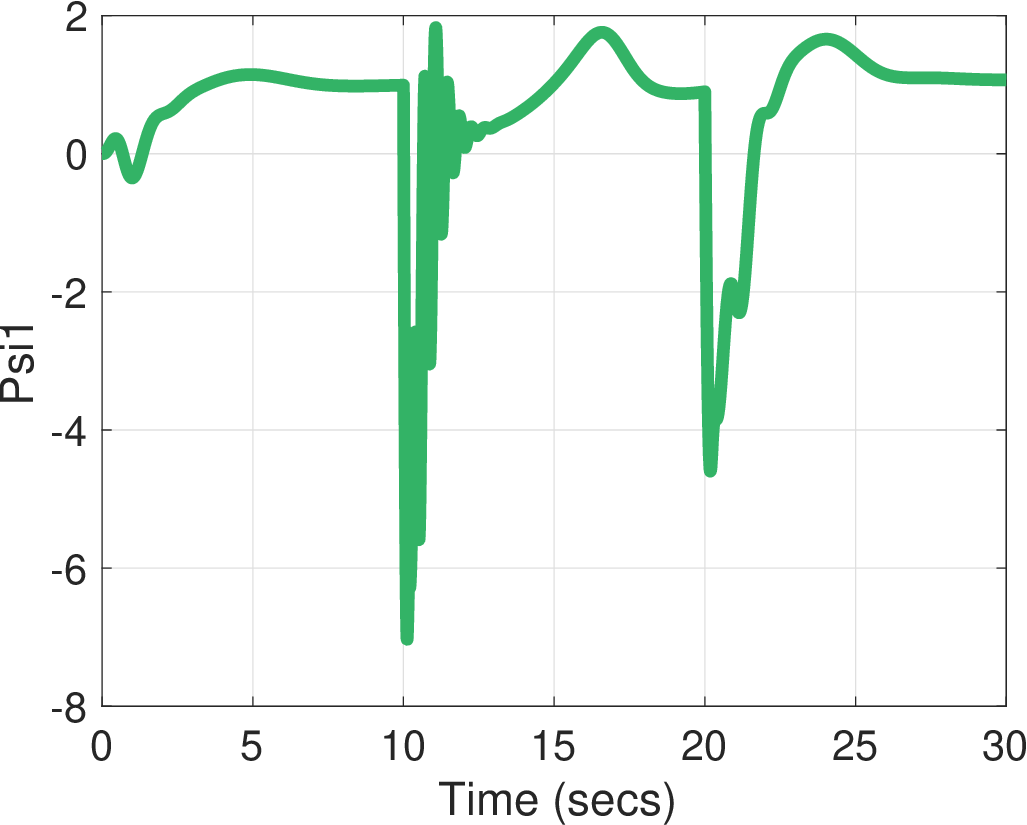}
\end{subfigure} \\
\begin{subfigure}[b]{.25\textwidth}
\centering
\includegraphics[width=.9\linewidth]{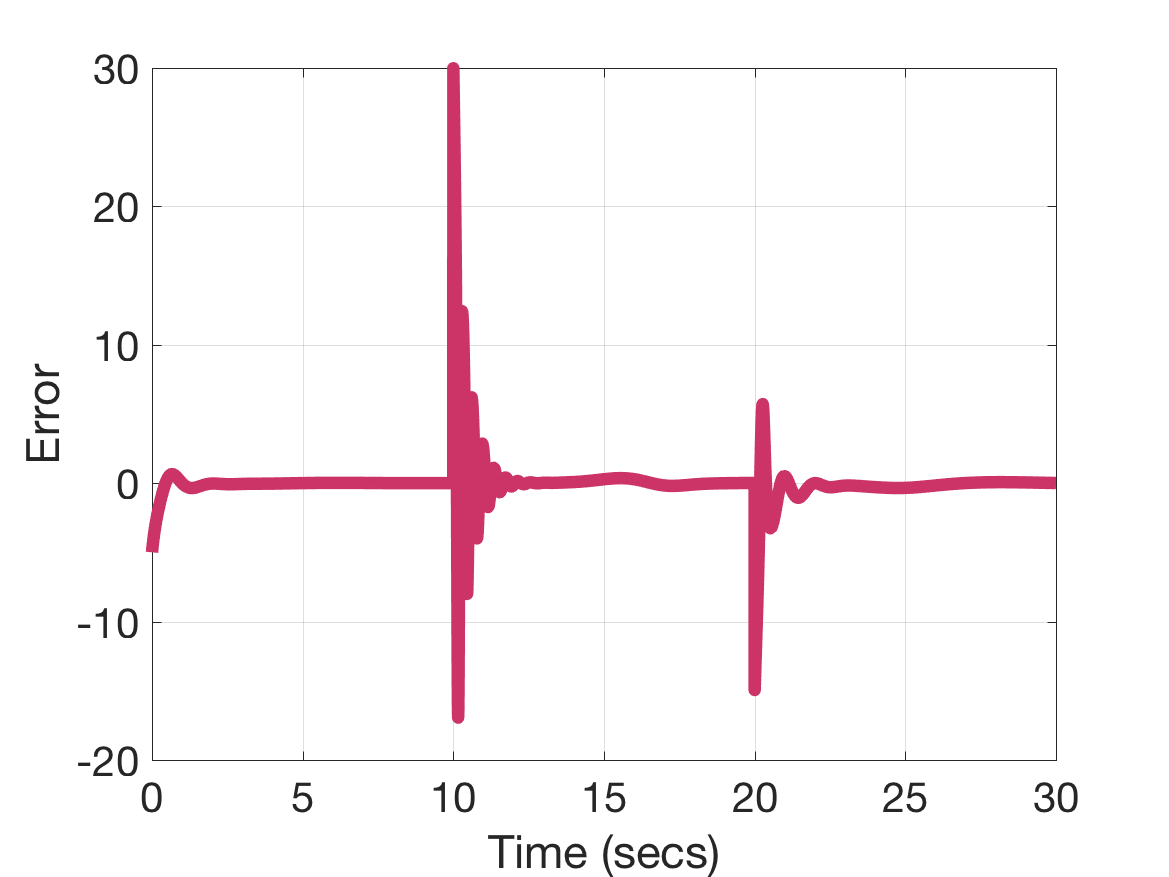}
\end{subfigure}
\begin{subfigure}[b]{.25\textwidth}
\centering
\includegraphics[width=.9\linewidth]{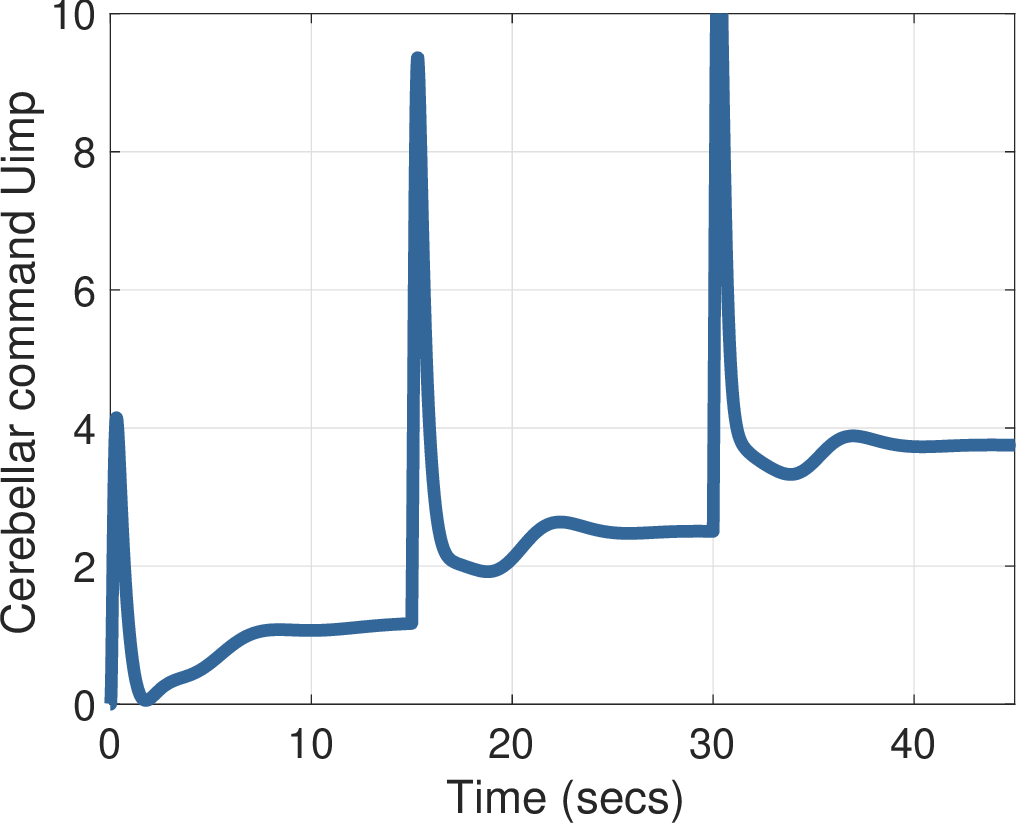}
\end{subfigure} 
\begin{subfigure}[b]{.25\textwidth}
\centering
\includegraphics[width=.9\linewidth]{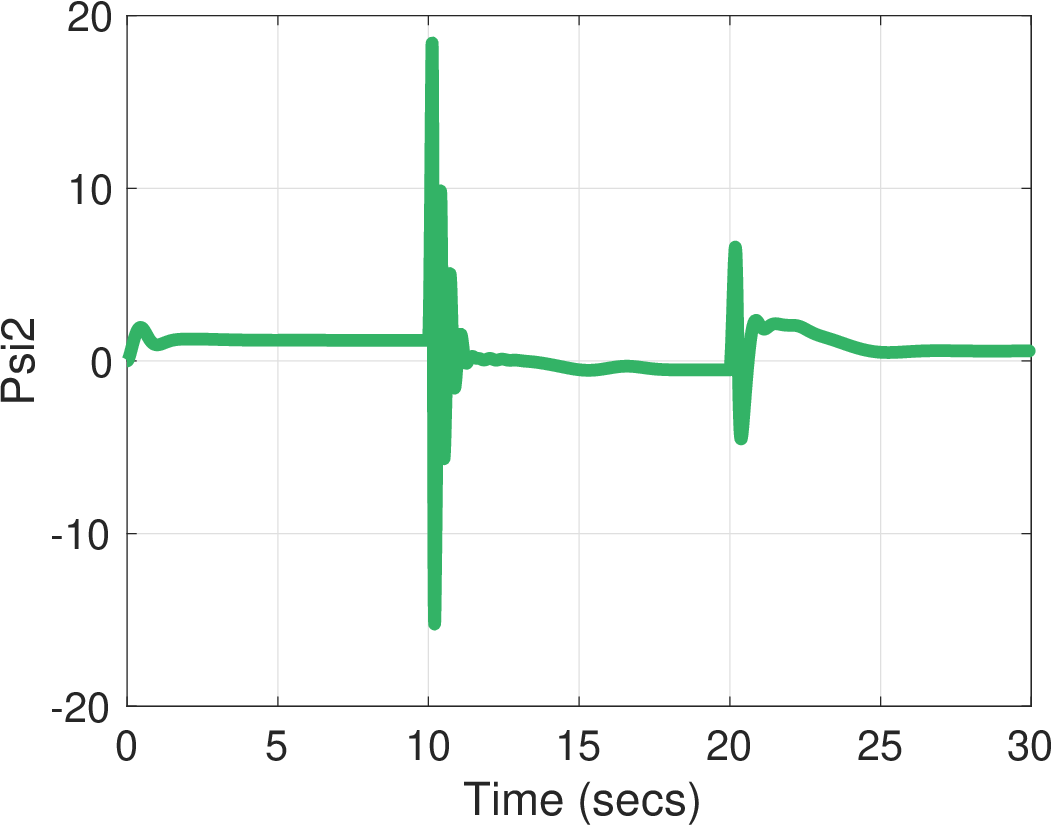}
\end{subfigure} 
\caption{Gaze holding. The signals are the same as in Figure~\ref{fig:VOR1}.}
\label{fig:GAZE1}
\end{figure}

Consider the problem of holding the horizontal gaze on a stationary target with an angle $r \neq 0$ 
while the head is stationary with angle $x_h = 0$. The error is given by $e = r - x$. Assuming that 
$\hat{x}(t) \simeq x(t)$, the error dynamics \eqref{eq:edot} take the form
\begin{equation}
\label{eq:errorGaze}
\dot{e} = - \Kxt e - u_c + \Kxt r \,. 
\end{equation}
We can see that the role of $u_{imp}$ is to estimate the disturbance $\Kxt r$. 
Figure~\ref{fig:GAZE1} shows the behavior for three target angles: $r(t) = 5^{\circ}$ 
for $t \in [0,15]$; $r(t) = 10^{\circ}$ for $t \in [15,30]$, and $r(t) = 15^{\circ}$ 
for $t \ge 30$. We observe that the output of the cerebellum is proportional to the eye angle, a 
behavior observed experimentally in many studies \cite{NODASUZUKI79}. It arises in our model 
because $u_{imp}$ must cancel a disturbance $\Kxt r$, which is proportional to the target position. 

\begin{figure}[t!]
\centering
\begin{subfigure}[b]{.25\textwidth}
\centering
\includegraphics[width=.9\linewidth]{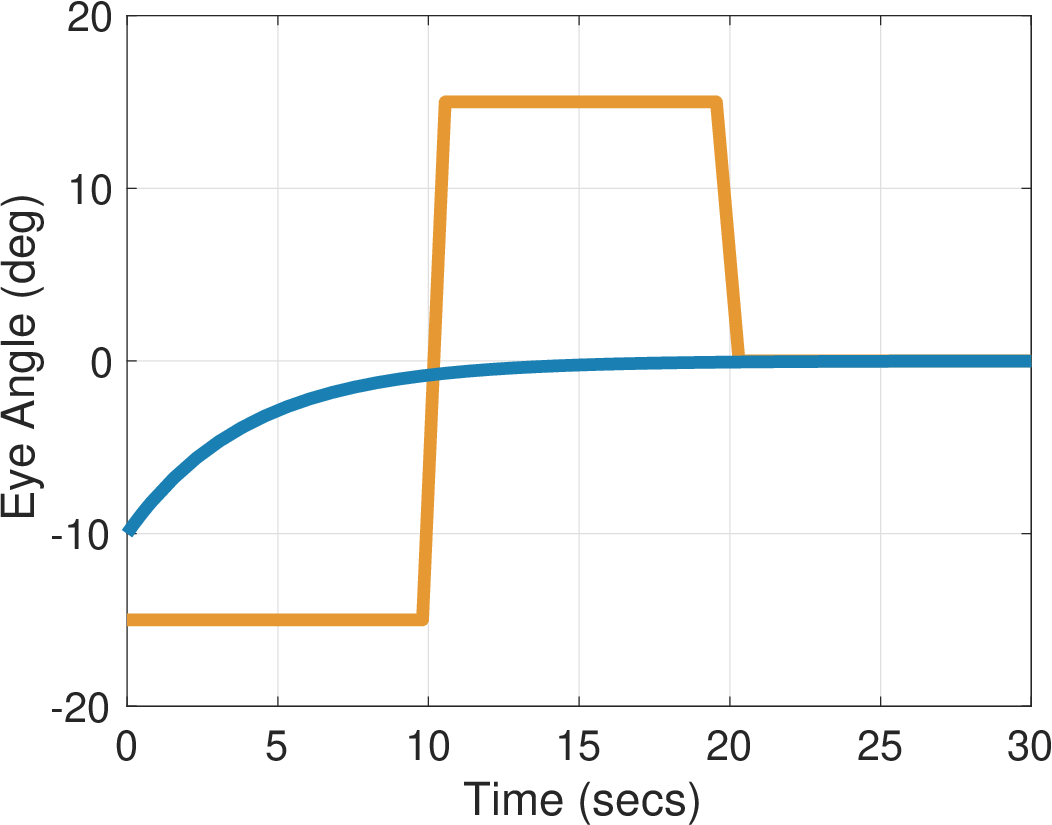}
\end{subfigure}
\begin{subfigure}[b]{.25\textwidth}
\centering
\includegraphics[width=.9\linewidth]{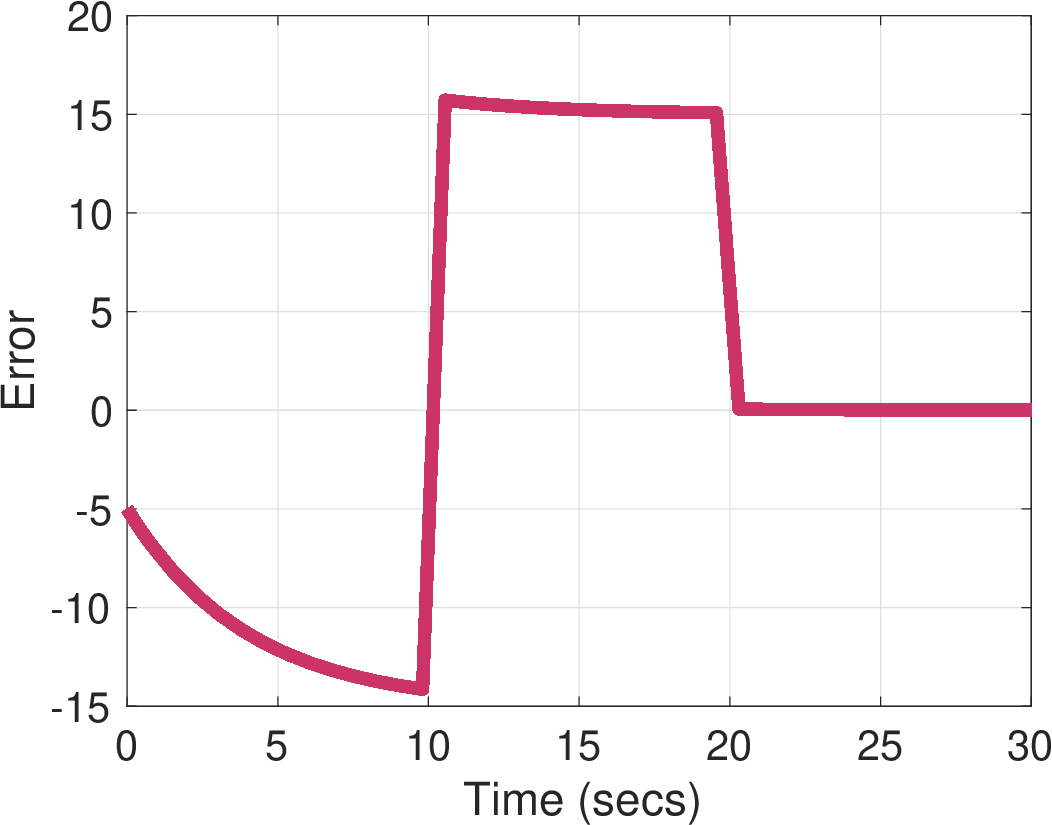}
\end{subfigure} 
\begin{subfigure}[b]{.25\textwidth}
\centering
\includegraphics[width=.9\linewidth]{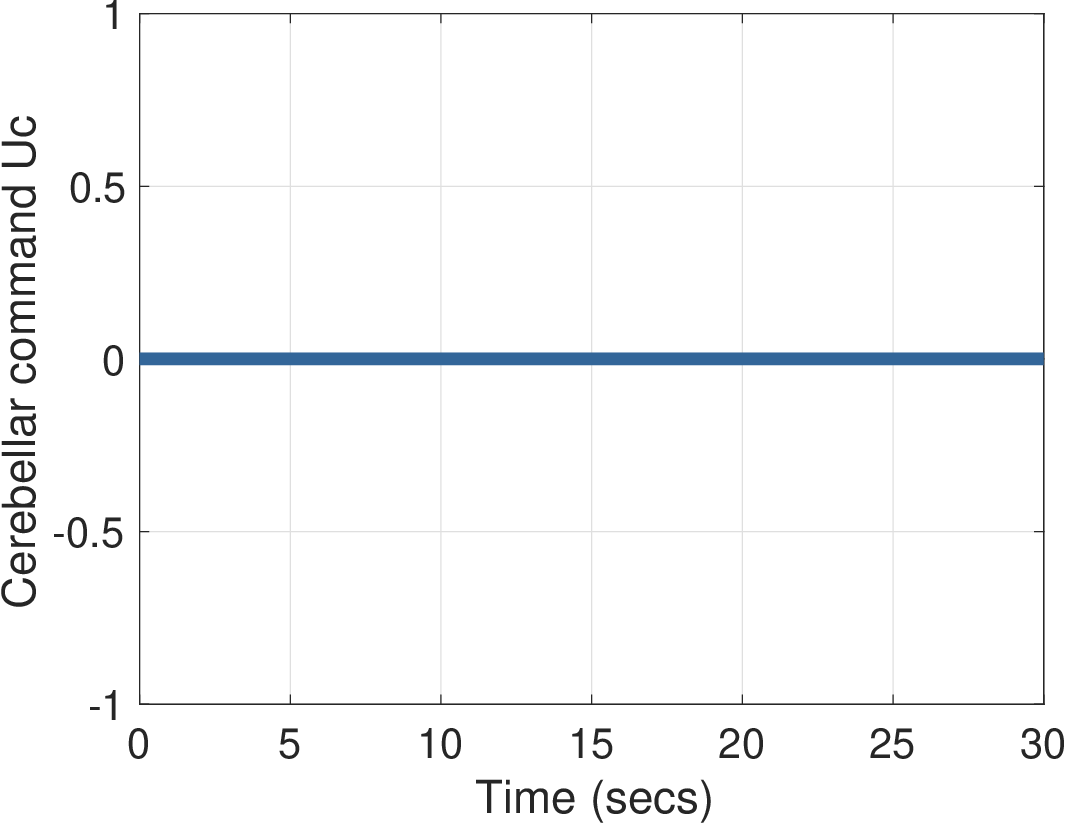}
\end{subfigure} 
\caption{Gaze holding with the cerebellum disabled.} 
\label{fig:GAZE2}
\end{figure}

Further evidence that $\Kxt \neq 0$ comes from studies in which the cerebellum is disabled,
either through ablation or disease. It is well known that in this case, the eye has a slow drift back to
the central position $x = 0$ \cite{CARPENTER72,NODASUZUKI79,ROBINSON74B,SKAVENSKI73,ZEE76}. 
For suppose $x_h = 0$ and $u_c = 0$. Then $u = u_b = \alpha_x \hat{x}$, and assuming 
$\hat{x}(t) \simeq x(t)$, the eye position evolves according to the dynamics
\[
\dot{x} = - \Kxt x \,.
\] 
That is, the eye drifts back to center at an exponential rate determined by $\Kxt$. 
Figure~\ref{fig:GAZE2} depicts this behavior for the same target angles as in Figure~\ref{fig:GAZE1}. 

\subsection{Smooth Pursuit}

We consider a task of the smooth pursuit system in which the eyes must track a horizontally moving 
target.  We assume that any head rotation is involuntary. Let $r(t)$ be the target angle and 
$x_h(t)$ the head angle. The error is given by $e = r - x_h - x$. Assuming that $\hat{x}(t) \simeq x(t)$, 
the error dynamics take the general form in \eqref{eq:edot1}. We observe that the role of 
$u_{imp}$ is to estimate the disturbance 
$\dot{r} + \Kxt r - (1 - \alpha_h) \dot{x}_h - \Kxt x_h$. 

\begin{figure}[t!]
\centering
\begin{subfigure}[b]{.25\textwidth}
\centering
\includegraphics[width=.9\linewidth]{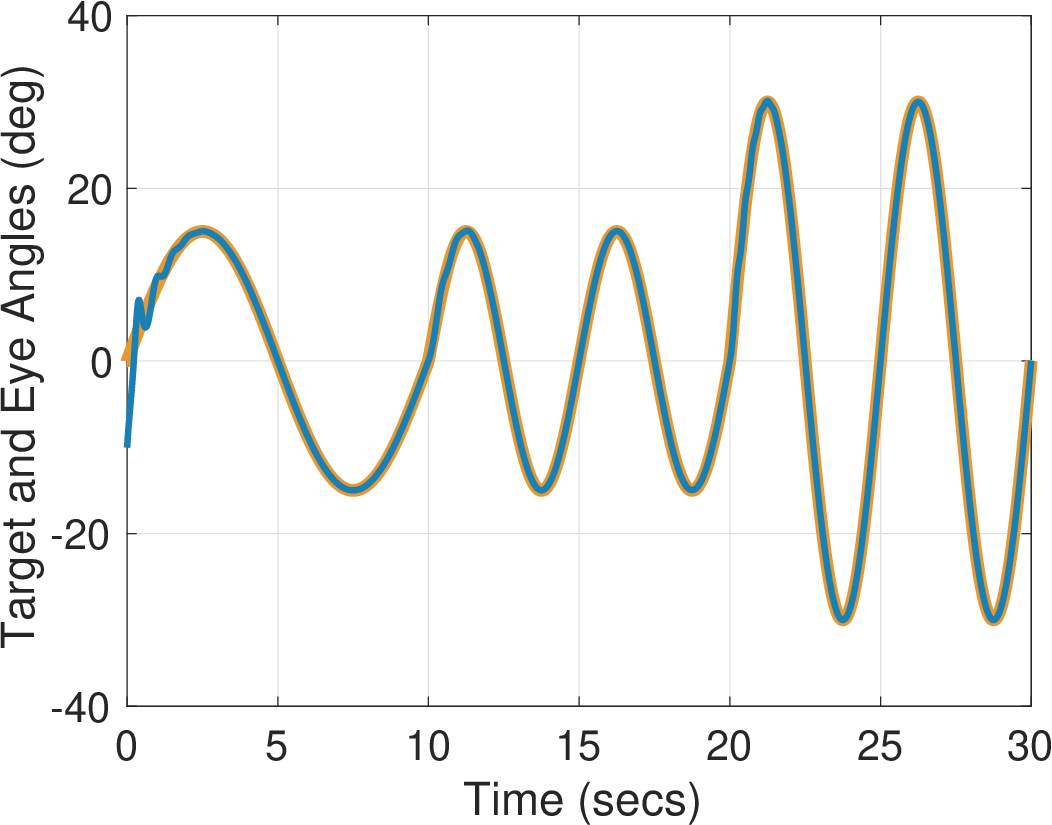}
\end{subfigure}
\begin{subfigure}[b]{.25\textwidth}
\centering
\includegraphics[width=.9\linewidth]{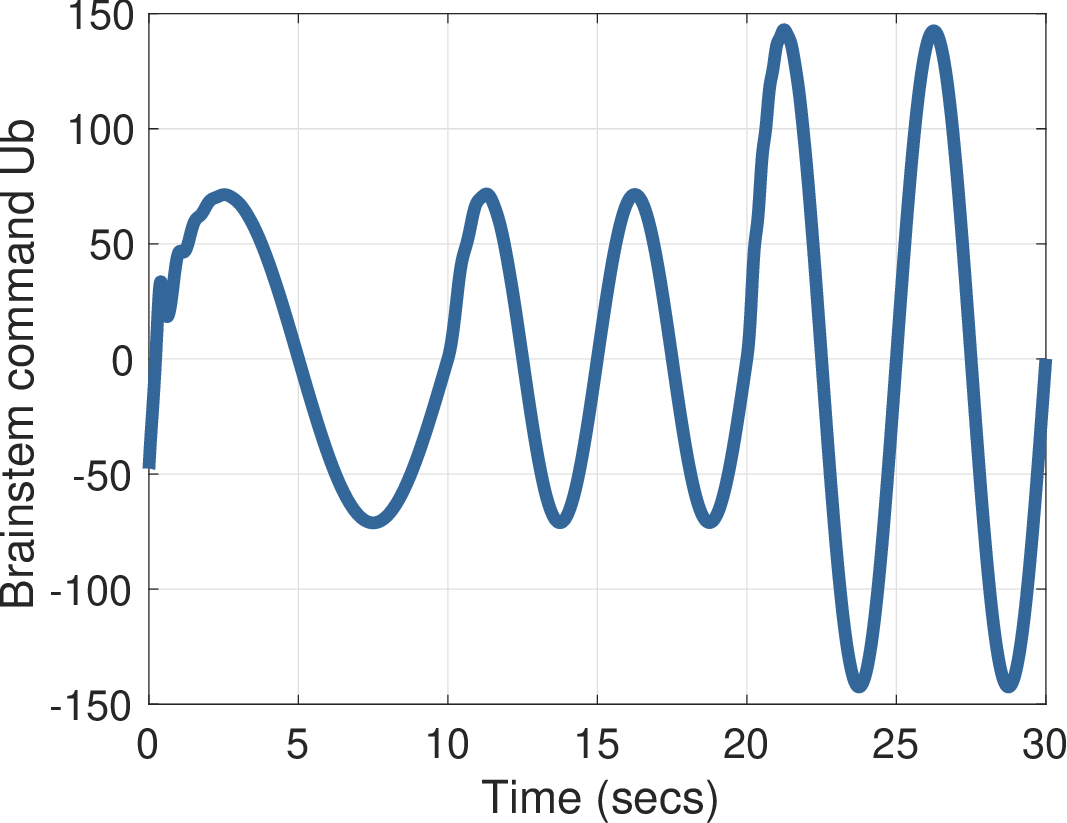}
\end{subfigure} 
\begin{subfigure}[b]{.25\textwidth}
\centering
\includegraphics[width=.9\linewidth]{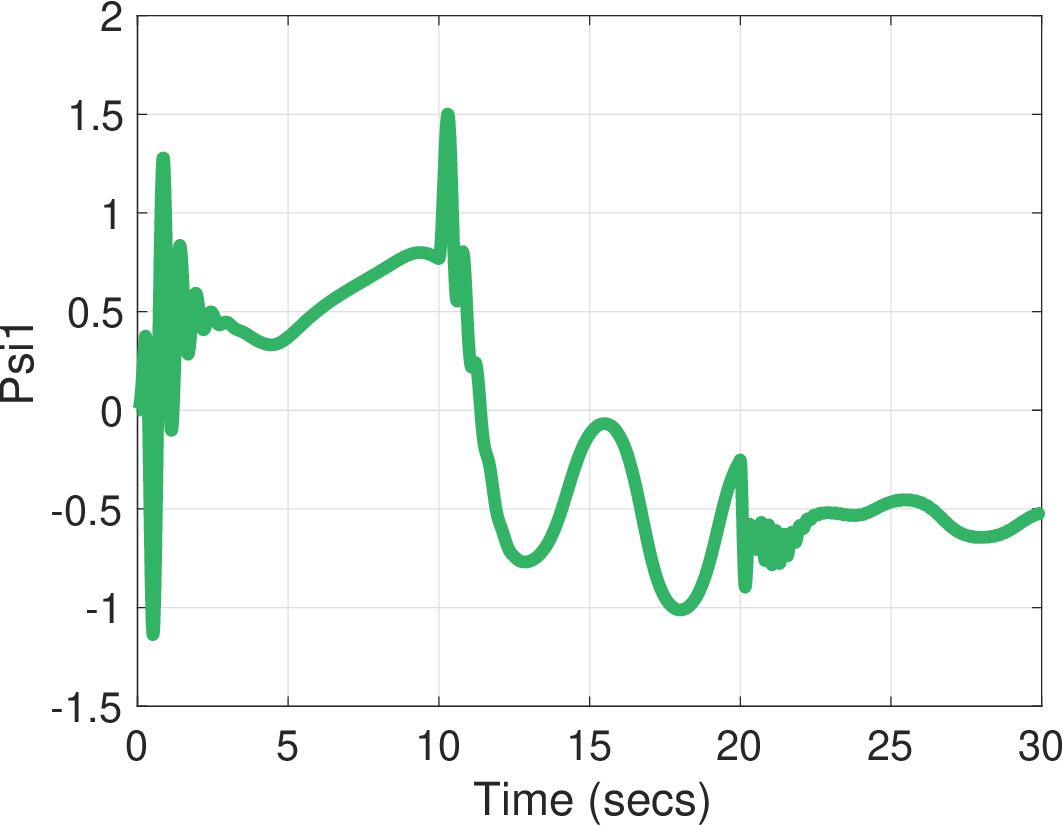}
\end{subfigure} \\ ~ \\
\begin{subfigure}[b]{.25\textwidth}
\centering
\includegraphics[width=.9\linewidth]{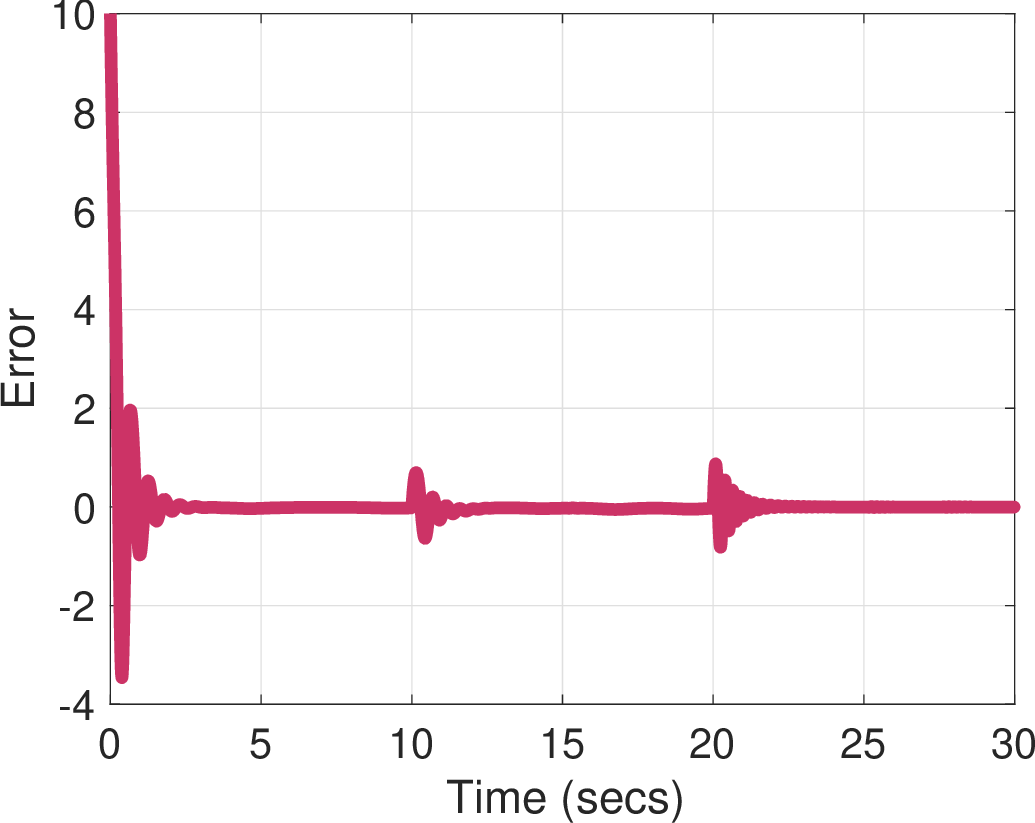}
\end{subfigure}
\begin{subfigure}[b]{.25\textwidth}
\centering
\includegraphics[width=.9\linewidth]{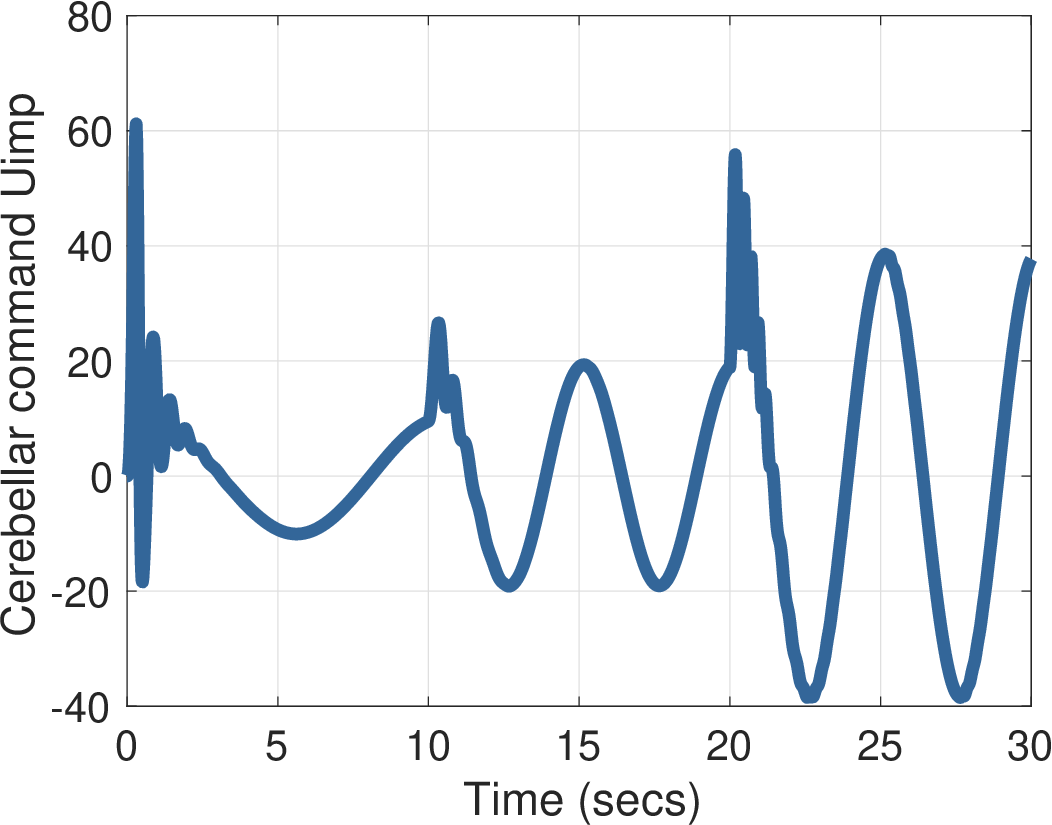}
\end{subfigure} 
\begin{subfigure}[b]{.25\textwidth}
\centering
\includegraphics[width=.9\linewidth]{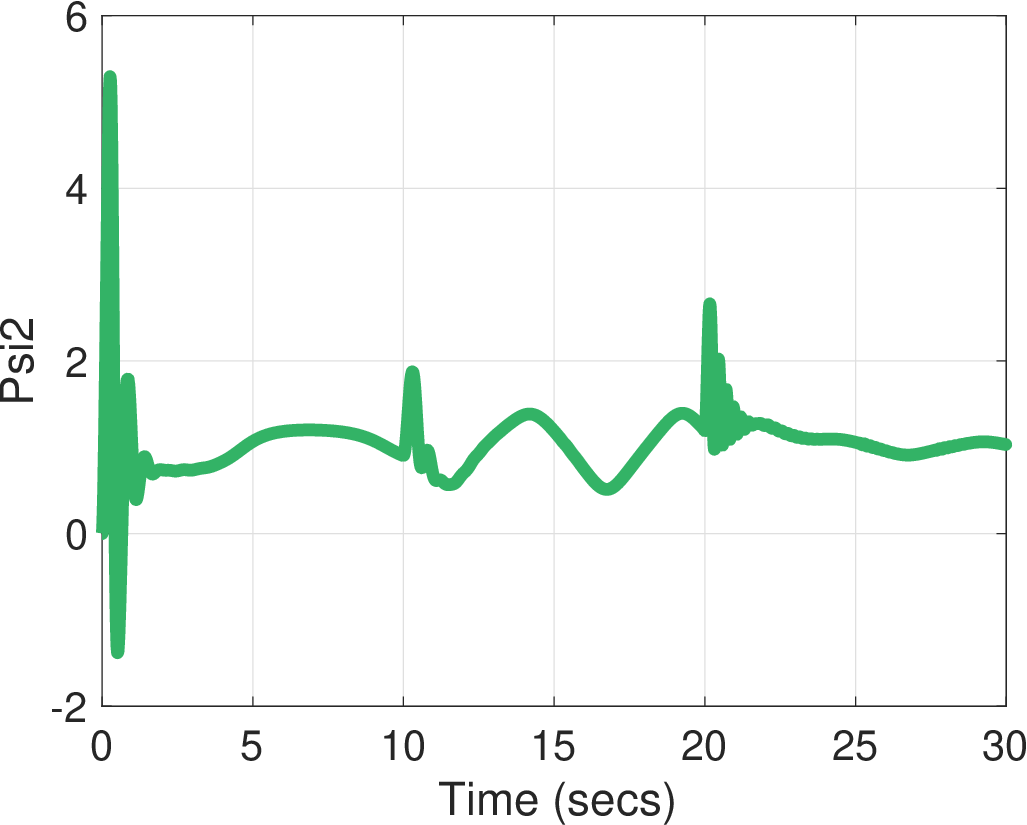}
\end{subfigure} 
\caption{Smooth pursuit of a sinusoidal target. The signals are the same as in Figure~\ref{fig:VOR1}.}
\label{fig:SP1}
\end{figure}

Figure~\ref{fig:SP1} depicts smooth pursuit with our model for a sinusoidal target 
$r(t) = a_h \sin ( \beta_h t )$, with $a_h = 15$, $\beta_h = 0.1$Hz for $t \in [0,10]$ and 
$\beta_h = 0.2$Hz for $t \in [10,20]$. We see that the cerebellar output $u_{imp}$ is strongly 
modulated during tracking of a sinusoidal target, as observed experimentally \cite{LISBERGER09}.

\begin{figure}[t!]
\centering
\begin{subfigure}[b]{.25\textwidth}
\centering
\includegraphics[width=.9\linewidth]{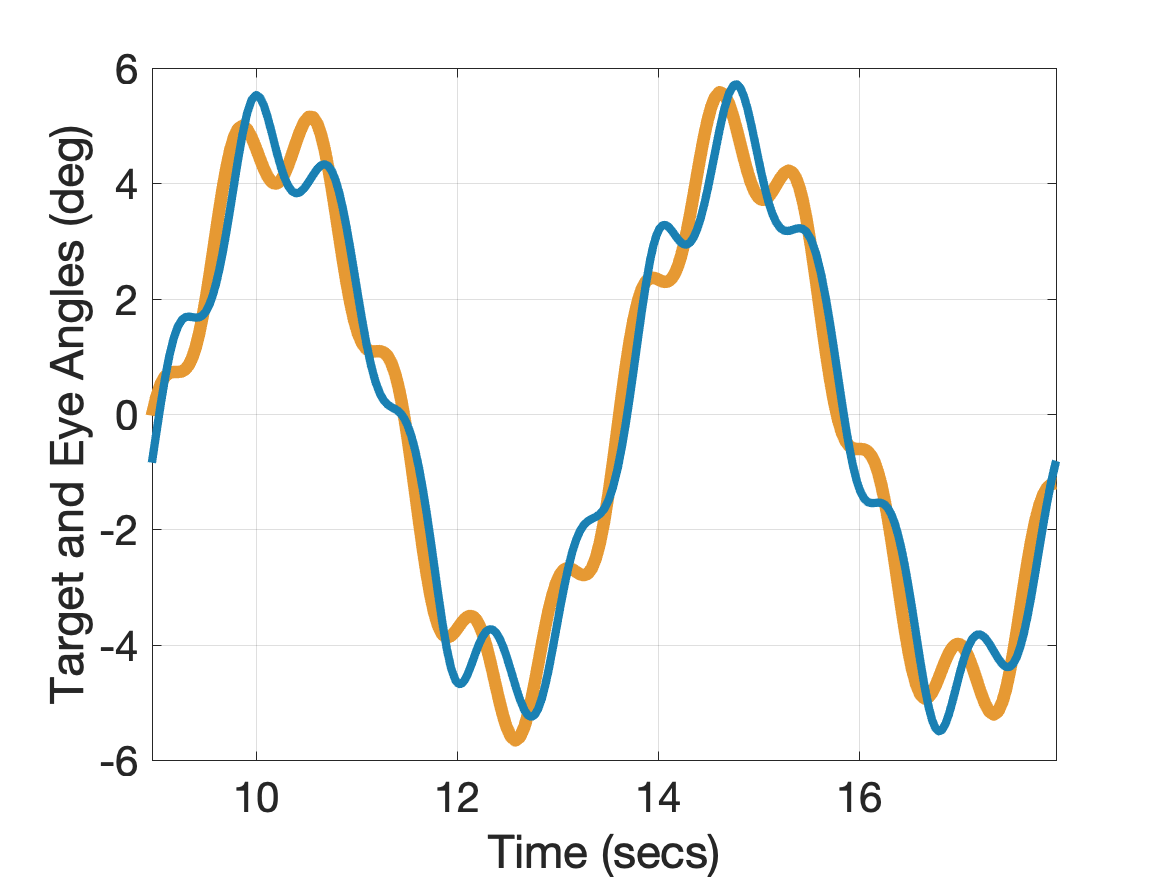}
\end{subfigure}
\begin{subfigure}[b]{.25\textwidth}
\centering
\includegraphics[width=.9\linewidth]{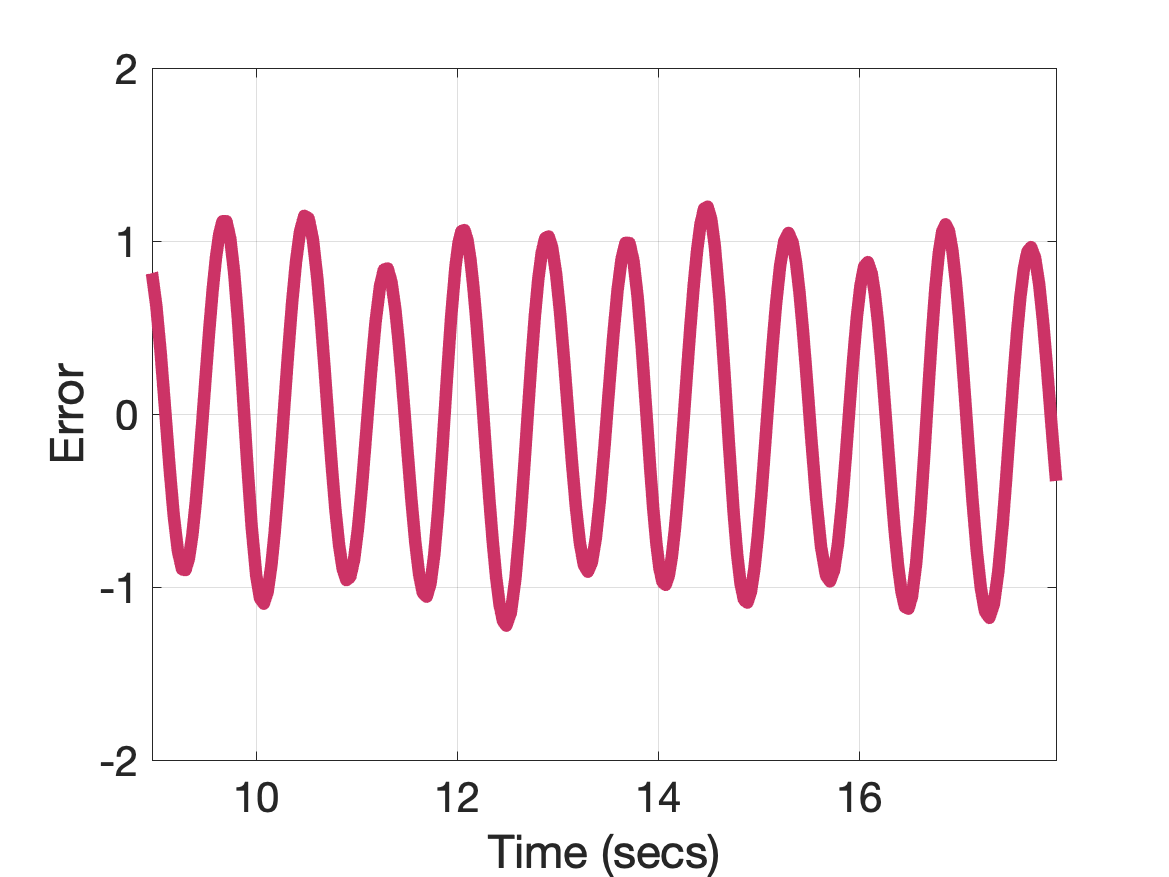}
\end{subfigure} 
\begin{subfigure}[b]{.25\textwidth}
\centering
\includegraphics[width=.8\linewidth]{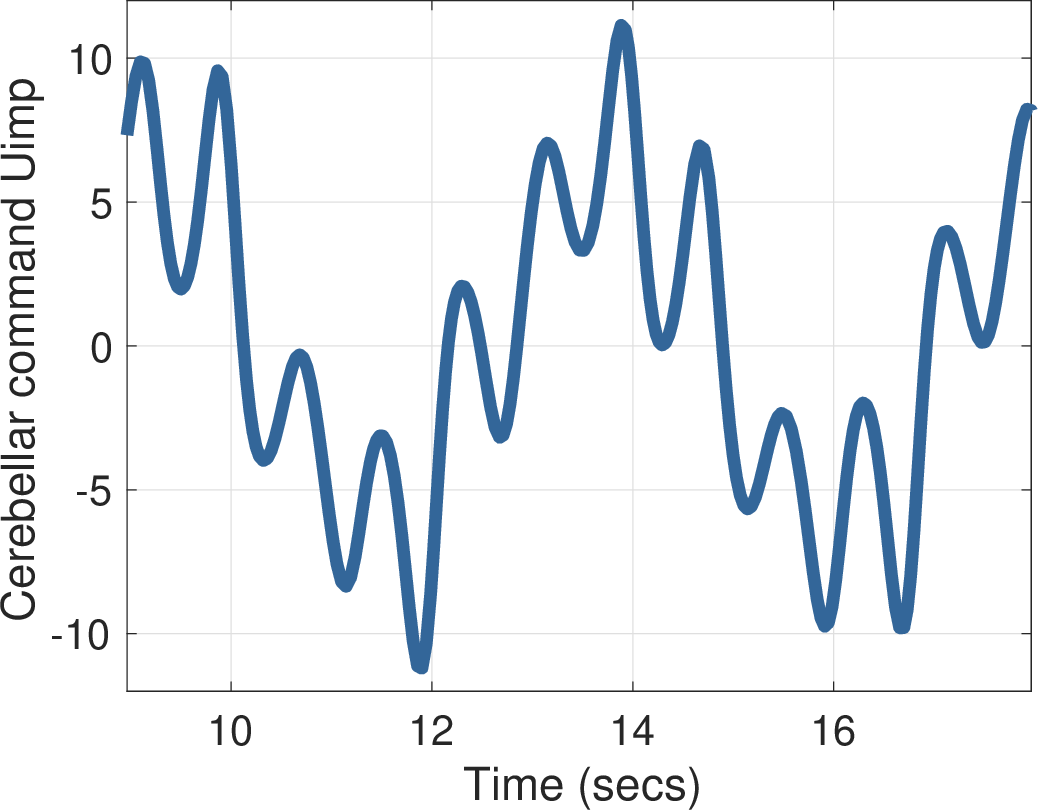}
\end{subfigure} 
\caption{Smooth pursuit of a sum of two sinusoids.
From left to right, the target angle (yellow) and eye angle (light blue), the error $e$ (red), 
and the cerebellar output $u_{imp}$ (blue).}
\label{fig:SP_SUM1}
\end{figure}

The perfect tracking capability of the smooth pursuit system has been well documented over the years; 
a small sampling includes \cite{BAHILL83A,COLLEWIJN84,DENO95,WYATT88}. 
This tracking capability improves as the targe motion becomes more predictable \cite{BAHILL83B}. 
Figure~\ref{fig:SP_SUM1} depicts the behavior of our model for smooth pursuit of a target 
$r(t) = a_1 \sin ( 2 \pi \beta_1 t ) + a_2 \sin ( 2 \pi \beta_2 t )$, with $a_1 = 4.85$, 
$\beta_1 = 0.22$Hz, $a_2 = 0.853$ and $\beta_2 = 1.25$Hz. The time interval $t \in [9,18]$ was 
chosen to match the data in Figure 1 of \cite{BARNES87}. This simulated behavior reproduces 
what is observed in experiments; namely, that while humans are not capable of perfect tracking 
of a sum of two or more sinusoids, nevertheless the smooth pursuit system performs reasonably well. 
The non-zero error displayed in the center of 
Figure~\ref{fig:SP_SUM1} is corroborated by experimental findings in \cite{BARNES87}.

\begin{figure}[t!]
\centering
\begin{subfigure}[b]{.25\textwidth}
\centering
\includegraphics[width=.9\linewidth]{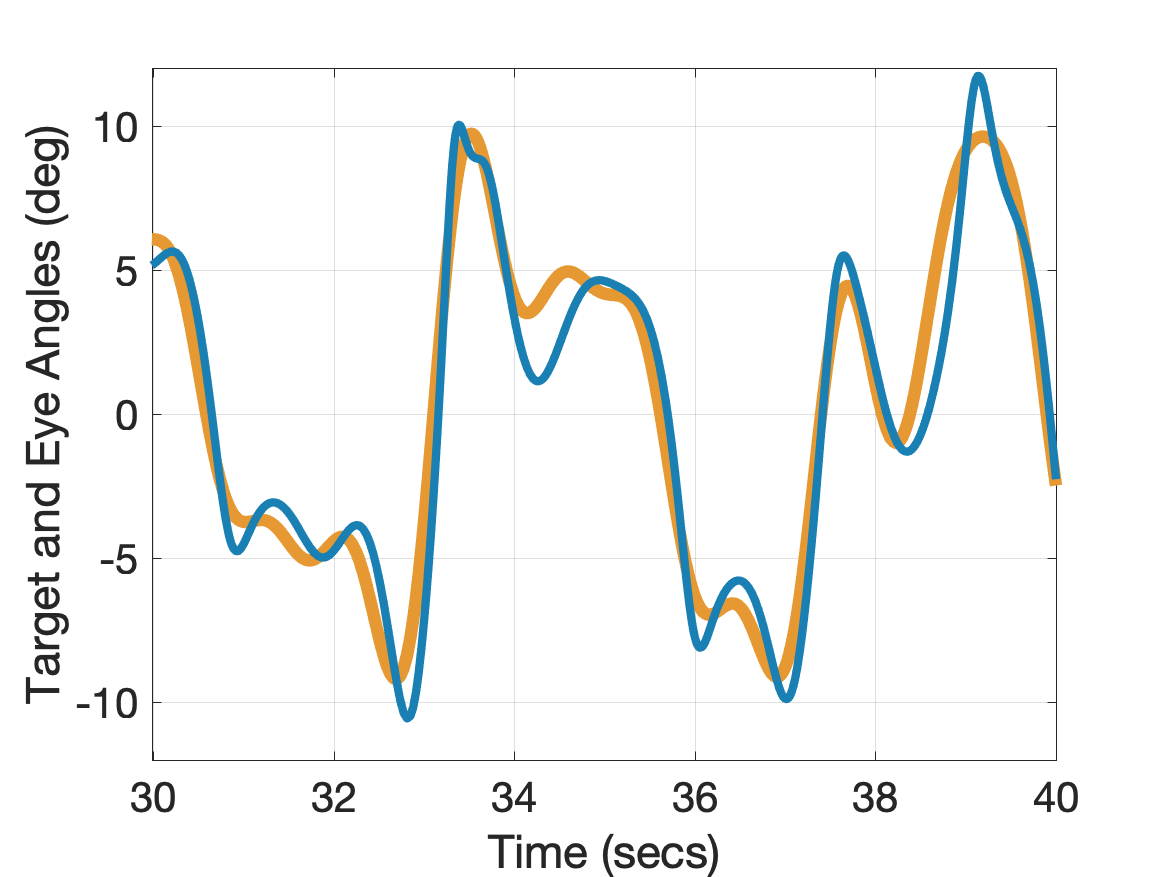}
\end{subfigure}
\begin{subfigure}[b]{.25\textwidth}
\centering
\includegraphics[width=.9\linewidth]{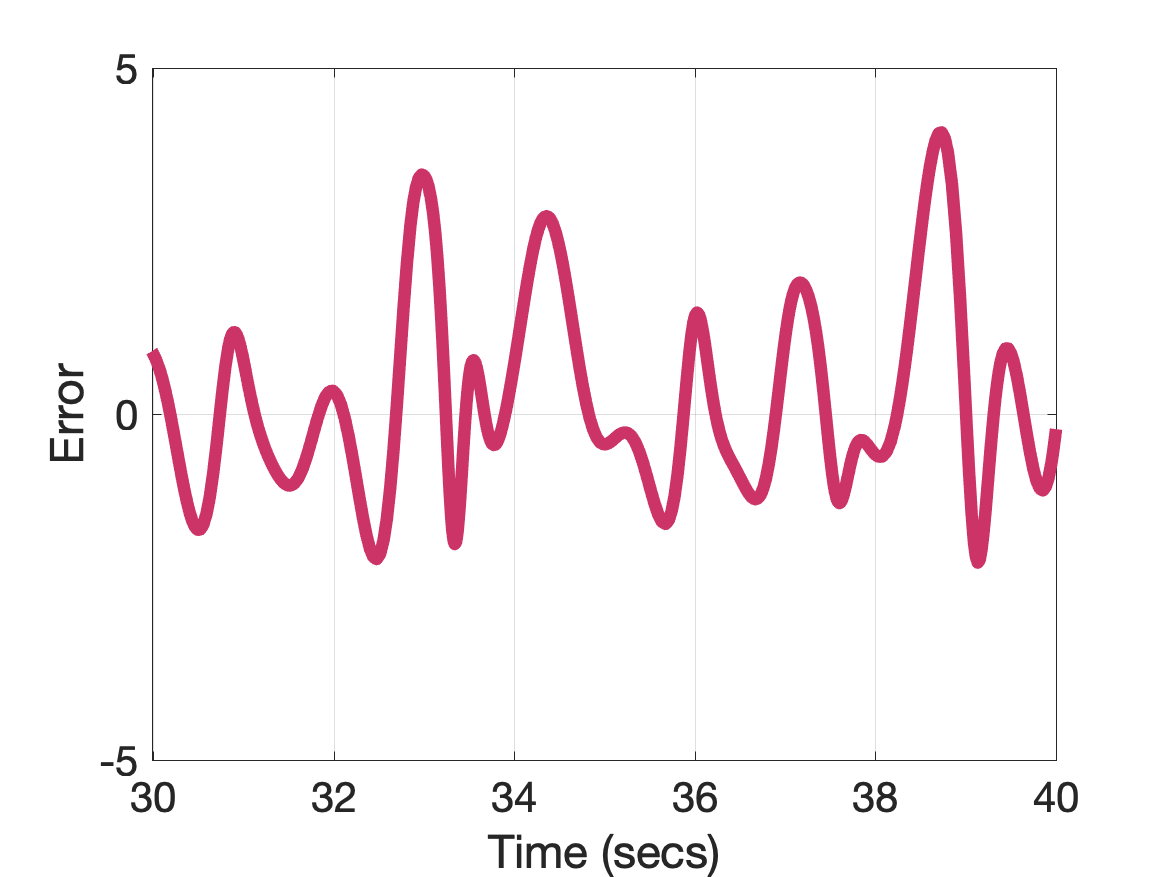}
\end{subfigure} 
\begin{subfigure}[b]{.25\textwidth}
\centering
\includegraphics[width=.8\linewidth]{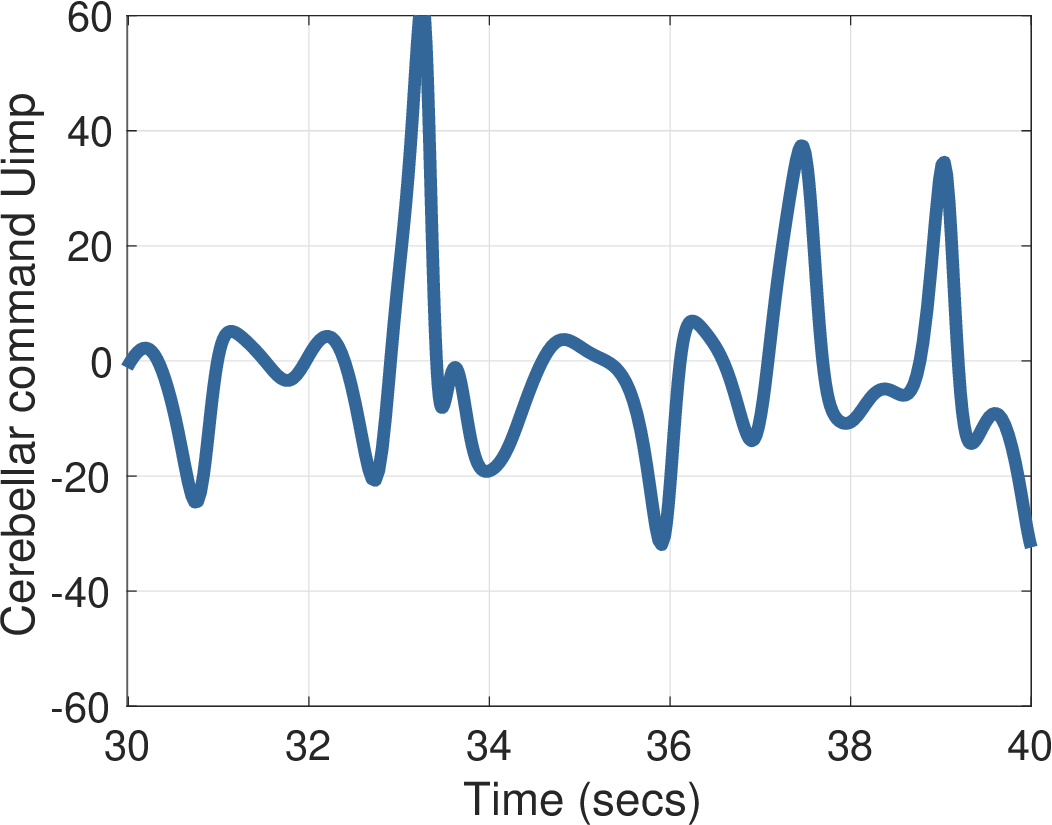}
\end{subfigure} 
\caption{Smooth pursuit of a sum of four sinusoids.
From left to right, the target angle (yellow) and eye angle (light blue), the error $e$ (red), and the 
cerebellar output $u_{imp}$ (blue).}
\label{fig:SP_SUM2}
\end{figure}

Figure~\ref{fig:SP_SUM2} depicts the behavior of our model for smooth pursuit of a target 
$r(t) = a_1 \sin ( 2 \pi \beta_1 t ) + \cdots + a_4 \sin ( 2 \pi \beta_4 t )$, with $a_1 = 6.94$, 
$\beta_1 = 0.214$Hz, $a_2 = 2.86$, $\beta_2 = 0.519$Hz, $a_3 = 2.11$, $\beta_3 = 0.702$Hz, $a_4 = 1.57$, 
and $\beta_4 = 0.946$Hz. The results are comparable to those obtained experimentally as shown in Figure~2 
of \cite{COLLEWIJN84}. 

\begin{figure}[t!]
\centering
\begin{subfigure}[b]{.25\textwidth}
\centering
\includegraphics[width=.9\linewidth]{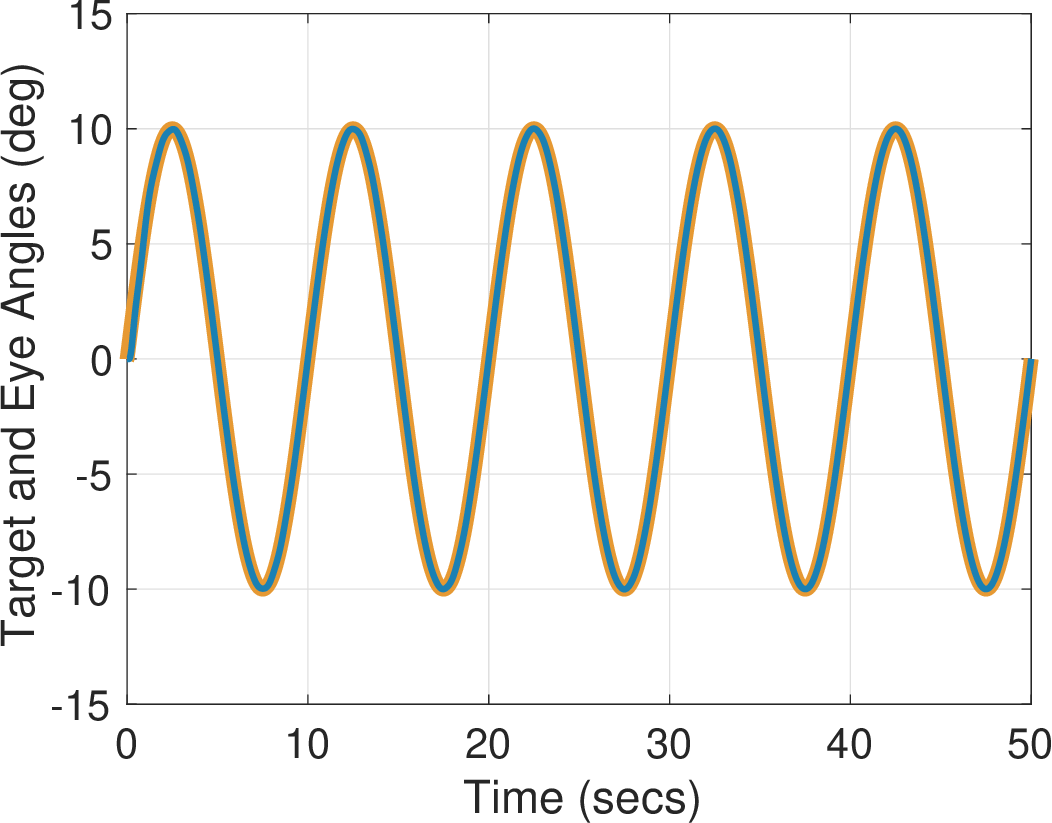}
\end{subfigure}
\begin{subfigure}[b]{.25\textwidth}
\centering
\includegraphics[width=.9\linewidth]{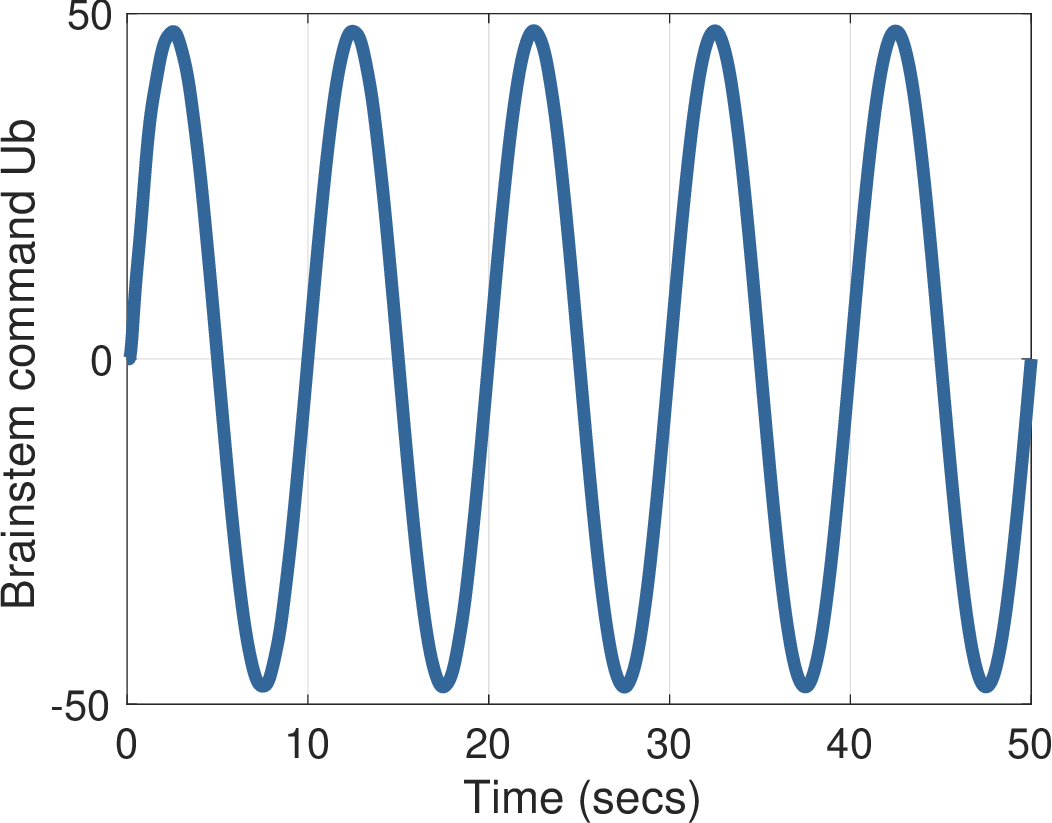}
\end{subfigure} 
\begin{subfigure}[b]{.25\textwidth}
\centering
\includegraphics[width=.9\linewidth]{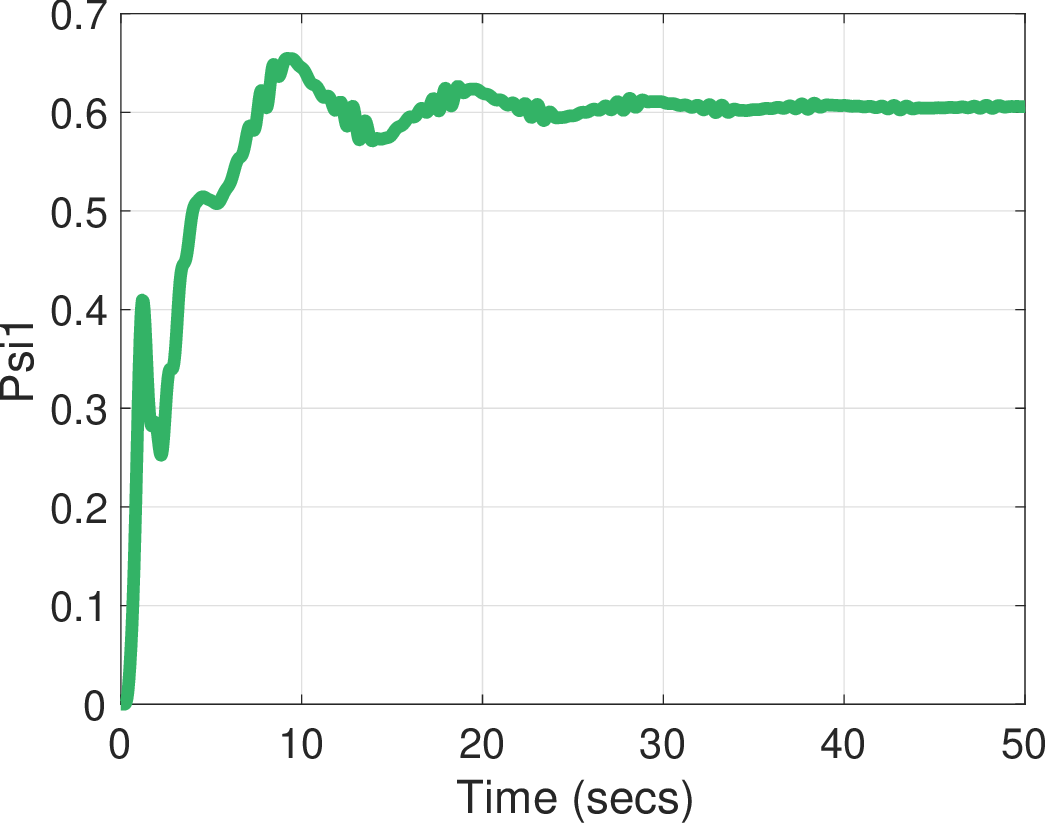}
\end{subfigure} \\ ~ \\
\begin{subfigure}[b]{.25\textwidth}
\centering
\includegraphics[width=.9\linewidth]{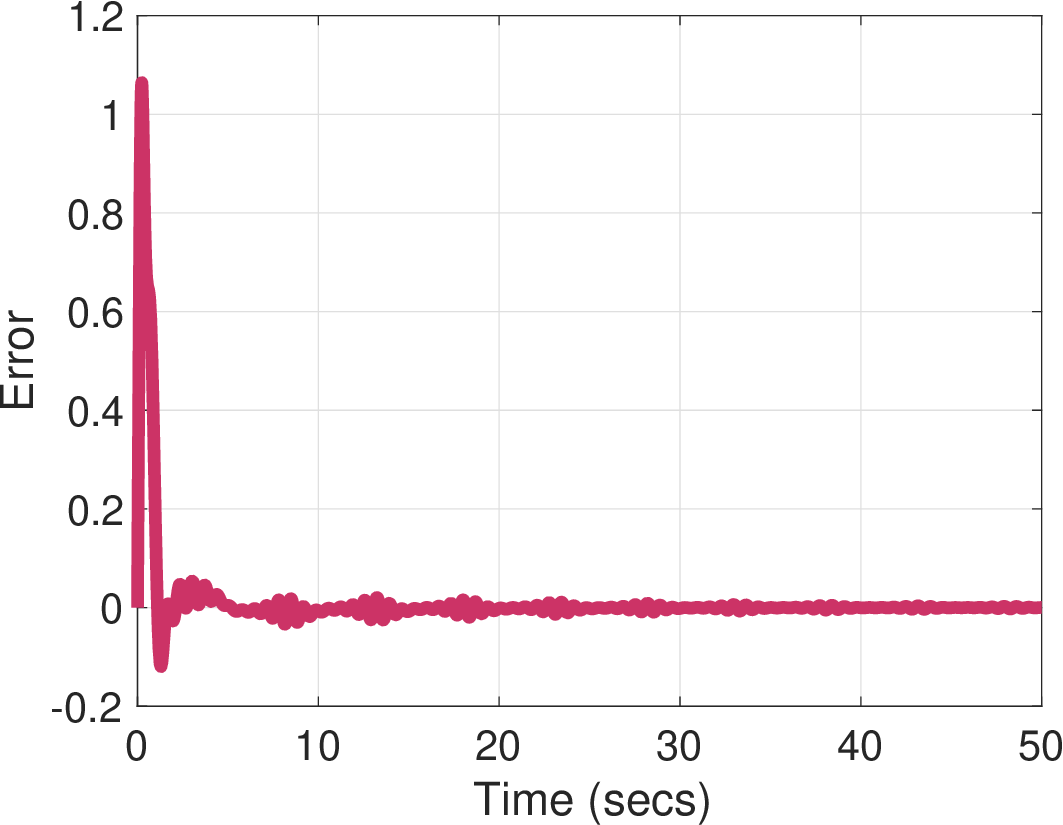}
\end{subfigure}
\begin{subfigure}[b]{.25\textwidth}
\centering
\includegraphics[width=.9\linewidth]{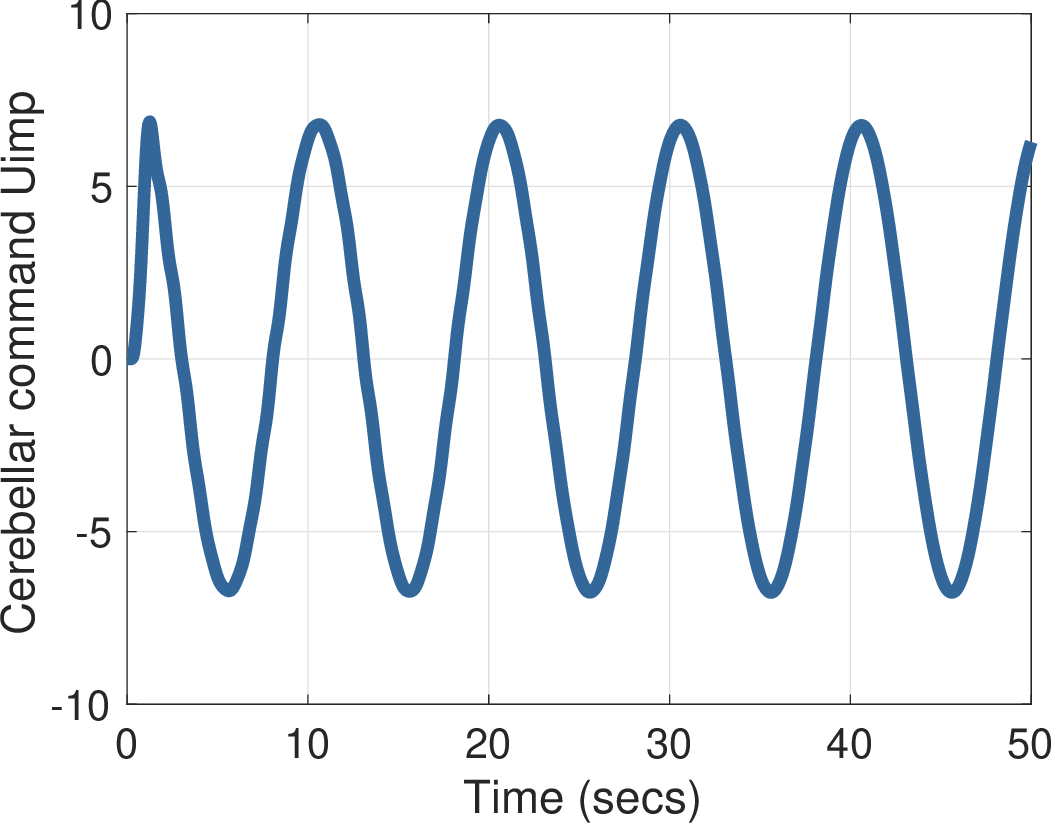}
\end{subfigure} 
\begin{subfigure}[b]{.25\textwidth}
\centering
\includegraphics[width=.9\linewidth]{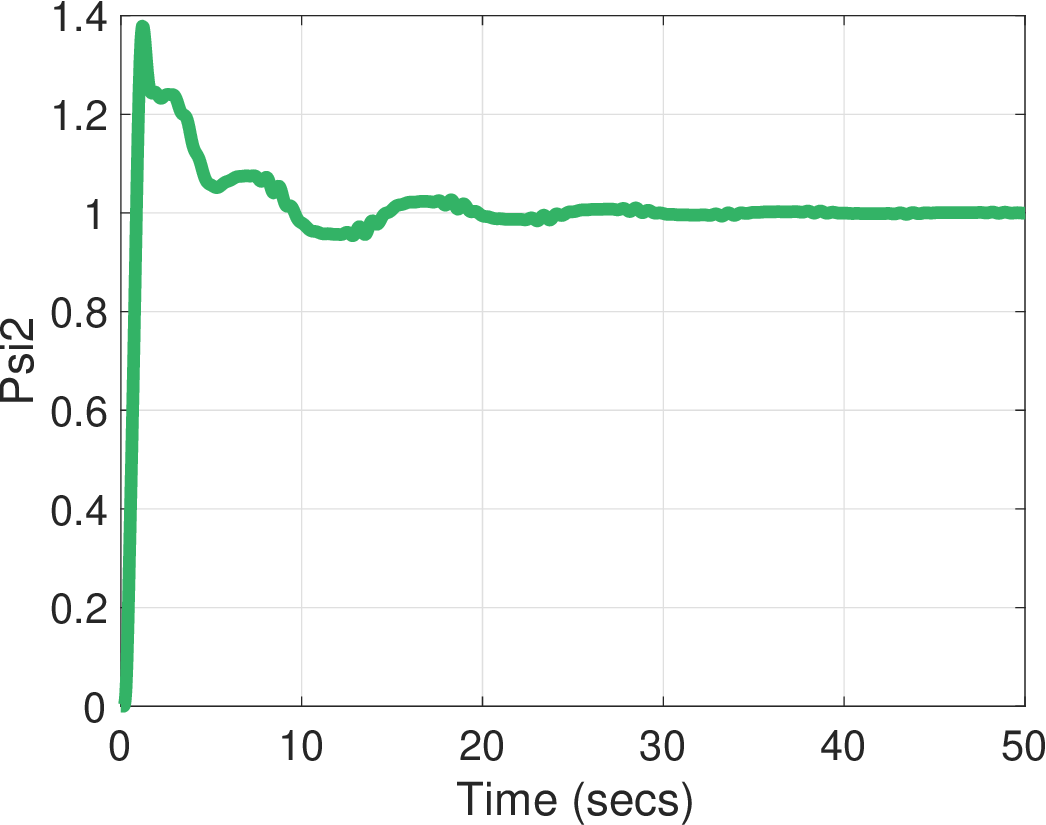}
\end{subfigure} 
\caption{Smooth pursuit of a sinusoidal target with a time delay of $107$ms in the retinal error signal.
The signals are the same as in Figure~\ref{fig:VOR1}.}
\label{fig:delay}
\end{figure}

It is known that the processing delay for the retinal error to arrive at the cerebellum is on the order 
of $100$ms. Nevertheless, the smooth pursuit system achieves nearly perfect tracking capability; its 
ability to do so in the face of this delay has been interpreted as a predictive capabability 
\cite{DENO95}. Our model does not impart any prediction to the smooth pursuit system, but the presence of the adaptive 
internal model aids in overcoming delays. Figure~\ref{fig:delay} depicts the behavior when tracking a 
sinusoidal target $r(t) = a \sin ( 2 \pi \beta t )$ with $a = 10$ and $\beta = 0.1$Hz. The error $e$ has been 
replaced by $e(t - \tau)$ in \eqref{eq:Psihatdot} and \eqref{eq:uc}, with a time delay of $\tau = 107$ms. 
The other parameter values are the same as before but we set $K_e = 8$ for closed-loop stability. We 
observe there is little degradation in the system's tracking capability.

\begin{figure}[t!]
\centering
\begin{subfigure}[b]{.2\textwidth}
\centering
\includegraphics[width=.9\linewidth]{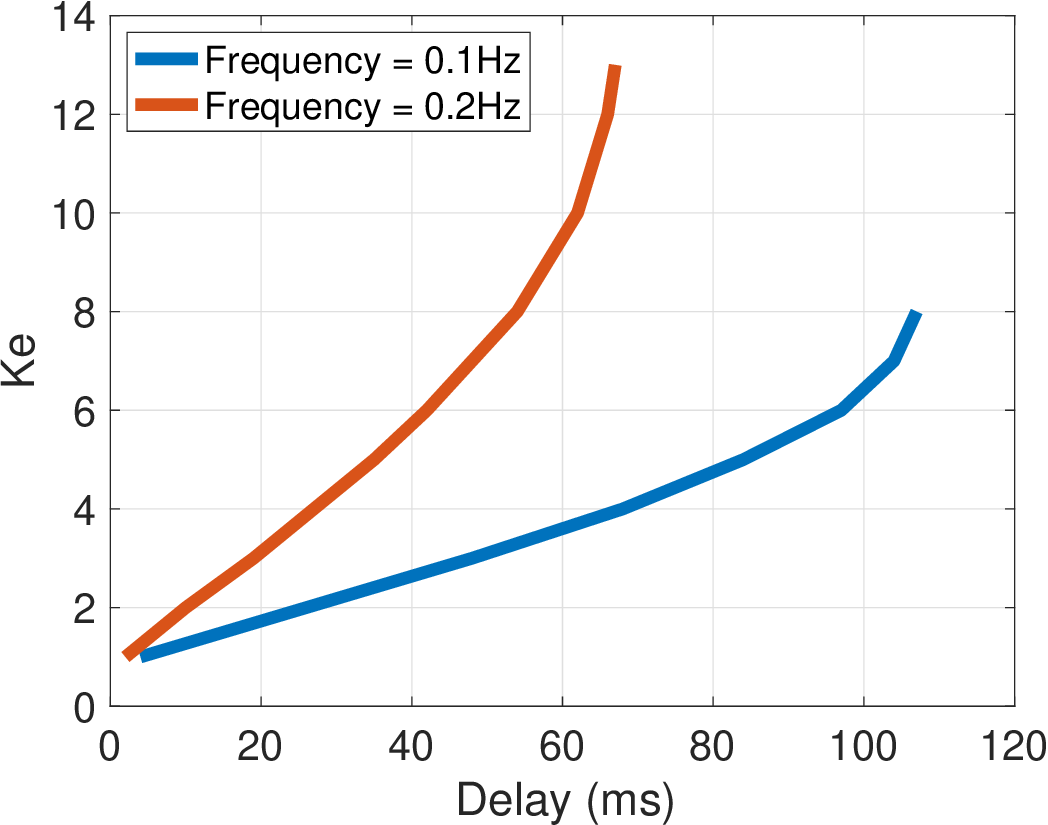}
\end{subfigure}
\begin{subfigure}[b]{.2\textwidth}
\centering
\includegraphics[width=.9\linewidth]{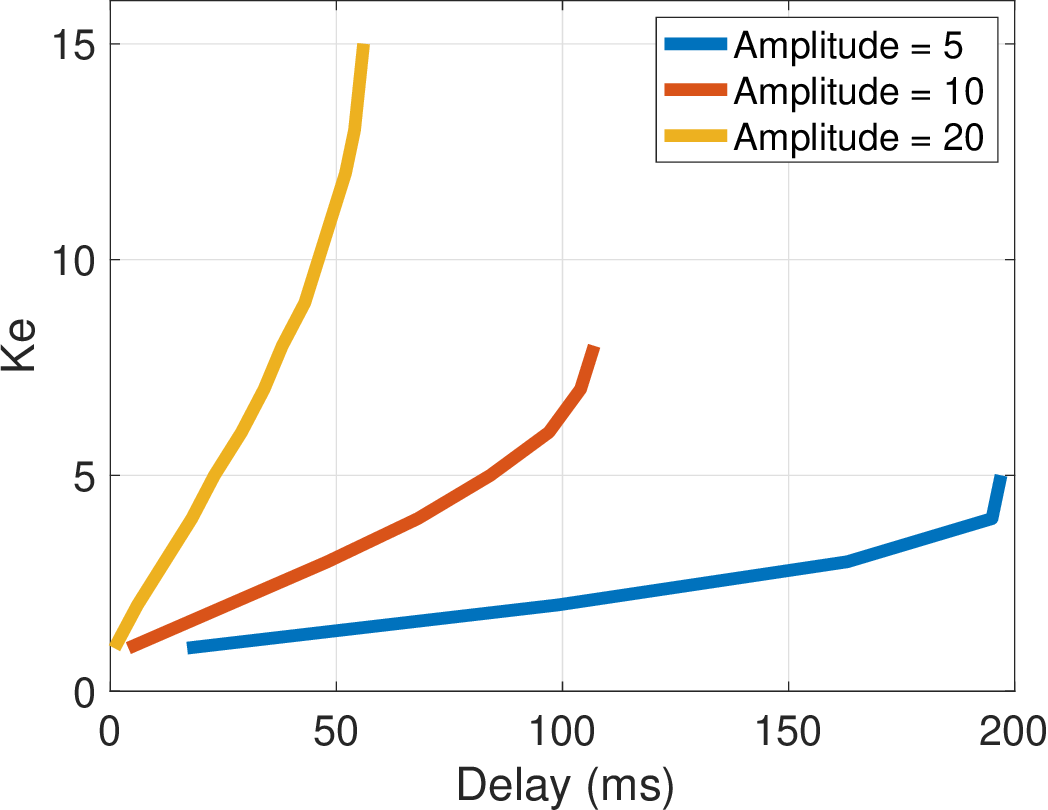}
\end{subfigure} 
\caption{Maximum time delay as a function of $K_e$.}
\label{fig:delay2}
\end{figure}

The choice of $K_e$ to achieve closed-loop stability is tied to the time delay and the magnitude of the 
reference $r(t)$. Figure~\ref{fig:delay2} depicts the largest delay attained with the smallest $K_e$ for 
varying frequencies and amplitudes of reference signals of the form $r(t) = a \sin ( 2 \pi \beta t)$. With 
$a = 10$ and $\beta = \{ 0.1, 0.2 \}$Hz, delays of $107$ms and $67$ms were achieved with $K_e$ equal to 
$8$ and $13$, respectively. Holding $\beta = 0.1$Hz but with $a = \{ 5, 10, 20 \}$, the model overcomes 
delays of $197$ms, $107$ms, and $56$ms with $K_e$ equal to $5$, $8$ and $15$, respectively.

\begin{figure}[t!]
\centering
\includegraphics[width=.4\linewidth]{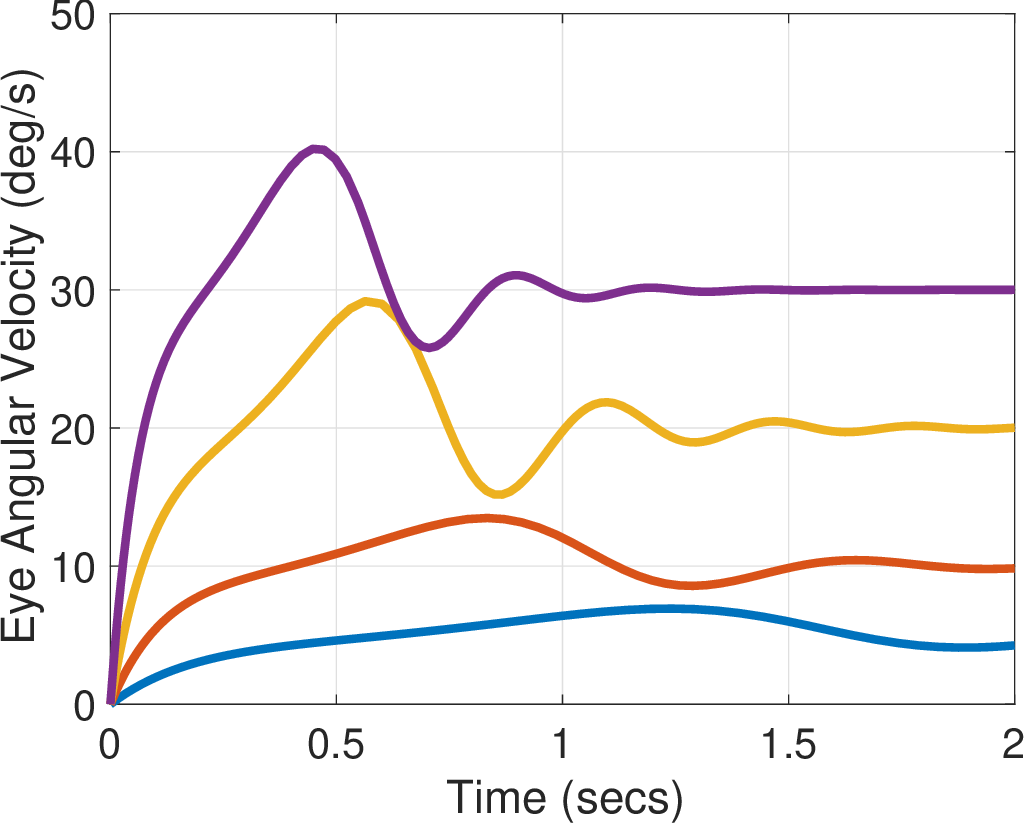}
\caption{Smooth pursuit of a ramp target with velocity $v = 5, 10, 20, 30$ (blue, red, yellow, purple).}
\label{fig:SP2}
\end{figure}

Figure~\ref{fig:SP2} depicts the transient response of our model for smooth pursuit of a 
ramp target $r(t) = v t$ with $v = 5, 10, 20, 30$. This transient response matches that 
reported in Figure~3 in \cite{ROBINSON86}. Similar behavior is reported in \cite{WYATT87}. 

In an experiment documented in \cite{LISBERGER94C}, monkeys were adapted to a new VOR gain by wearing 
goggles in their cages. It was found that changes in the VOR gain had no affect on the monkey's 
ability to track a moving target.  This behavior is explained in our model when we consider that 
the cerebellar output $u_{imp}$ compensates for whatever fraction of the vestibular signal 
entering the error that is not already cancelled by the brainstem component 
$-\alpha_h \dot{x}_h$. 

\begin{figure}[t!]
\centering
\begin{subfigure}[b]{.25\textwidth}
\centering
\includegraphics[width=.9\linewidth]{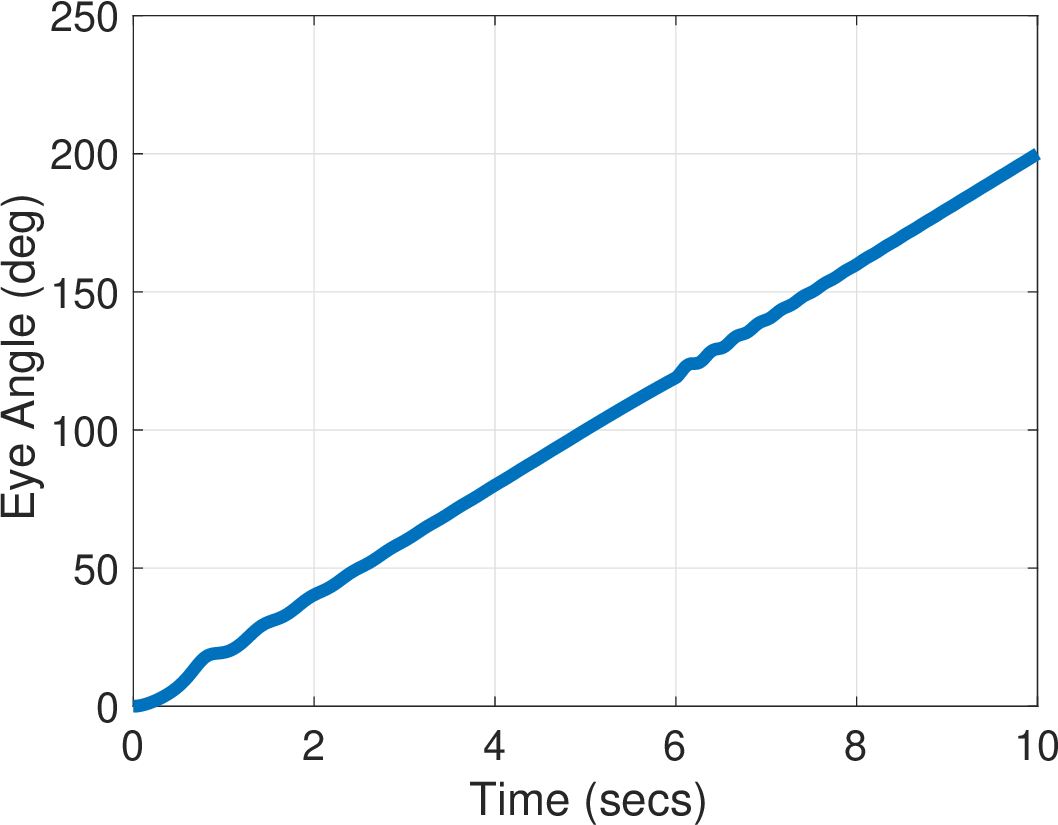}
\end{subfigure}
\begin{subfigure}[b]{.25\textwidth}
\centering
\includegraphics[width=.9\linewidth]{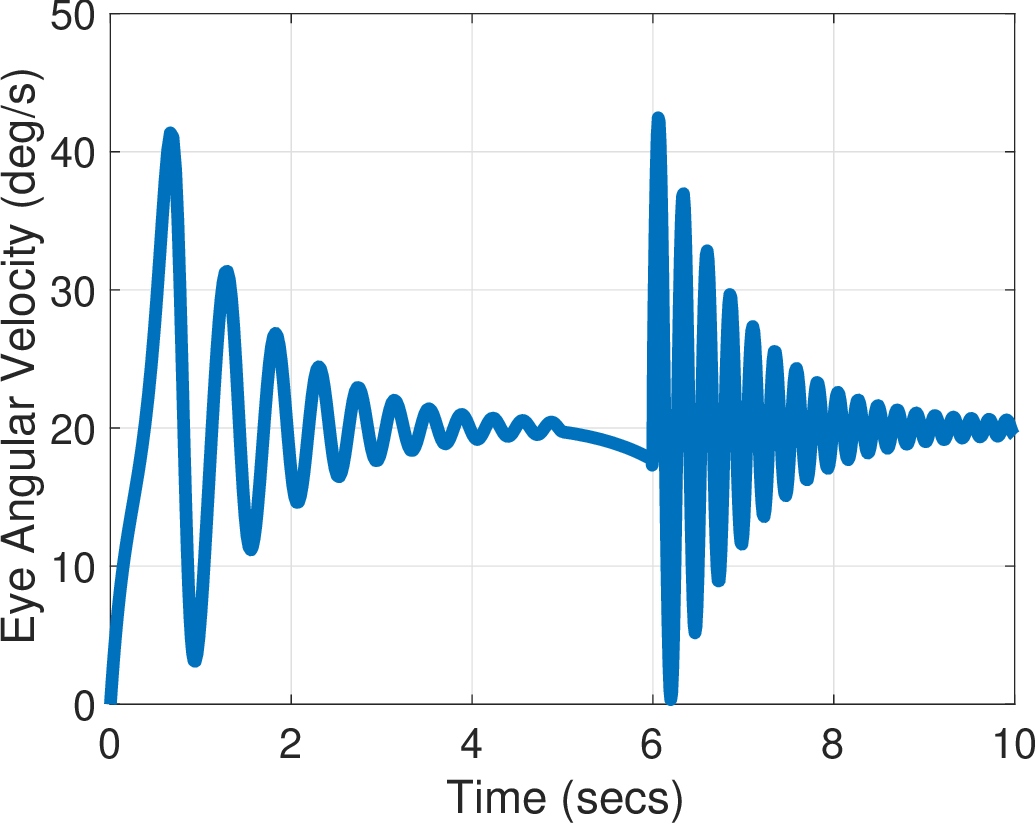}
\end{subfigure} 
\begin{subfigure}[b]{.25\textwidth}
\centering
\includegraphics[width=.9\linewidth]{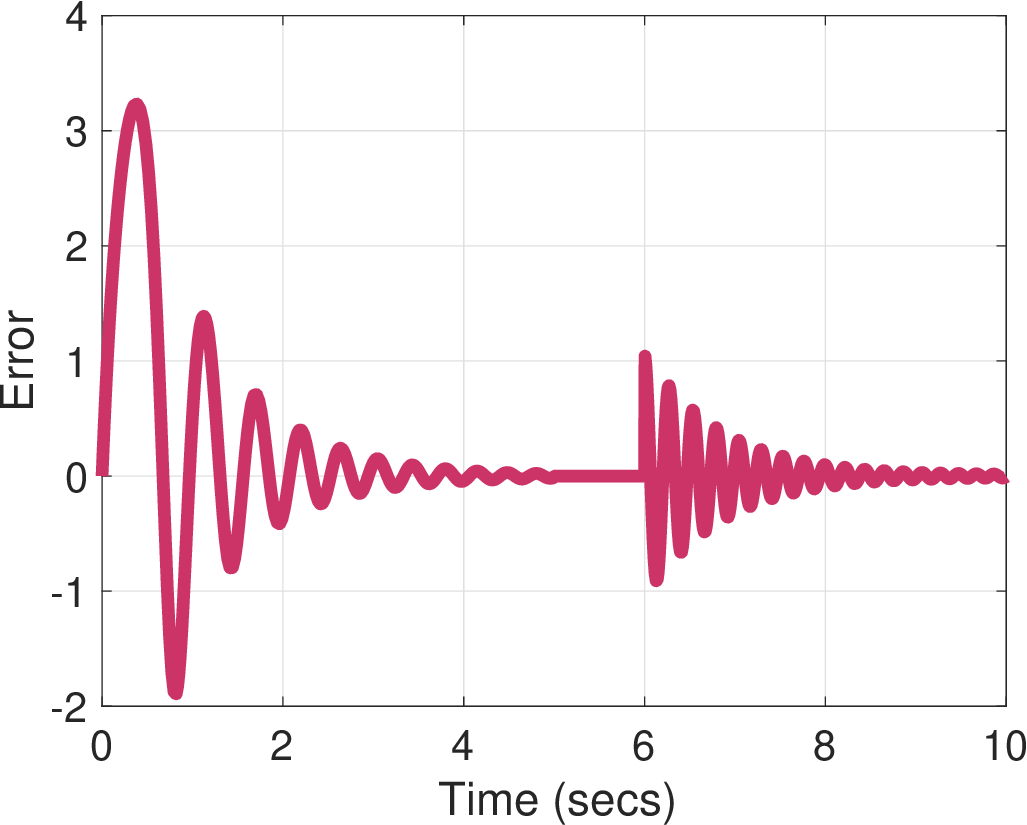}
\end{subfigure} 
\caption{Smooth pursuit with an error clamp during $t \in [5,6]$s. 
From left to right, the head angle, the head angular velocity, and the retinal error $e$.} 
\label{fig:SP3}
\end{figure}

The {\em error clamp} experiment explores the role of the error signal using a technique called 
{\em retinal stabilization} \cite{BARNES95,MORRIS87,STONE90}. A monkey is trained to track a visual 
target moving at constant speed.  After reaching steady-state, the retinal error is optically 
clamped at zero using an experimental apparatus that places the target image on the fovea. 
In experiments it is observed that the eye continues to track the target for some time after. 
Figure~\ref{fig:SP3} depicts the error clamp behavior with our model, showing that the eye 
continues to track the target despite the error being clamped at $e \equiv 0$ during the time interval 
$t \in [5,6]$. 

In another series of experiments researchers explored the difference between {\em target stopping} and 
{\em target blanking}. In target stopping, a target with a ramp position is abruptly stopped. It is 
demonstrated experimentally that during target stopping, the oculomotor system switches from smooth 
pursuit to gaze holding \cite{KRAUZLIS96,LUEBKE88,ROBINSON86}. In target blanking the target is 
blanked out or occluded, so that it is no longer visible. It is shown experimentally that with 
target blanking the eye continues to track for some time \cite{CERMINARA09,CHURCHLAND03}. 

\begin{figure}[t!]
\centering
\begin{subfigure}[b]{.25\textwidth}
\centering
\includegraphics[width=.9\linewidth]{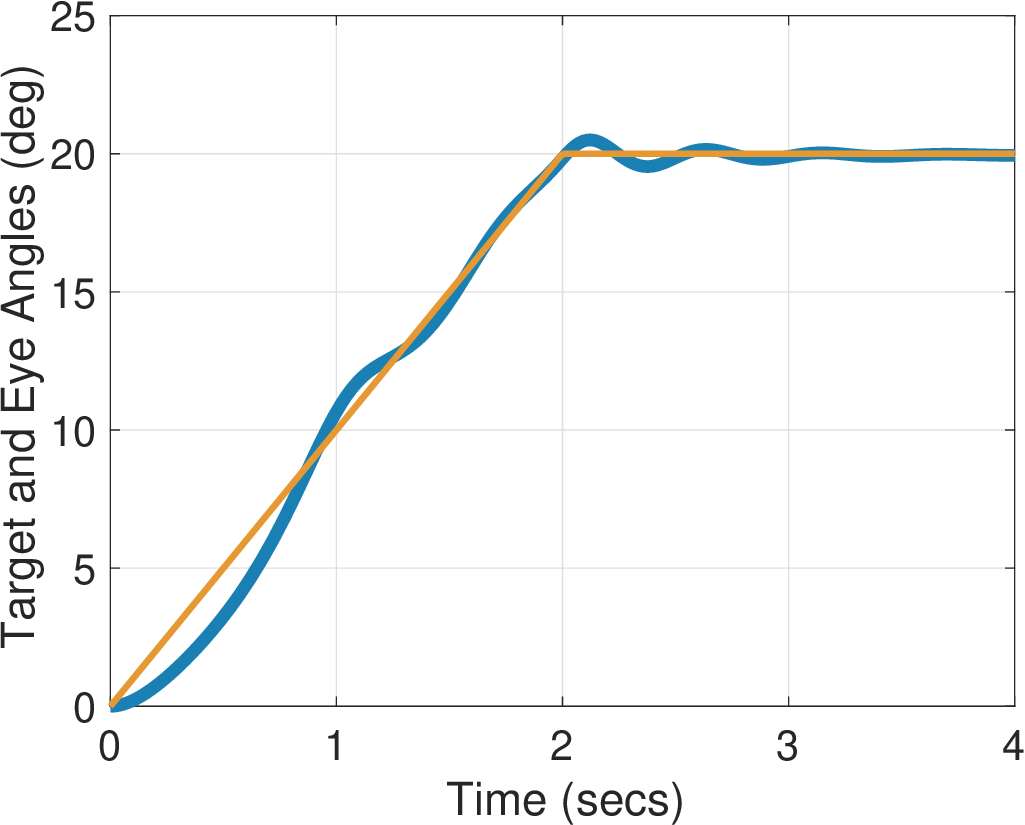}
\end{subfigure}
\begin{subfigure}[b]{.25\textwidth}
\centering
\includegraphics[width=.9\linewidth]{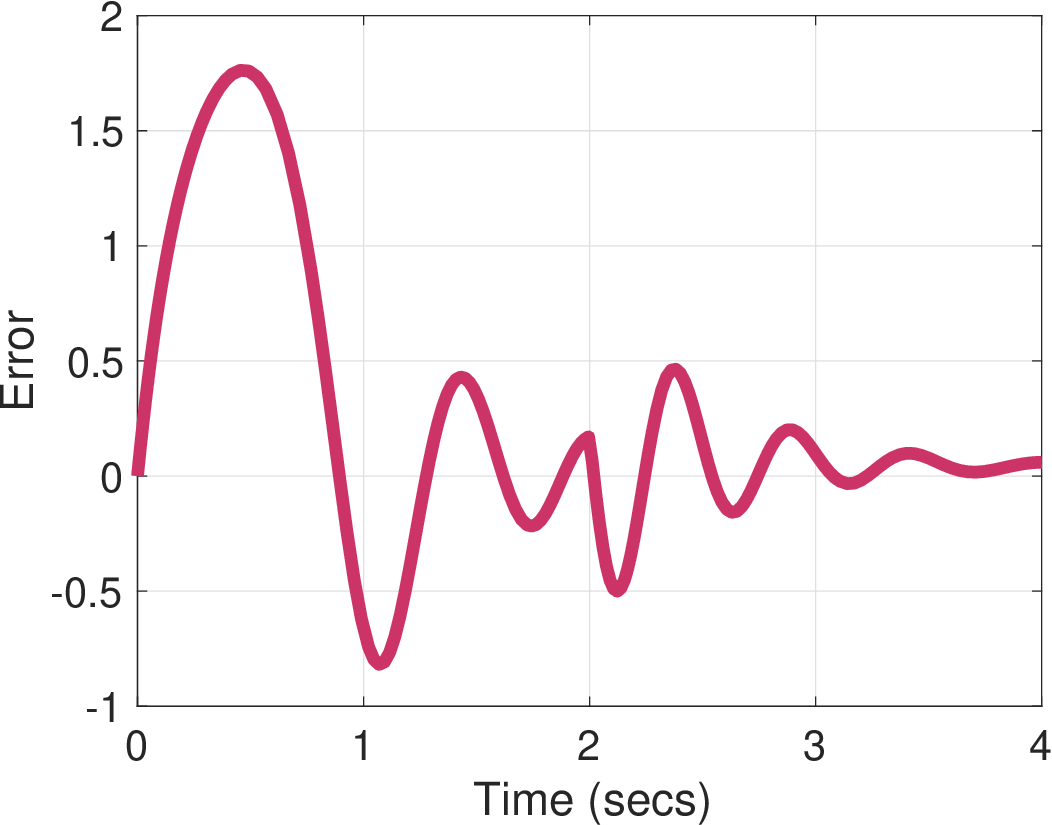}
\end{subfigure} 
\begin{subfigure}[b]{.25\textwidth}
\centering
\includegraphics[width=.9\linewidth]{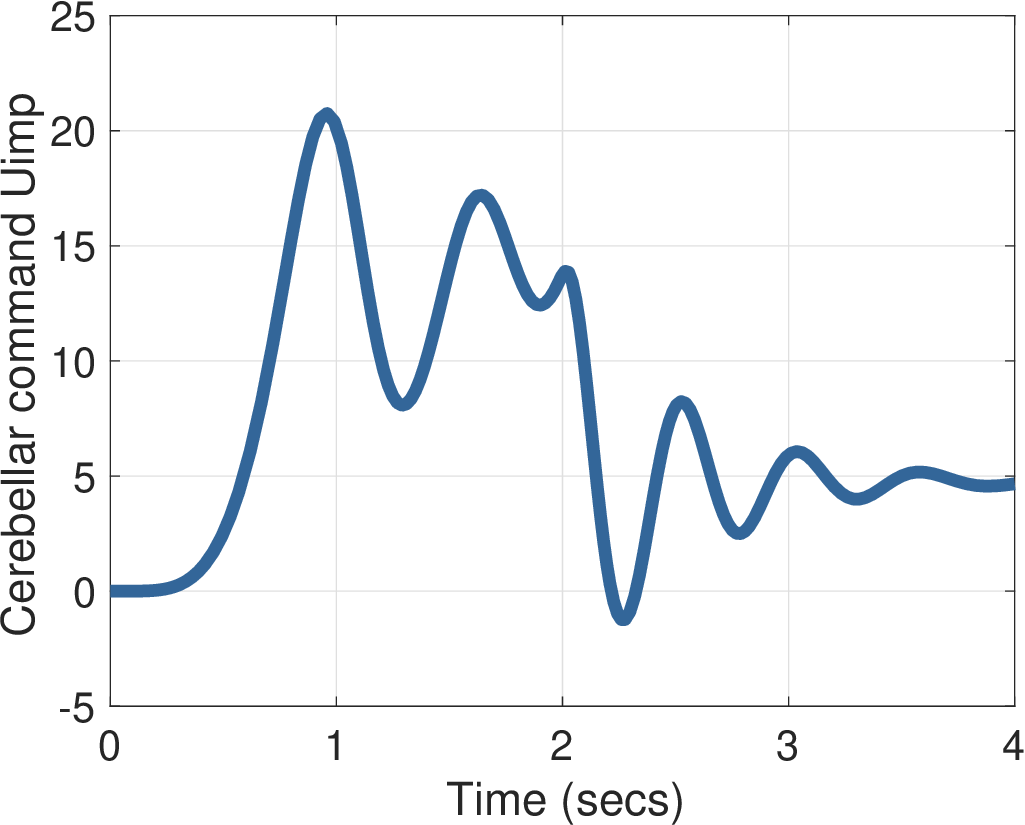}
\end{subfigure} 
\caption{Smooth pursuit with target stopping at $t = 2$s.
From left to right, the head angle, the retinal error $e$, and the cerebellar output $u_{imp}$.}
\label{fig:SP4}
\end{figure}

Figure~\ref{fig:SP4} depicts target stopping, in which $r(t) = 10 t$ for $t \in [0,2]$, and 
$r(t) = 20^{\circ}$ for $t \ge 2$. We observe that the error decays to zero with an exponential 
envelope after target stopping, as expected for the gaze holding system. Target blanking may be 
interpreted in our model as a zero error signal. As we have seen from the results of the error 
clamp experiment, depicted in Figure~\ref{fig:SP3}, the smooth pursuit system continues to track 
for some time. 

\section{Discussion}
\label{sec:discuss}

{\bf Architecture.} \
Our proposed architecture for the oculomotor system is shown in Figure~\ref{fig:MEB}. The symbol 
$\bf P$ denotes the oculomotor plant \eqref{eq:plant}; $\bf C$ is the cerebellum comprising 
\eqref{eq:whatdot}, \eqref{eq:Psihatdot}, and \eqref{eq:uc}; $\bf B$ is the brainstem comprising 
\eqref{eq:xhatdot}, \eqref{eq:ub} and \eqref{eq:u}. 

\begin{figure}[t!]
\centering
\begin{tikzpicture}[auto]
\tikzstyle{point}=[thick,draw=black,circle,inner sep=0pt,minimum width=3pt,minimum height=3pt]

\draw[->, line width = 0.4mm] (-1.8,0.5) -- (-0.28,0.5);
\node[label=above:{$r - x_h$}] (a) at (-1.3,0.5) {};
\node[label=above:{$+$}]             (b) at (-0.5,0.35) {};
\node[label=left:{$-$}]              (c) at (0.2,0.05) {};
\draw [line width = 0.4mm] (0,0.5) circle [radius = .12in];
\node[label=center:{$\Sigma$}] (c) at (0,0.5) {};
\node[label=above:{$e$}]             (d) at (0.8,0.5) {};

\filldraw [solid, fill={rgb:yellow,1;white,1}, fill opacity = 0.7] (2,0) -- (3,0) -- (3,1) -- (2,1);
\draw[thick] (2,0) -- (3,0) -- (3,1) -- (2,1) -- (2,0);
\node [label={[label distance = 0cm]15:${\bf B}$}] at (2.2,0.2) {};
\draw[->, line width = 0.4mm] (0.3,0.5) -- (2,0.5);
\draw[->, line width = 0.4mm] (0.5,0.5) -- (0.5,2) -- (1,2);
\node[label=above:{$u$}] (d) at (3.5,0.5) {};
\node[label=left:{$\dot{x}_h$}]     (e) at (2.7,-0.35) {};

\draw[->, line width = 0.4mm] (3,0.5) -- (4,0.5);
\filldraw [solid, fill={rgb:blue,1;green,1;white,2}, fill opacity = 1.0] (4,0) -- (5,0) -- (5,1) -- (4,1);
\draw[thick] (4,0) -- (5,0) -- (5,1) -- (4,1) -- (4,0);
\node [label={[label distance = 0cm]15:${\bf P}$}] at (4.2,0.2) {};
\draw[->, line width = 0.4mm] (5,0.5) -- (6,0.5);
\draw[->, line width = 0.4mm] (2.5,-0.5) -- (2.5,0);
\node[label=above:{$x$}]             (e) at (5.5,0.5) {};

\filldraw [solid, fill={rgb:red,1;yellow,1;white,2}, fill opacity = 0.9] 
(1,1.5) -- (2,1.5) -- (2,2.5) -- (1,2.5);
\draw[thick] (1,1.5) -- (2,1.5) -- (2,2.5) -- (1,2.5) -- (1,1.5);
\node [label={[label distance = 0cm]15:${\bf C}$}] at (1.2,1.7) {};
\draw[->, line width = 0.4mm] (5.5,0.5) -- (5.5,-1) -- (0,-1) -- (0,0.2);
\draw[->, line width = 0.4mm] (2,2) -- (2.5,2) -- (2.5,1);
\node[label=above:{$u_{imp}$}]           (e) at (2.5,1.86) {};
\draw[->, line width = 0.4mm] (2,0.75) -- (1.5,0.75) -- (1.5,1.5) ;
\node[label=right:{$u_c$}]           (e) at (1.35,1.15) {};

\end{tikzpicture}
\caption{Proposed architecture for the oculomotor system. $\bf P$ is the oculomotor plant, 
$\bf B$ is the brainstem, and $\bf C$ is the cerebellum.} 
\label{fig:MEB}
\end{figure}

\begin{figure}[t!]
\centering
\begin{tikzpicture}[auto]
\tikzstyle{point}=[thick,draw=black,circle,inner sep=0pt,minimum width=3pt,minimum height=3pt]

\draw[->, line width = 0.4mm] (-1.8,0.5) -- (-0.28,0.5);
\node[label=above:{$r - x_h$}] (a) at (-1.3,0.5) {};
\node[label=above:{$+$}]             (b) at (-0.5,0.35) {};
\node[label=left:{$-$}]              (c) at (0.2,0.05) {};
\draw [line width = 0.4mm] (0,0.5) circle [radius = .12in];
\node[label=center:{$\Sigma$}] (c) at (0,0.5) {};
\node[label=above:{$e$}]             (d) at (0.8,0.5) {};

\filldraw [solid, fill={rgb:yellow,1;white,1}, fill opacity = 0.7] (2,0) -- (3,0) -- (3,1) -- (2,1);
\draw[thick] (2,0) -- (3,0) -- (3,1) -- (2,1) -- (2,0);
\node [label={[label distance = 0cm]15:${\bf B}$}] at (2.2,0.2) {};
\draw[->, line width = 0.4mm] (0.3,0.5) -- (2,0.5);
\draw[->, line width = 0.4mm] (0.5,0.5) -- (0.5,1.75) -- (1,1.75);
\draw[->, line width = 0.4mm] (0.5,2.25) -- (1,2.25);
\node[label=left:{$r, \dot{r}, x_h, \dot{x}_h$}] (e) at (0.6,2.25) {};
\node[label=above:{$u$}] (d) at (3.5,0.5) {};
\node[label=left:{$\dot{x}_h$}]     (f) at (2.7,-0.35) {};

\draw[->, line width = 0.4mm] (3,0.5) -- (4,0.5);
\filldraw [solid, fill={rgb:blue,1;green,1;white,2}, fill opacity = 1.0] (4,0) -- (5,0) -- (5,1) -- (4,1);
\draw[thick] (4,0) -- (5,0) -- (5,1) -- (4,1) -- (4,0);
\node [label={[label distance = 0cm]15:${\bf P}$}] at (4.2,0.2) {};
\draw[->, line width = 0.4mm] (5,0.5) -- (6,0.5);
\draw[->, line width = 0.4mm] (2.5,-0.5) -- (2.5,0);
\node[label=above:{$x$}]             (g) at (5.5,0.5) {};

\filldraw [solid, fill={rgb:red,1;yellow,1;white,2}, fill opacity = 0.9] (1,1.5) -- (2,1.5) -- (2,2.5) -- (1,2.5);
\draw[thick] (1,1.5) -- (2,1.5) -- (2,2.5) -- (1,2.5) -- (1,1.5);
\node [label={[label distance = 0cm]15:${\bf C}$}] at (1.2,1.7) {};
\draw[->, line width = 0.4mm] (5.5,0.5) -- (5.5,-1) -- (0,-1) -- (0,0.2);
\draw[->, line width = 0.4mm] (2,2) -- (2.5,2) -- (2.5,1);
\node[label=above:{$u_c$}]           (h) at (2.4,2) {};

\end{tikzpicture}
\caption{Feedback error learning architecture.} 
\label{fig:FEL}
\end{figure}

An alternative architecture \cite{GOMI92,KAWATO92} called {\em feedback error learning} (FEL) is depicted in
Figure~\ref{fig:FEL}. The primary difference between our architecture and FEL is that in FEL the signals 
$r, \dot{r}, \ddot{r}, \ldots$, which arise from exogenous disturbance and reference signals, are assumed to be directly 
measurable by the cerebellum. These signals are used to estimate model parameters in order to obtain an inverse model of 
the plant. In contrast, our proposal is that the cerebellum receives only (sensory) error signals, which it uses 
to reconstruct both persistent, exogenous disturbance and reference signals, as well as model parameters. Indeed, for 
the oculomotor system, an inverse model is not strictly necessary, since the brainstem neural integrator provides a 
model of the oculomotor plant. A recent architecture for the computations of the cerebellum emphasizes its role as an 
adaptive filter \cite{PORRILL04}. Our model aligns with this interpretation in the sense that we include the standard 
parameter adaptation law \eqref{eq:Psihatdot}. On the other hand, we explicitly account for the internal model principle,
while the architecture in \cite{PORRILL04} does not.

{\bf Limitations.} \
We have already mentioned that we only consider horizontal movement of a single eye. Second, we have not taken 
explicit account for differences between species. When we cited experimental results for a particular eye movement system, 
implicitly we restrict to those species that possess such a system. Third, we do not model those signals in the brain that 
trigger a particular eye movement system; recognizing that trigger signals may be different from driving signals. Fourth, 
we do not consider detailed models of the semicircular canals of the ear which transmit the vestibular signal, and we do 
not consider a detailed model of the muscles of the eye. Fifth, we do not consider the role of attention or fatigue of the 
subject. Sixth, we have not included time delays inherent in the oculomotor system. Finally, we only consider involuntary
head movements; voluntary head movements may require a model that bypasses the cerebellum. These modelling omissions were 
calculated to best illuminate the basic operations of the oculomotor system and the cerebellum. 

{\bf Open Problems.} \
An important question not directly addressed in our work is: what is the value of $q$? We have chosen $q = 2$ based on the fact 
that disturbance and reference signals are typically steps, ramps, or sinusoids. Additionally, 
experiments show that
humans are able to achieve near perfect tracking of a single sinusoidal reference signal, while tracking the sum of two 
sinusoids is degraded \cite{WYATT88}. Further experimentation is needed to determine the value of $q$. 

A second important question not addressed by our model is long term adaptation (over days and weeks) 
of system parameters such as the VOR gain. The cerebellum mediates short term adaptation (over the 
timespan of a single experiment), for instance, by increasing the effective value of the VOR gain to 1. 
But such short term adaptation is not retained in our model. On the other hand, VOR gain adaptation 
may be stored in the term $\alpha_h$ in $u_b$. How is an effective change in VOR gain due to cerebellar 
{\em training} transferred to a more permanent change of the parameter $\alpha_h$ in the brainstem? 
A next step would be to examine the role of persistency of excitation in the parameter adaptation to
address this question. Indeed, monkeys deprived of sufficiently rich visual experience following 
long-term VOR adaptation do not relearn a normal VOR gain \cite{MILESFULLER74}.

\section{Conclusion}
\label{sec:conclude}

We have proposed a new model of the oculomotor system, particularly the VOR, OKR, gaze fixation, and 
smooth pursuit systems. Our key insight is to exploit recent developments on adaptive internal models. 
Our model recovers behaviors from a number of oculomotor experiments. Additionally, we make a proposal 
about the function of the cerebellum: the cerebellum embodies internal models of all persistent, 
exogenous reference and disturbance signals acting on the body. 

\bibliographystyle{IEEEtran}

\begin{appendix}
\section{~~}

In this section we prove that the controller \eqref{eq:control} solves Problem~\ref{prob1};
the proof closely mimicks that of \cite{SERRANI00}. 
First we state the result for transforming the exosystem \eqref{eq:exosystem} into the form \eqref{eq:wdot}. 
\begin{lem}[\cite{NIKIFOROV98}]
\label{lem:nikiforov}
Let $F \in \RR^{q \times q}$ and $G \in \RR^q$. Suppose that $(F,G)$ is a controllable pair and $(E,S)$ is an 
observable pair. Also suppose that $F$ is Hurwitz, and $F$ and $S$ have disjoint spectra. Then the Sylvester 
equation
\begin{equation}
\label{eq:M}
M S = F M + G E 
\end{equation}
has a unique solution $M \in \RR^{q \times q}$ which is nonsingular.
\end{lem}

Consider the error dynamics in \eqref{eq:edot} and the estimation error $\xt = x - \hat{x}$.
If we take the states of the closed-loop system to be $(e, \hat{w}, \xt, \Psih )$, then the closed-loop 
system is
\begin{subequations}
\label{eq:closedloop1}
\begin{eqnarray}
\dot{e}       & = & - (\Kxt + K_e) e + \alpha_x \xt - \Psih \hat{w} + \Psi w \\
\dot{\hat{w}} & = & (F + G \Psih ) \hat{w} + G K_e e \\
\dot{\xt}     & = & - K_x \xt \\
\dot{\Psih}   & = & e \hat{w}^{\Tr} \label{eq:Psihdot} \,,
\end{eqnarray}
\end{subequations}
Define the exosystem and parameter estimation errors:
$\wt := \hat{w} - w + G e$ and $\Psit := \Psih - \Psi$. In terms of these errors we have 
\begin{subequations}
\label{eq:closedloop2}
\begin{eqnarray}
\dot{e}   & = & - K e + \alpha_x \xt - \Psi \wt - \Psit \hat{w} \\
\dot{\wt} & = & F \wt - H e + \alpha_x G \xt \\ 
\dot{\xt} & = & - K_x \xt \,,
\end{eqnarray}
\end{subequations}
where $K:= K_x - \alpha_x + K_e - \Psi G$ and $H := F G + G \Kxt$. 
Suppose that $\Psit = 0$ in \eqref{eq:closedloop2}, and let $\xit := ( \wt, \xt )$. 
Then \eqref{eq:closedloop2} becomes 
\begin{subequations}
\label{eq:closedloop3}
\begin{eqnarray}
\dot{e}    & = & - K e + \widetilde{G} \xit \\
\dot{\xit} & = & \widetilde{F} \xit + \widetilde{H} e 
\end{eqnarray}
\end{subequations}
where $\widetilde{F} = \begin{bmatrix} F & \alpha_x G \\ 0 & - K_x \end{bmatrix}$, 
$\widetilde{G} = \begin{bmatrix} -\Psi & \alpha_x \\ \end{bmatrix}$, and 
$\widetilde{H} = \begin{bmatrix} -H \\ 0 \end{bmatrix}$. 
By assumption $F$ is Hurwitz and $K_x > 0$, so $\widetilde{F}$ is Hurwitz. Given any $\gamma > 0$, 
there exists a symmetric, positive definite matrix $P \in \RR^{(q+1) \times (q+1)}$ such that 
$P \widetilde{F} + \widetilde{F}^{\Tr} P = - \gamma I$. 
Define the Lyapunov function for the system \eqref{eq:closedloop3}:
\[
V := \| e \|^2 + {\xit}^{\Tr} P \xit \,.
\]
Then along solutions of \eqref{eq:closedloop3}, we have
\begin{eqnarray*}
\dot{V} & =  & - 2 K \| e \|^2 + 2 e \widetilde{G} \xit + 2 \xit^{\Tr} P \widetilde{H} e - \gamma \| \xit \|^2 \\  
        & =  & \begin{bmatrix} e^{\Tr} & \xit^{\Tr} \end{bmatrix}
               \begin{bmatrix} - 2 K & \widetilde{H}^{\Tr} P + \widetilde{G} \\
                                P \widetilde{H} + \widetilde{G}^{\Tr}  & - \gamma I 
               \end{bmatrix} \begin{bmatrix} e \\ \xit \end{bmatrix} \\
        & =: & \begin{bmatrix} e^{\Tr} & {\xit}^{\Tr} \end{bmatrix} \widetilde{Q} \begin{bmatrix} e \\ \xit 
               \end{bmatrix} \,.
\end{eqnarray*}
Since the unknown parameters $(\Kxt, \Psi^{\Tr})$ belong to a compact set $\cP$, the off-diagonal 
elements of $\widetilde{Q}$ are bounded. Then by a standard argument we can choose $K > 0$ sufficiently large 
(by choosing $K_e > 0$ sufficiently large) such that $\widetilde{Q}$ is negative definite for all 
$(\Kxt, \Psi^{\Tr}) \in \cP$.

Now consider \eqref{eq:closedloop2} with $\Psit \neq 0$ and define the Lyapunov function 
\[
V_{\Psi} := V + \Psit \Psit^{\Tr} \,.
\]
Let $\dot{V}_{\eqref{eq:closedloop3}}$ denote the Lie derivative of $V$ along solutions of \eqref{eq:closedloop3} 
(with $\Psit = 0$). Evaluating the derivative of $V_{\Psi}$ along solutions of \eqref{eq:closedloop2} and invoking 
\eqref{eq:Psihdot}, we obtain 
\[
\dot{V}_{\Psi} = \dot{V}_{\eqref{eq:closedloop3}} - 2 e \Psit \hat{w} + 2 \Psit \dot{\Psit}^{\Tr} 
               = \dot{V}_{\eqref{eq:closedloop3}} \,,   
\]
which is again negative definite at $(e,\wt,\xt) = (0,0,0)$. Finally, applying the LaSalle Invariance Principle, 
we obtain that $\lim_{t \rightarrow \infty} e(t) = 0$, as required.

\end{appendix}
\end{document}